\documentclass[fleqn,usenatbib]{mnras}

\usepackage[T1]{fontenc}

\DeclareRobustCommand{\VAN}[3]{#2}
\let\VANthebibliography\thebibliography
\def\thebibliography{\DeclareRobustCommand{\VAN}[3]{##3}\VANthebibliography}

\usepackage{graphicx}	% Including figure files
\usepackage{amsmath}	% Advanced maths commands
\usepackage{amssymb}	% Extra maths symbols
\usepackage[normalem]{ulem}
\usepackage[dvipsnames]{xcolor}

\usepackage{xspace}
\usepackage{enumitem}% http://ctan.org/pkg/enumitem
\newcommand{\sage}{{\sc SAGE}\xspace}
\newcommand{\sagesh}{{\sc SAGE}$_{\rm sh}$\xspace}
\newcommand{\unit}{{\sc UNIT}\xspace}

\newcommand{\ha}{H$_{\alpha}$\xspace}
\newcommand{\OII}{$\left[\mathrm{O\,\textrm{\sc{ii}}}\right]$\xspace}

\usepackage{newtxtext,newtxmath}
\title[HOD for \ha ELGs]{An improved Halo Occupation Distribution prescription from UNITsim \ha Emission Line Galaxies: conformity and modified radial profile}

\author[Reyes-Peraza et al.]{Guillermo Reyes-Peraza $^{1,2}$\thanks{guillermo.reyes@uam.es}, 
Santiago Avila  $^{1,2,3}$\thanks{savila@ifae.es},
Violeta Gonzalez-Perez $^{2,4}$\thanks{violetagp@protonmail.com}, 
Daniel Lopez-Cano $^{2,5}$,
\newauthor
Alexander Knebe $^{2,4,6}$,
Sujatha Ramakrishnan$^{2,4}$,
Gustavo Yepes $^{2,4}$
\\
$^{1}$ Instituto de F\'isica Teorica UAM-CSIC, c/ Nicolás Cabrera 13-15, , 28049 Madrid \\
$^{2}$ Departamento de F\'isica Te\'orica,  Facultad de Ciencias M-8,   Universidad Aut\'onoma de Madrid, 28049 Madrid (Spain) \\
$^{3}$ Institut de física d’altes energies (IFAE), The Barcelona Institute of Science and Technology campus UAB, 08193 Bellaterra Barcelona, Spain \\
$^{4}$Centro de Investigaci\'{o}n Avanzada en F\'isica Fundamental (CIAFF), Facultad de Ciencias, Universidad Aut\'{o}noma de Madrid, 28049 Madrid, Spain \\
$^{5}$ Donostia International Physics Center (DIPC), Paseo Manuel de Lardizabal, 4, 20018 Donostia-San Sebastián, Spain \\
$^{6}$International Centre for Radio Astronomy Research, University of Western Australia, 35 Stirling Highway, Crawley, Western Australia 6009, Australia\\
}

\date{Accepted XXX. Received YYY; in original form ZZZ}

\pubyear{2015}

\begin{document}
\label{firstpage}
\pagerange{\pageref{firstpage}--\pageref{lastpage}}
\maketitle

\begin{abstract}
Emission line galaxies (ELGs) are targeted by the new generation of spectroscopic surveys to make unprecedented measurements in cosmology from their distribution.
Accurately interpreting this data requires understanding the imprints imposed by the physics of galaxy formation and evolution on galaxy clustering.
In this work we utilize a semi-analytical model of galaxy formation (\sage) to explore the necessary components for accurately reproducing the clustering of ELGs. We focus on developing a Halo Occupation Distribution (HOD) prescription able to reproduce the clustering of \sage galaxies. Typically, HOD models assume that satellite and central galaxies of a given type are independent events. We investigate the need for conformity, i.e. whether the average satellite occupation depends on the existence of a central galaxy of a given type. Incorporating conformity into HOD models is crucial for reproducing the clustering in the reference galaxy sample. Another aspect we investigate is the radial distribution of satellite galaxies within haloes. The traditional density profile models, NFW and Einasto profiles, fail to accurately replicate the small-scale clustering measured for \sage satellite galaxies. To overcome this limitation, we propose a generalization of the NFW profile, thereby enhancing our understanding of galaxy clustering.
\end{abstract}

\begin{keywords}
methods: numerical – galaxies: formation – cosmology: large-scale structure of
Universe
\end{keywords}

%%%%%%%%%%%%%%%%%%%%%%%%%%%%%%%%%%%%%%%%%%%%%%%%%%

%%%%%%%%%%%%%%%%% BODY OF PAPER %%%%%%%%%%%%%%%%%%
%Comentar en color 
\section{INTRODUCTION}

The study of the large-scale structure in the Universe plays a crucial role in contemporary cosmology. It serves as a powerful tool for understanding the fundamental properties of our Universe, including the values of different cosmological parameters and the nature of dark matter and dark energy. Galaxy clustering is an observable that allows us to probe both the cosmological parameters and the physics of galaxy formation. 

Over the past few decades, numerous %projects 
endeavours have been dedicated to creating large galaxy maps, such as the 2dFGRS \citep{2dFGRS}, SDSS \citep{York,Eisenstein}, WiggleZ \citep{Drinkwater,Parkinson}, BOSS \citep{Dawson2012,Alam2017}, eBOSS \citep{Dawson2016,eBoss}, and DES \citep{DES,Abbott2018b}. These endeavors have enhanced our understanding of the Universe, by providing measurements of its expansion history and increasing the constrains on dark energy and alternative theories of gravity. These cosmological surveys have provided the community with large 3-D maps of galaxies up to $z\sim1$, allowing for a better understanding of the evolution of galaxies. Despite significant progress, these areas of investigation remain open and ongoing.

New generation surveys, including Euclid \citep{Laureijs,Amendola}, Dark Energy Spectroscopic Instrument \citep[DESI,][]{DESI2016}, the Nancy Grace Roman Space Telescope \citep[from now Roman,][]{Spergel2013,Spergel2015} and the 4-metre Multi-Object Spectroscopic Telescope \citep[4MOST,][]{DeJong}, are set to map galaxies beyond $z\sim1$ and to increase the density of QSOs at higher redshifts. 
%These missions will provide the scientific community with more comprehensive, deeper, and more accurate data, leading to stronger constraints on theoretical models and improved estimates of cosmological parameters.
One of the main tracers beyond $z\sim1$ are galaxies with strong spectral emission lines or emission line galaxies (ELGs). In particular, Euclid utilizes near-infrared grisms to observe galaxies with a strong \ha spectral line. eBOSS and DESI target ELGs with a strong \OII emission. The properties of \ha and \OII emitters can be slightly different, due to a different number of lines coming from narrow line regions in AGNs and shocks \citep[e.g.][]{favole2023}. Nevertheless, these galaxies are dominantly star-forming galaxies and results for one type will be broadly adequate for the other.

%In the past (e.g. BOSS), galaxy clustering measurements have primarily targeted Luminous Red Galaxies (LRGs) at z $\approx <1$, the focus of these new surveys has shifted towards Emission Line Galaxies (ELGs) in order to observe galaxies at higher redshifts. 
From both hydrodynamical and semi-analytical models (SAMs) of galaxy formation, we expect the connection between galaxies selected by their star-formation rate and dark matter haloes to be more complex than %when stellar mass is used 
for stellar-mass selected samples \citep[e.g.][]{Zheng2005,Orsi_2018,VGP2020}. SAMs of galaxy formation and evolution populate  dark matter haloes at early cosmic times with gas and let them evolve with a set of coupled differential equations \citep[e.g.][]{cole00,Croton_2016,Hirschmann2016,cora2018,lagos2019}. These equations aim to encapsulate the physical processes that are understood to be relevant during the formation and evolution of galaxies, such as the cooling of gas, the formation of stars, the interplay between super massive black holes and the intergalactic medium, etc \citep[see the reviews on modelling galaxies by][]{Baugh2006,somerville15,Wechsler2018}. SAMs are faster than hydrodynamical simulations \citep[e.g.][]{schaye2015,Bahamas,TNG_method_2,TNG_clustering}, as they are ran on dark matter only N-body simulations, at the expense of loosing the effect that baryons have over dark matter haloes \citep{baryons_schneider, baryons_arico2020}. 

In this work we use \sage, the SAM introduced in \citet{Croton_2016}, 
to explore how ELGs populate dark matter halos. SAMs have been previously used to produce Euclid and Roman-like \ha emitter catalogues that have been used to understand their clustering and relation to their host halos ~\citep{merson2018,zhai2019,zhai_hod,zhai2021, Knebe2022}.
%\SA, this reference did not make sense in this sentence:~\citep[e.g][]{mccarthy2023}.
%\SA{If we use SAGE, why do you talk (more) about galacticus?} \DLC{agree, maybe only cite a list of SAMs when introducing them}
%Galacticus \citep{Galacticus3} is a free and open-source semi-analytic model of galaxy formation. The Galacticus model was designed to be highly modular to facilitate expansion and the exploration of alternative descriptions of key physical ingredients. With Galacticus, they demonstrate results from an example model that is in reasonably good agreement with several observational datasets. %\vgp{aquí introducción a lo que ya se ha hecho sobre ELGs utilizando Galacticus (Andrew Benson es co-autor de los 3 o 4 artículos que hay, aunque creo que no lidera ninguno).} 
Here we use the \sage run on the \unit dark-matter only simulation \citep{2019Unitsims}, as described and released in the work by \citet{Knebe2022}. This simulation uses the fixed and paired technique to increase its effective volume, allowing more precise statistical analyses \citep{2016Angulo}. The results for \ha ELGs can be generalised to galaxies selected using other spectral emission lines, so throughout this paper, we will refer to our target galaxies, simply as ELGs.

SAMs can produce model galaxies accounting for a large range of physical phenomena in simulation volumes comparable to those required to interpret observations from current cosmological surveys. Nevertheless, they are still too slow to be able to produce many realisations and they require the construction of merger trees tracing the evolution of halos in fine time slices, which is not available in many simulations due to computing/storing limitations. The Halo Occupation Distribution (HOD) model provides a simplified way to describe the relation between galaxies and haloes and can be constructed from a single halo catalogue. These simplifications makes HOD valid for the target observables they are constructed to match, but sometimes will show limitations when trying to extend it to other observables ~\citep[e.g.][]{chaves2023}.
%The simplifications make these models useful mostly for clustering analysis but biased results can arise if used for other observables such as lensing \SA{<-no me termina de convencer esta frase, ¿la quitamos de momento? No es 100\% necesaria.}.

HOD models are computationally efficient. They have demonstrated their capability to reproduce the observed clustering across a range of galaxies, including the dependency with luminosity and colour~\citep{zehavi2005,christodoulou2012,Carretero}.
In their basic form, HOD models assume that the mass of haloes determine the average number of galaxies they can host \citep[e.g.][]{benson2000,berlind2003,Zheng2005}. Catalogues of model galaxies produced with HOD models can be produced fast enough as to choose the best fit to a set of clustering observables using statistical techniques that require hundreds or thousands of realisations in large simulation volumes~\citep[e.g.][]{alam2021,Yuan2023,Rocher2023}. HOD models have become a popular tool for producing realistic mock catalogs in large simulation volumes to match the clustering of a target galaxy sample \citep[e.g., in BOSS, DES and eBOSS:][]{Manera,S.Avila2018,avila2020}.

The objective of this work is to construct an HOD model that accurately reproduces the galaxy clustering of the ELGs modelled by \sage. In particular, we explore if two of the usual assumptions made by HOD models should be challenged when compared with model ELGs. One of this usual assumptions is that central and satellite galaxies of a certain type are independent events within a halo, a lack of {\it conformity}. The other is the modelling of the radial profile of satellite galaxies.

%Studies such as \citet{Lacerna} also examine galactic conformity and its effect on galaxy clustering using semi-analytic models. 
%\SA{Mira si de esto de abajo hay algo que se relevante relamente para motivar el estado de la cuestión, pero conciso ->}
%These studies consider the correlation between star formation in central galaxies and their neighboring galaxies at both small and large separations. However, the reproduction of this signal in semi-analytic models varies, and a greater understanding of the involved physical processes is required. 

One of the main focus of this paper is to study the effect that galactic conformity has on the clustering of ELGs. The term  galactic conformity was first coined by \citet{Weinmann} to describe strong correlations between the properties of satellite galaxies and their central galaxies in data from the Sloan Digital Sky Survey (SDSS). A strong 1-halo conformity for ELGs has recently been suggested from the comparison of DESI data with model galaxies produced with different HOD models \citep{DESIonehaloterm, desi_conformity}. Overall, galactic conformity remains an active research topic and a significant puzzle in the formation and evolution of galaxies.

%It is worth noting that in standard \textit{vanilla} HOD models, the halo occupation function does not assume the phenomenon of galactic conformity. That is, central galaxies and satellite galaxies are assumed to be independent events, with no connection between them. In this article, we will show that 1-halo conformity is a necessary ingredient to reproduce small scale ELG clustering. 
%introduce the concept of one-halo conformity, which is essential to characterize the behavior of the \sage satellite galaxy sample and demonstrate its effect on clustering at small scales.

The second main focus of this work is to explore whether the typical assumptions made by HOD models for the radial profiles of satellite galaxies are upheld. Commonly, HOD models assume that the distribution of satellite galaxies within the dark matter halo follows the distribution of the dark matter itself, or often an NFW \citep{NFW} or an Einasto \citep{Einasto} profile. We will try different ways to sample these types of profiles and also propose an extension.
%and find them to fail to reproduce the ELG clustering predicted by \sage. Hence, we will propose an extended NFW profile, which will be shown to improve significantly the ELG small scale clustering.

%\vgp{haloes -> haloes throughout the paper}
%\vgp{I would not state here the conclusions}

%This assumption involves applying the NFW profile associated with $R_{\rm s}$, internal property that is related to the concentration of its corresponding halo, to each satellite galaxy. In other words, sampling an NFW profile using the concentration provided by each halo. In other words, we sample an NFW profile using the concentration provided by each halo.

%\SA{Reducir y comentar por encima. Aquí no tienes que hablar de resulltados, si no motivar qué vamos a estudiar y por qué consideramos que era necesario estudiar esto. ->}
%Furthermore, we also fit the radial profile of our entire sample of SAGE satellite galaxies to classical density profiles, such as NFW and Einasto \citep{Einasto}. However, in terms of clustering, sampling an NFW profile using the concentration given by each halo and fitting classical density profiles (Einasto and NFW) do not accurately reproduce the small-scale clustering of SAGE satellite galaxies. Consequently, we adopt an analytical expression to reproduce the radial distribution profile of satellite galaxies. This approach yields clustering results that are very similar and can be interpreted as an extension of the NFW profile. It can be easily implemented in future works using different simulations.

The plan of this paper is as follows. In section (\S~\ref{sec:UnitSims}) we introduce UNIT DM simulation. In section (\S~\ref{sec:sage}) we describe the model galaxy sample of reference. First (\S~\ref{subsec:euclidnd}), we introduce UNITsim-\sage, how our reference galaxy sample is built and we describe the average halo occupation distribution that we find. We then describe in (\S~\ref{subsec:shuffle}) how we build the final sample of reference by applying the {\it shuffling} method to the previous model sample. In section (\S~\ref{sec:HODm}), we detail the various halo occupation models and properties that we employ to generate mock catalogues. Section (\S~\ref{sec:conformity}) discusses conformity, the method we employ to model it and its effect in galaxy clustering. Section  (\S~\ref{sec:radialprofile}) is devoted to studying the radial profile of \sage and how to reproduce it in order to recover an accurate galaxy clustering. Finally, in Section (\S~\ref{sec:conclusion}), we conclude and discuss our findings and future prospects.

%It is thought that these measurements could be especially relevant in Redshift Space Distortions (RSD) measurements.For this reason, working groups need to carry out realistic studies based on simulations beforehand.\\

%\section{Input simulations and models}
%In this section we discuss the parameters and specifications of the simulations and the Galaxy formation model, SAGE, that we have employed.\\

\section{UNIT DM simulation} \label{sec:UnitSims}
The \unit suite is a set of N-body cosmological simulations \citep{2019Unitsims} designed to cover the expected halo masses for emission line galaxies (ELGs), in particular [OII] emitters targeted by DESI \citep[$\sim10^{11} h^{-1}{\rm M}_{\odot}$][]{VGP2020} and Euclid ELGs \citep[$\sim4\times10^{11} h^{-1}{\rm M}_{\odot}$][]{Cochrane}. %The cosmic variance in the \unit simulations is reduced with the Fix and Pair method \citep{2016Angulo} 
To reduce the cosmic variance in the \unit simulations, we employ the Fix and Pair method \citep{2016Angulo}. Following this method, two pairs of simulations are generated with initial conditions with a power spectrum with an amplitude fixed to the %average of the set. 
expected value, given by the input power spectrum. 
The initial conditions of 
one of the simulations from each pair is set with the opposite phases than the other one from the pair \citep{2019Unitsims}. These simulations were generated using \textsc{GADGET2} \citep{Gat}, which fully solves the gravitational evolution of the continuous distribution of dark matter. The phase-space halo finder  \textsc{ROCKSTAR} \citep{Bezi} was used to identify haloes from the $129$ existing snapshots. Subsequently, the \textsc{ConsistentTrees} software \citep{Bezi} was employed to calculate their merging histories. The software tracks dark matter halos (regions of the universe where gravity has made matter collapse into denser structures) as they evolve over time and move in space, constructing a \textsc{\it tree}-structure that illustrates how these halos merge together to form larger halos. These merger trees are 'consistent' in the sense that the mass and position of a halo at a given moment in the tree are informed by the mass and position of its progenitors and its descendants.

For this work we use simulations with boxes of side $1 h^{-1} \text{Gpc}$ and $4096^{3}$ dark matter particles. The cosmological parameters of these simulations are: $\Omega_{m,0} = 1 - \Omega_{\Lambda,0} = 0.3089, \hspace{0.25cm} h = 0.6774, \hspace{0.25cm} n_{\rm s} = 0.9667, \hspace{0.25cm} \sigma_{8} = 0.8147$. 
%SA: por que ya no se mencionan los merger trees?

We perform our analysis over the simulation snapshot at $z = 1.321$, which approximately corresponds to the mean redshift expected for Euclid ELGs, $0.9 < z < 1.8$. 
%In our catalogues, haloes with 100 particles have masses greater than $1.2 \times 10^{11} h^{-1} {\rm M}_{\odot}$. \SA: frase cortada?
%\grp{Consistencia en las unidades /h o h-1}

\section{UNITsim-sage: Reference galaxy sample}\label{sec:sage}

We aim to provide an improved Halo Occupation Distribution model for ELGs, focusing on conformity and satellite radial profiles. Our reference catalogue of galaxies is generated from the UNITsim-Galaxies catalogues described in \citet{Knebe2022} and released along it \footnote{\url{http://www.unit sims.org}}. This catalogue is based on the \sage semi-analytical model of galaxy formation and evolution \citep{Croton_2016} by populating the dark-matter-only UNITsim simulations (\autoref{sec:UnitSims}). 
For each model galaxy, spectral emission lines associated with star formation events are obtained with the \textsc{get\_emlines} model described in \citet{orsi_2014}. One of these spectral lines is the \ha, targeted by Euclid and Roman. The spectral emission lines are then attenuated by dust using a \citet{Cardelli1989} law following the code developed in \citet{favole2020}. 
All these models were implemented in the catalogue in \citet{Knebe2022} and we refer the reader to that publication for a more detailed description.

In order to single out the modelling of both conformity and the satellite radial profiles, we use a shuffled sample, \sagesh (\S~\ref{subsec:shuffle}). To construct this sample, we remove the galaxy assembly bias: the effect that other properties beyond halo mass have on the distribution of galaxies within haloes.

%\subsection{Model ELGs with an Euclid-like number density}
\subsection{Model ELGs with an Euclid-like number density}
\label{subsec:euclidnd}
%%%%%%%%%%%%%%%%%%%%%%%%%%%%
\begin{table}
\begin{center}

\begin{tabular}{ c c c c |}
Sample & $\frac{d^2N(z)}{dzdA}$ & Number density & Flux cutoff  \\ 
 &  {\scriptsize $ \rm Galaxies\hspace{1mm}deg^{-2}$ } & $\rm Mpc^{-3} h^{3}$ & $\rm ergs\rm s^{-1}\rm cm^{-2}$  \\ \hline
Pozzetti nº1 & $4377$ & $1.299\times10^{-3}$ & $1.041\times10^{-16}$ \\
{\bf Pozzetti nº3} & $\mathbf{2268}$ & $\mathbf{6.731\times10^{-4}}$ & $\mathbf{1.325\times10^{-16}}$  \\ 
Flux cut  & $580.7$ & $1.723\times10^{-4}$ & $2\times10^{-16}$ \\ \hline
\end{tabular}
\caption{ 
Properties of the three model galaxy catalogues described in \autoref{sec:sage}. The first column on the left shows the name given to the samples. In the second column we quote the galaxies per square degree per redshift interval from \citet{Pozzetti} for the first two samples and the one we obtain applying the \ha flux Euclid-expected limit to the \sage sample. The third column contains the corresponding numerical volume densities ($n$). The fourth column shows the \ha flux limit needed to recover the target $n$ for the \citeauthor{Pozzetti} models and the flux limit expected for Euclid. Our reference model from \autoref{subsec:shuffle} onwards is {\bf Pozzetti nº3} (or the {\it shuffled} version of it, see \autoref{subsec:shuffle} ), which will be referred to as \sage (or \sagesh, when {\it shuffled}).}
\label{tab:nd_models}
\end{center}
\end{table}
%%%%%%%%%%%%%%%%%%%%%%

In order to produce catalogues of ELGs, we aim to reproduce the mean number density forecasted by \cite{Pozzetti} over the entire redshift range observed by Euclid: $0.9 < z < 1.8$.
For this, we use a simulation snapshot at $z= 1.321$, which approximately corresponds to the middle of that redshift range. 
This approach is slightly different to the one taken in \citet{Knebe2022}, where we matched the \citet{Pozzetti} differential luminosity function, at the redshift of the snapshot. 

\citet{Pozzetti} provides a range of forecasts, with the extreme number densities being $n_{1} = 1.299\times10^{-3}$ $\rm Mpc^{-3} h^{3}$ and $n_{3} = 6.731\times10^{-4}$ $\rm Mpc^{-3} h^{3}$, which are tabulated in the first column of \autoref{tab:nd_models}. These volume number densities, $n$, are obtained from the angular ones, $\eta = \frac{d^2N(z)}{dzdA}$, provided by \citet{Pozzetti} (third column in \autoref{tab:nd_models}), as follows:
\begin{equation}
n = \frac{\eta \hspace{1mm} 180^{2}}{\pi ^{2} \hspace{1mm} D_{\rm c}^{2}(z) \hspace{1mm} \frac{{\rm d}D_{\rm c}(z)}{{\rm d}z}} \, ,
\label{eq:density}
\end{equation}
where $D_{\rm c}(z)$ is the comoving distance to the object at redshift $z$. We apply the flux cuts indicated in \autoref{tab:nd_models} to get the extreme number densities forcasted by \cite{Pozzetti}. These are well below $2 \times 10^{-16}{\rm ergs\, s^{-1}\,cm}^{-2}$, the expected \ha flux limit for Euclid. If we apply this cut to the UNITsim-Galaxies catalogue, we obtain a sample with a number density $1.723\times10^{-4}$ $\rm {Mpc^{-3}}$$ h^{3}$, far below those from \citeauthor{Pozzetti}'s models. 
The exact shape of the \sage \ha luminosity function is very sensitive to the dust attenuation modelling, in particular at the tails of the distribution, sampled by the flux limits. 
An alternative approach would be to fine tune the dust attenuation model until we match the \citet{Pozzetti} luminosity function as done in \citet{zhai2019}. However, this approach is not clear to be more physical than the one taken here, where we are following the approach described in \citet{Knebe2022}. A more physical alternative would be to fit simultaneously the dust attenuation parameters, the emission line modelling and the SAM to different available observations with luminosity functions from different lines or even clustering measurements. However, this is beyond the scope of this paper and deserves and stand-alone study. 

%%%%%%%%%%%%%%%%%%%%%%%%%%
\begin{figure}
    \centering
	\includegraphics[width=0.95\columnwidth]{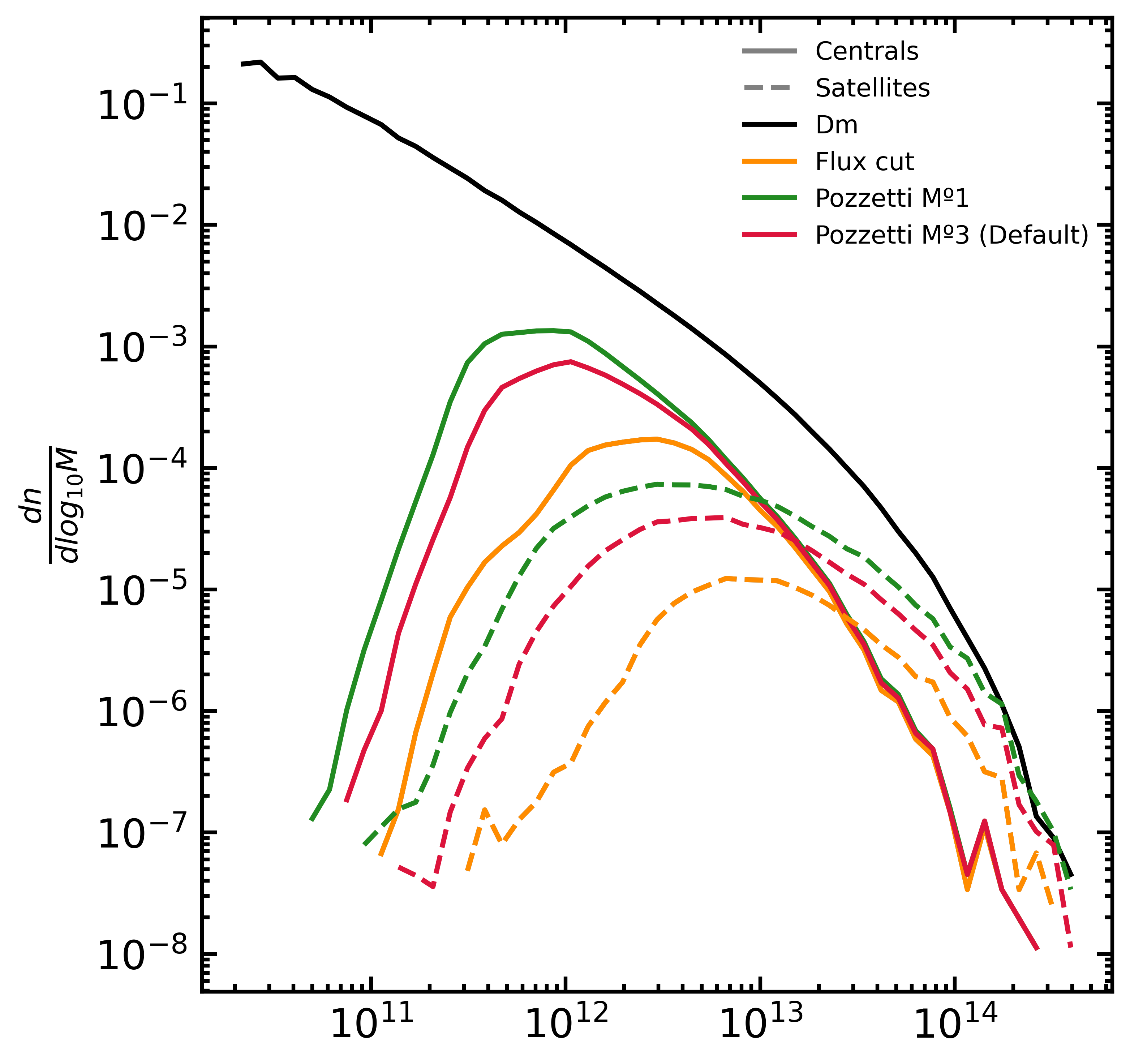}
	\includegraphics[width=0.95\columnwidth]{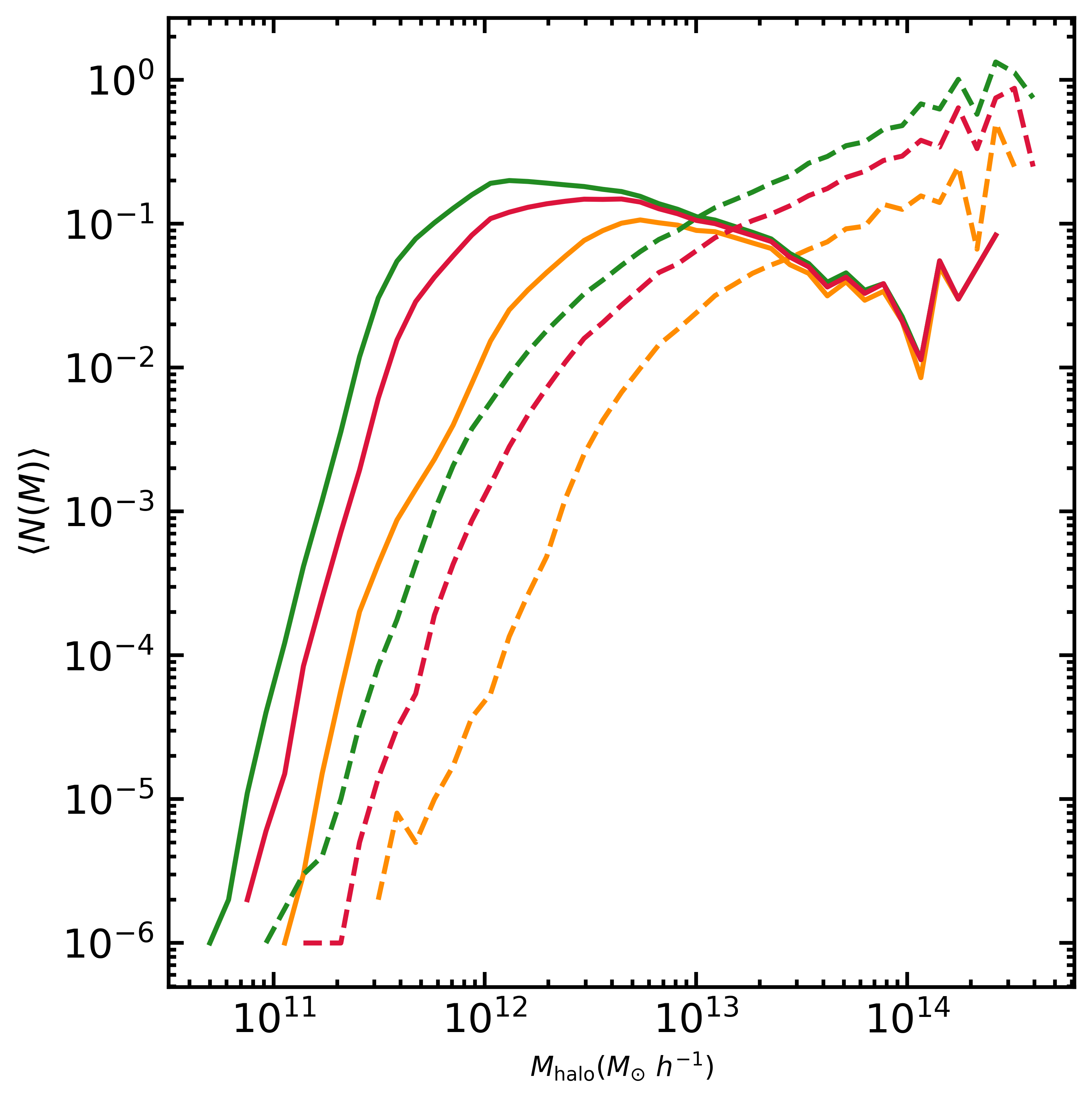}
    \caption{{\it Top:} The mass function of our catalogues of dark matter haloes (dark matter haloes , solid black \autoref{sec:UnitSims}) and of the different samples of ELGs considered: the flux-cut raw sample (gold) together with the optimistic (green) and reference (red) samples based on \citet{Pozzetti} luminosity functions (see \autoref{tab:nd_models}). The solid curves correspond to the central galaxies and the dashed curves to the satellite galaxies. The mass function corresponds to the density of haloes/galaxies for a given halo mass divided by the {\bf halo mass} bin size. To achieve this, histograms of objects in the halo/galaxy catalogues are created as a function of $M_{\rm 200c}$, and they are normalized by their volume ($1h^{-1} \text{Gpc}$) and the size of the corresponding log$M_{\rm halo}$ bin. {\it Bottom:} the mean halo occupation distribution as a function of halo mass of the haloes. This is compute as the ratio of the galaxy to halo mass functions.}
    \label{fig:HMF}
\end{figure}
%%%%%%%%%%%%%%%%%%%%%%%%%%%%%%%%%%%
In the upper panel of \autoref{fig:HMF}, we present the halo mass function of main dark matter haloes and galaxies for the three samples summarised in \autoref{tab:nd_models}. In this figure, we separate the contributions of central (solid curves) and satellite galaxies (dashed curves). The halo mass functions for dark matter haloes follow a typical shape close to a power law, with a change in the slope at the massive end. Central ELGs follow this trend above a mininimum halo mass. The numbers of both central and satellite galaxies decrease rapidly below a minimum halo mass. The number of ELGs is lower than that for haloes for all halo masses. As expected, satellite galaxies populate more massive haloes than centrals.

The average halo occupation distribution (HOD) for model ELGs selected with different number densities is shown in the lower panel of \autoref{fig:HMF}. Satellite ELGs follow the typical (truncated) power law seen for model satellite galaxies \citep[e.g.][]{Zheng2005,Avila22}.

The average HOD for central model galaxies do not reach unity, as it would do in a complete sample sample of galaxies without an particular type selection. This was expected from previous theoretical works on star-forming and ELGs \citep[e.g.][]{Zheng2005,Violeta2018, zhai_hod} and also found for DESI ELGs \citep{DESIonehaloterm}. This implies that we do not expect to find a \ha emitter in the center of every halo above a certain mass threshold. 
In \citet{avila2020}, a similar shape for the mean HOD of centrals galaxies, although with a steeper decay, was described as a asymmetric Gaussian. 

From this point onward, we use \citeauthor{Pozzetti}'s Model 3 by default, labelled as \sage. This model has been used as the baseline for several  Euclid studies, as it is expected to be the most realistic. 

\subsubsection*{Evolution with redshift}\label{subsec:evoz}
\begin{figure}
    \centering
    \includegraphics[width=0.95\columnwidth]{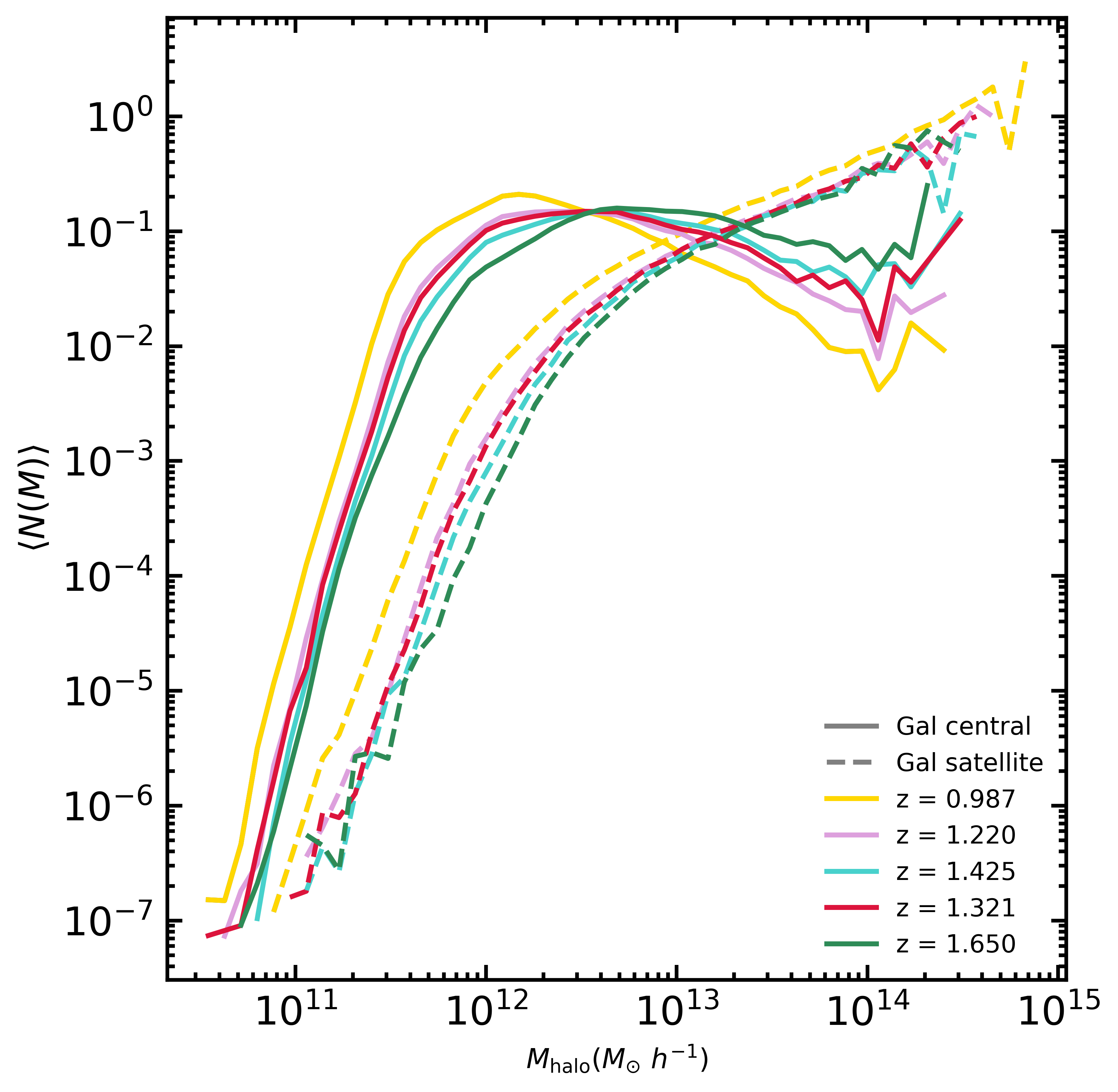}
    \caption{Mean halo occupation distribution for galaxies generated with HOD models at different redshift, as indicated in the legend. The number densities of this samples correspond to those from the diferential luminosity function number 3 from \citeauthor{Pozzetti} and are summarised in \autoref{tab:densityz}. The contribution from central galaxies is shown as solid lines and that from satellite galaxies as dashed ones.}
    \label{fig:HODz}
\end{figure}

\begin{table}
\begin{center}
\begin{tabular}{ c c c }
{\it dPozzetti nº3} & Number density & Flux cutoff \\ 
Redshift &  $\rm Mpc^{-3} h^{3}$ & $\rm erg\, \rm s^{-1}\rm cm^{-2}$ \\ \hline
$0.987$  & $1.381 \times10^{-3}$ & $1.88 \times10^{-16}$\\
$1.220$ &  $7.731 \times10^{-4}$ & $1.50 \times10^{-16}$\\
$1.425$ &  $5.185 \times10^{-4}$ & $1.22 \times10^{-16}$\\
$1.650$ &  $3.371 \times10^{-4}$ & $1.00 \times10^{-16}$\\ \hline
\end{tabular}
\caption{The columns in this table contain from left to right: the redshift of each sample; the number densities ($n$) derived from the differential luminosity function number 3 from \citeauthor{Pozzetti} at the given redshifts ({\it dPozzetti nº3}, with the {\it d} indicating the use of {\it differential} luminosity function); and the \ha flux cut used in the \sage to recover each $n$. Note that for the case of $z=1.3$ we consider a range in redshift, instead of the single redshifts considered here (\autoref{tab:nd_models}).}
\label{tab:densityz}
\end{center}
\end{table}

%In this section we discuss how the clustering changes  across the Euclid redshift range, 
%

We now describe how we also prepare the same galaxy sample for different redshifts within the range explored by Euclid, $0.9<z<1.8$. The goal is to study how the findings for our reference sample are not unique to the particular redshift selected above. 
For this, we follow the approach by \citet{Knebe2022}, where we considered 4 different snapshots in this range ($z=0.99$, $1.22$, $1.42$ and $1.65$) with a number density matched to the differential luminosity function number 3 from \citet{Pozzetti}.
The number density and flux cuts resulting at these four redshifts are listed in \autoref{tab:densityz}. 
Note that this procedure is somewhat different to our default $z=1.3$ analysis where we are targeting the average number density over the entire redshift range, according to \citeauthor{Pozzetti} model number 3 (\autoref{tab:nd_models}).

\autoref{tab:densityz} shows that the number densities increase by a factor of $4$ from the lowest at $z=1.65$ to the highest at $z=0.99$. The number density for the sample at $z=1.321$ (\autoref{tab:nd_models}), agrees with this trend, despite being calculated differently. The variation in number density is smaller than the factor of $7$ found among the samples at $z=1.3$ described by \autoref{tab:nd_models}. 

The average halo occupation distribution (HOD) for ELGs selected with our Pozzetti Model 3 at different redshifts shown in the \autoref{fig:HODz}, showing smaller variations of the mean HOD than those seen in \autoref{fig:HMF} at fixed redshift. The typical mass of haloes hosting ELGs increases slightly with redshift, as expected by having selected brighter (in \ha) galaxies to obtain the lower number density predicted by the luminosity function. In these lines, the largest difference in mass is observed for $z=0.987$, which also shows the larges difference in number density.

\subsection{\texorpdfstring{\sagesh}{SAGEsh}: the shuffled reference galaxy sample}
\label{subsec:shuffle}
At first order, galaxy clustering at large scales only depends on the mass of the host haloes. However, galaxy clustering can be affected by other halo properties, what is usually known as assembly bias. Assembly bias has been measured from simulated haloes and galaxies in different ways \citep[e.g.][]{Natureassembly,Cosmologicalassemblybias,jimenez2021}. %.\vgp{las listas de referencias tienen que estar en orden cronológico}. 
Moreover, there is compelling observational evidence of galaxy assembly bias \citep{2020JCAP...10..058O,2021MNRAS.502.3582Y,2022MNRAS.516.4003W,2023arXiv230501266A}. 

Properties encapsulating the environment of haloes have been found to be the most relevant secondary property \citep[e.g.][]{GAB}. However, in previous articles, one can observe how other internal properties of the halo, such as $V_{\rm max}$ (which is the maximum circular velocity of the halo) or the haloes own concentration, could explain at least part of the assembly bias. However, we attempted to construct an HOD prescription based on $V_{\rm max}$ instead of $M_{\rm halo}$ ({\it \`a la} \autoref{fig:HMF}) and we found that it captures less clustering than the $M_{\rm halo}$ based HOD.
The way to quantify the effect was 
to check the goodness in the recovery of the 2-halo term (large scales) clustering of \sage\ (see discussion of \autoref{fig:vanilla2PCF} below).

This work is focused on the exploration of both conformity and the satellite radial profile. To isolate the effect that varying these properties have on clustering we build a reference galaxy sample for which the effect of assembly bias is removed, \sagesh. 
We can achieve this by {\it shuffling} all galaxies within a given halo to the position of another halo of similar mass. This technique was first proposed in \citet{Shuffled}. In this proccess, the relative positions (and velocities) of galaxies to the center of a halo as well as internal properties of galaxies are preserved. It is worth noting that the bin size for shuffling coincides with the same binning used to calculate the mean galaxy occupation value (\autoref{fig:HMF}), here the range of the mass is 10.5 < $\rm log_{10}(M_{\rm 200c}/M_{\odot}h^{-1})$ < 14.5 with a binning of $\Delta$log$M$ = 0.057.

Throughout this paper we quantify the clustering by calculating the real-space two-point correlation function, $\xi$, with the publicly available code {\sc Cute}  \citep{Alonso2012}\footnote{https://github.com/damonge/CUTE} with logarithmic binning, where the number of bins is equal to 80, the bins in r per decade is $8$ and the $r_{\rm max}$ is $140 h^{-1}{\rm Mpc}$. 
The clustering of model ELGs before and after the shuffling process is shown in \autoref{fig:vanilla2PCF}. By construction these two agree for the 1-halo term contribution to the clustering (pairs of galaxies within the same halo). From $\sim 0.2{\rm Mpc}/h$, the clustering of the \sage and \sagesh ELGs start to differ, and stabilises at larger scales to a difference of about 10 per cent\footnote{This is the difference that was reduced but not eliminated when using a $V_{\rm max}$ based HOD, as we discussed at the beginning of this subsection}.  This difference in the scales dominated by the 2-halo term (pairs of galaxies from different haloes) is due to the history of formation and environments of dark matter haloes, the assembly bias. While for scales smaller than 0.22, the clustering is correctly repopulated, that is, the shuffling does not affect the 1-halo. Note that the shuffling procedure does not alter the two properties we are investigating here, the 1-halo conformity and the distribution of satellite galaxies within haloes as, by construction, the positions and properties of galaxies within haloes are preserved.

In what follows we attempt to reproduce clustering of the shuffled UNITsim-\sage \ha catalogue, that we label in short \sagesh.  
In \autoref{fig:vanilla2PCF} we can see our starting point, labeled as {\it Vanilla} HOD. In that figure, we can also see the model proposed ({\it default}, see \autoref{sec:DefaultHOD}) after the improvements developed in \autoref{sec:conformity} and \autoref{sec:radialprofile}. We see that the gain in recovering our reference clustering from \sagesh is very remarkable, with our {\it default} model closely following the clustering of \sagesh. We describe the main properties of these two models in the next section.

\begin{figure}
    \centering
	\includegraphics[width=1\columnwidth]{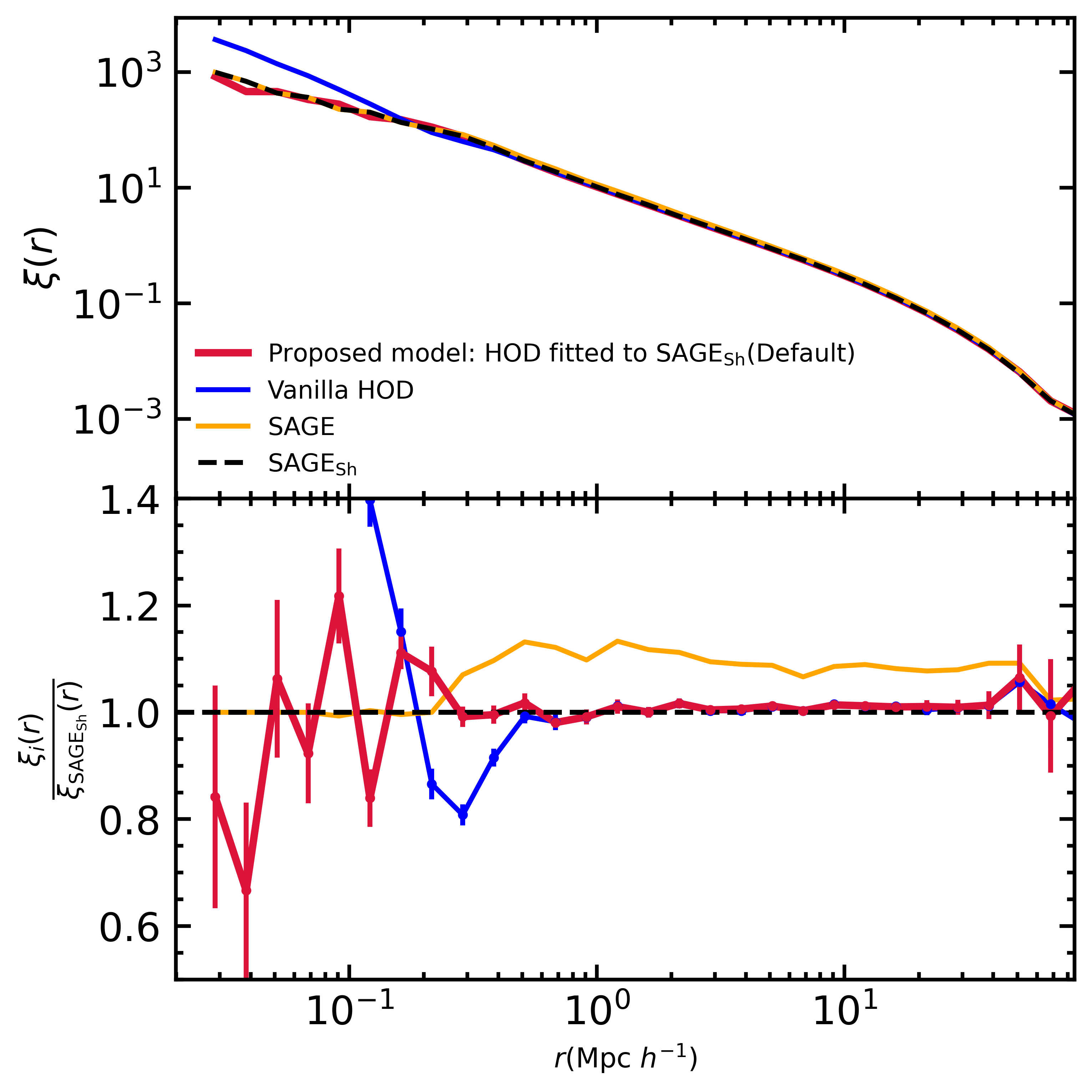}
    \caption{{\it Top:} Two-point correlation function in real space of different model galaxy catalogues: \sage (solid orange), shuffled \sage (\sagesh, dashed black) and two HOD catalogues fitted to \sage. The {\it Vanilla} HOD (\autoref{sec:VDefaultHOD}, solid blue) uses NFW radial profile and assumes that central and satellite \ha galaxies are independent events (no conformity). The {\it default} HOD uses the concept of conformity and the analytical $N(r)$ function implemented for the satellite galaxies fitted to \sage (\autoref{sec:DefaultHOD}, solid red).
    {\it Bottom:} Ratios with respect to \sagesh clustering. The error bars are calculated as the standard deviation of 100 realisations. While studying \sagesh ELGs, we have eliminated the assembly bias, to focus on other properties.}    
    \label{fig:vanilla2PCF}
\end{figure}

%%%%%%%%%%%%%%%%%%%%%%%%%%%%%%%%%%%%%%%%%%%%%%
%%%%%%%%%%%%%%%%%%%%%%%%%%%%%%%%%%%%%%%%%%%%%%
\section{Halo occupation distribution model}
\label{sec:HODm}
%%%%%%%%%%%%%%%%%%%%%%%%%%%%%%%%%%%%%%%%%%%%%%

In this section we summarise the halo occupation distribution (HOD) model we use to populate with \sagesh-like galaxies the \unit dark matter only simulation. 
%SA: I'd leave this out at the momment
%For this purpose we develop a code that can fit the level of 1-halo conformity and radial distribution of satellites in haloes of given mass, for any input catalogue of model galaxies. This code is publicly available at \grp{poner enlace}.

Halo Occupation Distribution (HOD) models are widely used to assign galaxies of different types to dark matter haloes \citep[e.g.][]{avila2020}. In their basic form, the mass of the halo determines the average number of galaxies that it hosts \citep[e.g.][]{benson2000,berlind2003,Zheng2005}. Central and satellite galaxies are modelled separately in HOD models. 
%In the computational implementation of HOD models first it is chosen
In order to implement an HOD galaxy assignment prescription, we first need to choose the shape of the mean HOD (\S~\ref{subsec:mean_hod}), then a probability distribution function determines if a halo contains or not a central galaxy and how many satellites (\S~\ref{sec:pdf}). The radial positions (\S~\ref{subsec:radial}) and velocities of satellite galaxies within haloes are chosen next. In this work, we have disregarded the velocities of satellites as our focus is solely on the two-point correlation function statistic in real space. 

%Throughout this work all the model galaxy catalogues are generated at $z = 1.321$ and have fixed  number density, $6.731\times10^{-4}$ $\rm Mpc^{-3} h^{3}$.
%, and linear bias, $1.86$. 
%These are what we refer to as the {\it ELG} samples.

We start with a {\it vanilla} HOD model (\S~\ref{sec:VDefaultHOD}), with assumptions commonly employed in the literature and propose a new model that better fits the characteristics of the model ELGs from \sagesh, our {\it default} HOD model (\S~\ref{sec:DefaultHOD}). The {\it default} model incorporates 1-halo conformity and utilizes the analytical function proposed in \autoref{subsec:analytical} for the radial profile of our satellite galaxy distribution. The clustering of mock galaxies produced with these two models, are compared to the \sage and \sagesh in \autoref{fig:vanilla2PCF}. This Figure shows how the {\it vanilla} HOD model departs from the expectation of \sagesh ELGs for the 1-halo term. 
We see that this model under-predicts the clustering by $\sim20\%$ at $0.3$Mpc$h^{-1}$ and over-predicts it by more than $50\%$ at $0.1$Mpc$h^{-1}$. On the other hand, the {\it default} model reproduces much better the galaxy clustering at small scales. 
%The differences surpasses a factor of $1.4$ by $0.1 {\rm Mpc}/h$. %\vgp{(40\% where??)with an error of over $40\%$ in the comparison.} 

%It is worth noting that the {\it vanilla} model and our Default model are consistently depicted in the figures throughout the entire article. The {\it vanilla} model is represented by light blue lines, while our Default model is represented by red lines. 

\subsection{Ingredients of our HOD models}
\label{sec:HOD_steps}

Below, we detail the different aspects that make up the HOD models we use for our analysis. We do not incorporate the velocity profile of the satellites, %$\phi(v_{r})$, <- removed by SA
in this study, as we focus only on the clustering in real space.

\subsubsection{Mean HOD shape}\label{subsec:mean_hod}

%\vgp{For this work we fixed the shape of the mean HOD for centrals and satellites following \cite{avila2020} HOD-3: y sigues con la explicación de Gaussiana asimétrica y power law (mira el texto que hay antes) o si no haces esto, explica las ecuaciones que utilizas, o ¿estás ajustando tb esto?}

For this work we need to know the expected number of central and satellite galaxies within a halo of mass $M$ is represented by $\langle N_{\rm cen}(M)\rangle$ and $\langle N_{\rm sat}(M)\rangle$, respectively. The \autoref{fig:HMF} (bottom panel) illustrates the outcomes obtained for the halo occupation function as a function of halo mass, noting again that we will focus in model 3.
%specifically within the context of the pessimistic model. 
As mentioned in \autoref{subsec:euclidnd}, to obtain the corresponding mean values, we first count the number of \ha galaxies of each type that we have in a series of mass intervals and determine the average occupation values by dividing by the number of haloes in each interval. These mass intervals range from 10.5 to 14.8 in terms of log(M). 

%SA: You have already discussed this (S 3.1). We need to decide where we want to dicuss the shape, but not in both places:
%We can observe how the shape of the mean value of central galaxies, $\langle N_{\rm cen}(M)\rangle$, resembles more a Gaussian function plus a step function or even a power law. Meanwhile, the predicted mean value of satellite galaxies, $\langle N_{\rm sat}(M)\rangle$, closely follows a power law above a minimum halo mass, which is typically an order of magnitude larger than the minimum halo mass required to host a central galaxy with the same selection. It is worth noting that this shape of the mean values we find in our \ha galaxy sample is similar to those found in the literature \citep{Violeta2018,Jimenez2019}. It is also important to mention that in this work, we will use this occupation law as observed for model nº 3 without modifications.
%\SA{Esta parte es más interesante. describe brevemente cómo calculas <N>. Describe un poco lo que ves en la figura. Di que la forma es parecida a otras en la literatura (Gonzalez-Perez 2018... )}

%Este comentario nunca lo implementaste?
%\SA{explica que en este trabajo, usaremos esa ley de ocupación tal cual se observa para el modelo 3.}
%Aqui creo que es más relevante que explique cómo implementas el HOD desde el halo catalogue. 

In this work, we do not fit a smooth curve to the measured mean HOD, but simply use it as measured. This is order to have something as close as possible to \sage\ and avoid having other sources of contribution to differences in the clustering and focusing on the main properties of study in this work (conformity and radial profiles). %When implementing our HOD, for each halo, we read its mass according to that assign to it an expected number of central galaxies $\langle N_{\rm cen}\rangle$ and a expected number of satellite galaxies $\langle N_{\rm sat}\rangle$. 
When applying our HOD model to a  halo catalogue, for each main halo, we read its mass, which will fall in one of the mass-bins defined in our HOD. We then assign to this halo a expected number of $\langle N_{\rm cen}\rangle$ central galaxies and $\langle N_{\rm sat}\rangle$ satellite galaxies, according to the mass bin.
In order to sample the mean HOD, for each main halo in our simulation
Nevertheless, the latter one ($\langle N_{\rm sat}\rangle$) will be modified by a factor ($K_1$ or $K_2$, see \autoref{sec:conformity}) whenever we are considering conformity.

\subsubsection{Conformity}

In this work we define the 1-halo conformity, or just conformity, as considering that central and satellite galaxies are not independent events for the HOD model. Conformity turns out to be of great relevance for recovering the target clustering of \sagesh ELGs. 

Reproducing the count of satellite \sagesh galaxies, both with and without a central galaxy of the same type, requires to include 1-halo conformity in our HOD models. In practice, this implies changing the $\langle N_{\rm sat}\rangle$, depending on whether we have assigned a central (\ha) galaxy or not. In \autoref{sec:conformity}, we explain in detail the concept of conformity and how we implement it. 

%%%%%%%%%%%%%%%%%%%%%%%
\begin{figure}
	\includegraphics[width=\columnwidth]{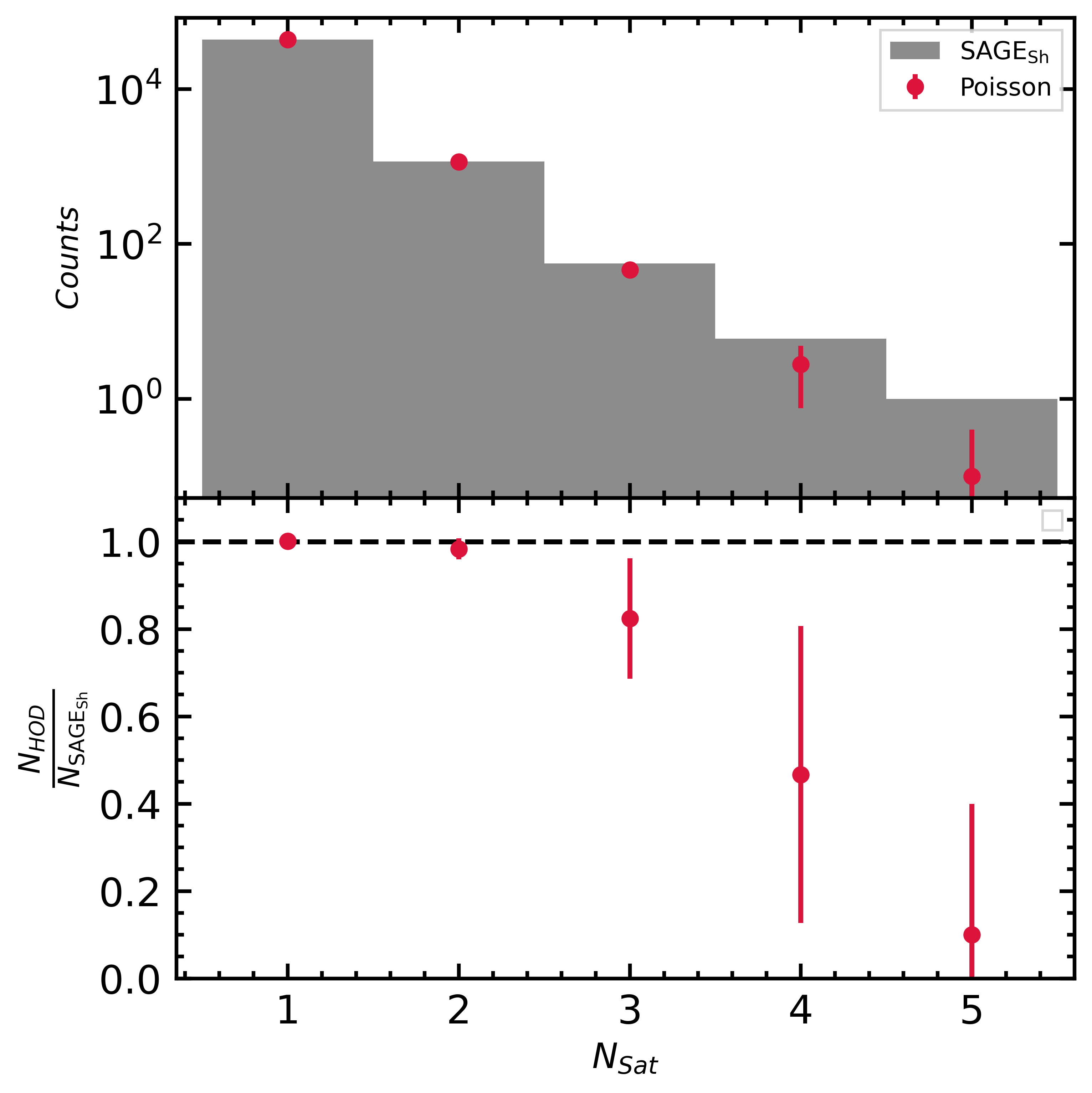}
    \caption{{\it Top:} The galaxy probability distribution function (PDF) as a function of the mean number of satellite galaxies, $N_{\rm sat}$. Results are shown for the \sagesh (black bars) model galaxies and those from our {\it default} HOD model (filled symbols). {\it Bottom:} Ratio between the number of model galaxies found in both catalogues. The error bars are calculated as the standard deviation of $100$ realisations.}

    \label{fig:Poisson}
\end{figure}
%%%%%%%%%%%%%%%%%%%%%%%
\subsubsection{Galaxy probability distribution function}\label{sec:pdf}

The probability distribution function (PDF) describes the sampling from the mean number of galaxies $\langle N_{\rm i}\rangle$ to a specific realization $N_{\rm i}$ in a given halo mass, ${P}(N_{\rm i}|\langle N_{\rm i}\rangle)$. In this work, we fix the PDF for both central and satellite galaxies. 

For central galaxies, the mean number of  galaxies, $N_{\rm cen}$, can only take the values of 0 or 1, following the Bernoulli distribution. 

For satellite galaxies, we assume the PDF of a Poisson distribution. This is the most common assumption for HOD models in the literature \citep[e.g.][]{Zheng2005,Rocher2023}. However, previous theoretical models have shown that the PDF of satellite ELGs might be better described by a non-Poissonian distribution. \citep{Jimenez2019} found model ELGs to be better described when a super-Poisson variance was assumed. Meanwhile, \citet{avila2020} and \citet{VosGines24} have found hints for eBOSS ELGs following either sub-Poisson  or supper-Poisson distribution, depending of the HOD model assumptions.

\sage satellite ELGs are well described by a Poisson distribution. %SA I found this unclear/unaccurate: We have reached this conclusion by comparing the number of satellites \SA{<-unclear} we obtain from our HOD models compared to those measured from the \sagesh catalogue. In \autoref{fig:Poisson} we compare these numbers as a function of the mean \SA{<-no} number of satellite galaxies, $N_{\rm sat}=i$ with $i$ being a natural number.
We have reached this conclusion by counting in \autoref{fig:Poisson} the number of times that a halo hosts a given number of satellite galaxies ($N_{\rm sat}$) in our HOD models compared to those measured from the \sagesh catalogue. 

By assuming a Poisson distribution, we are able to recover very well, within $1 \sigma$\footnote{For this comparisons, we assume the noise in counts follow a Poisson distribution.}, the number of haloes with no satellites, $N_{\rm sat}=0$ (not shown in \autoref{fig:Poisson}), and up to two satellites. 
For haloes with 3 or 4 satellites, we find slightly larger differences, but still within $1.5 \sigma$. 

\autoref{fig:Poisson} shows that larger differences appear for higher $N_{\rm sat}$ values, which are very rare.
In particular, haloes with $N_{\rm sat}= 5$ appear to be poorly described by a Poisson distribution. However, in the \sagesh catalogue we only find one halo with $N_{\rm sat}=5$. For our HOD galaxy catalogues we find no halo with $N_{\rm sat}=5$ in $90$\% of the cases run and a single halo in $10$\% of the cases. Thus, we consider the differences found for massive haloes to be statistically negligible. Even if they were not, we expect such small differences to have a negligible impact on the clustering, compared to the effect of assuming conformity or a different radial profile for satellite galaxies. 

Assuming a Poisson distribution is not expected  to affect the conclusions drawn in this paper.

%SA: I think it will be worth putting back the plot. 
%, but this
%This is due to the very limited number of haloes in our \sagesh sample with over three ELG satellites.%: 56 haloes \vgp{(better quote percentages) with 3 satellites and 6 haloes with 4 satellites, compared to 45 haloes with 3 satellites and 2 haloes with 4 satellites that we are able to reproduce.}

%By comparing the number of generated satellite galaxies with the number obtained from the \sage survey, we achieve a percentage match of $\sim 100\%$, specifically 46114 out of 46105 satellite galaxies. 

% \begin{table}[t]
% \begin{center}
% \begin{tabular}{ c c c c c c}
%  $N_{\rm sat}$& 1 & 2 & 3 & 4 & 5\\ \hline
% \sage & 43574 & 1167 & 56 & 6 & 1\\\hline
% Poisson  & 43624.46 & 1148.04 & 45.2 & 2.12 & 0.14\\
% $\%$ & 100.12 & 98.38 & 80.71 & 35.3  & 14 \\ 
% $\sigma$ & 234.62 & 30.56 & 6.58 & 1.41 & 0.35 \\\hline
% G.factor  &  43603.92 & 1171.7 &  46.79 & 2.57 & 0.12\\
% $\%$ & 100.07 & 100.40 & 83.55 & 42.83 & 11 \\ 
% $\sigma$ & 226.90 & 33.81 & 5.92 & 1.60 &  0.34\\\hline
% \end{tabular}
% \caption{}
% \label{tab:nd_models}
% \end{center}
% \end{table}
\subsubsection{Radial profile for satellite galaxies}
\label{subsec:radial}

While central galaxies are always located at the center of their host halo, a model is needed to place satellite galaxies within haloes. In this work we explore several different prescriptions for positioning satellite ELGs within their host haloes that will be discussed in detail in \autoref{sec:radialprofile}. The prescriptsions we have considered in this work are:

\begin{itemize}[wide, labelwidth=!,itemindent=!,labelindent=0pt, leftmargin=0em,parsep=0pt]
                \item Sampling individual halo profiles assuming an NFW given by the individual halo concentration, as detailed  in \autoref{sec:radialhalo}. 
                \item Adjusting the NFW or Einasto curves to the $r/R_{\rm s}$-stacked profile from all our \ha galaxies, as detailed in \autoref{subsec:r/Rs}.
                \item Adjusting the NFW or Einasto  curves to the $r/R_{\rm vir}$-stacked profile from all our \ha galaxies, as detailed in \autoref{subsec:r/Rvir}.
                \item The inherent $N(r)$ \sagesh distribution profile: We make use of the discrete (in a histogram) distribution profile of satellite galaxies in \sagesh as a function of the distance $r$ and sample from there.  
                \item {\bf Modified NFW profile}. We introduce a generalized version of the NFW density profile that effectively models the stacked profile of all our \ha galaxies as a function of $r$. We find an excellent fit with our modified NFW curve (\autoref{eq:NFWM}). In \autoref{subsec:analytical}, we describe in detail how we use this continuous function to accurately fit the distribution profile obtained from \sagesh. 
\end{itemize}
\subsection{{\it Vanilla} HOD model}
\label{sec:VDefaultHOD}

The benchmark model for this work, our {\it Vanilla} HOD, makes the following assumptions:

%\SA{This is not a standard HOD:}• We fix the mean HOD shape $\langle N_{\rm cen}\rm(M)\rangle$ and $\langle N_{\rm sat}\rm(M)\rangle$ to that of \sage galaxies selected as described in \autoref{sec:sage}.

\begin{itemize}[wide, labelwidth=!,itemindent=!,labelindent=0pt, leftmargin=0em,parsep=0pt]
    \item Mean HOD as described in \autoref{subsec:mean_hod} as read-off from our \sage-UNITsim catalogues. 
    %SA: This is was true: asymmetric Gaussian for centrals and truncated power law for satellite galaxies.
    \item Central and satellite \ha galaxies are independent events (no conformity).
    \item Poisson distribution for the satellite PDF.
    \item The radial profiles for satellite galaxies are obtained sampling a NFW profile given the concentration of each halo.
    %The standard approach used in the literature consists of applying the NFW profile associated with the concentration of its corresponding halo to each satellite galaxy. In other words, sampling a NFW profile using the concentration given by each halo.
\end{itemize}

Whereas the first point differs slightly from the most common practice, which would rely on assuming an analytical formula (see discussion in \autoref{subsec:mean_hod} and in \autoref{subsec:euclidnd}), the rest of points matches what is commonly assumed to generate HOD catalogues \citep[e.g.][]{avila2020}.
As previously noted, the {\it Vanilla} HOD model produces a clustering at small scales very different from that for \sagesh ELGs, as shown in \autoref{fig:vanilla2PCF}.

\subsection{Default HOD model}
\label{sec:DefaultHOD}

The {\it default} HOD model is our proposal to best fit the clustering of \sagesh ELGs (\S~\ref{sec:sage}). As a summary, these are the choices made for the {\it default} HOD model:

\begin{itemize}[wide, labelwidth=!,itemindent=!,labelindent=0pt, leftmargin=0em,parsep=0pt]
    \item Mean HOD as described in \autoref{subsec:mean_hod}: as read-off from our \sage-UNITsim catalogues.
    %Asymmetric Gaussian for centrals and truncated power law for satellite galaxies.
    \item We assume conformity, i.e. central and satellite ELGs are not modelled as independent events. We implement the conformity as a global factor independent of mass. This factor modifies the average random numbers of satellite galaxies, $\langle N_{\rm sat}\rangle$, depending on whether the central is another \ha\-selected galaxy or not, to match the numbers found in \sagesh.
    %, we achieve a notable improvement of galaxy clustering. satellite distribution in \sage as described in \autoref{sec:conformity}.
    \item Poisson distribution for the satellite PDF.
    \item Satellite galaxies are placed in haloes following an analytical function that encapsulates the number of model ELGs found as a function of their relative distance to the center of the halo. This function is presented in \autoref{subsec:analytical}, and can be considered as an extension of the NFW profile equation.
\end{itemize}

\autoref{fig:vanilla2PCF} shows the improvement achieved at small scales when using our proposed model, with respect to utilizing a {\it vanilla} HOD.

The HOD {\it default} model summarised in the previous paragraph was obtained from one of the 4 existing \sage-UNITsim boxes. The level of noise we find in the clustering when applying the {\it default} HOD model to the other simulation boxes, is similar to that found for the clustering of galaxies modelled by running \sage on different simulation boxes. Hence, we conclude that our {\it default}-HOD model works well for the other simulation boxes and that any noise in our inferred parameters would not affect our conclusions.

\begin{figure}
    \centering
    \includegraphics[width=0.95\columnwidth]{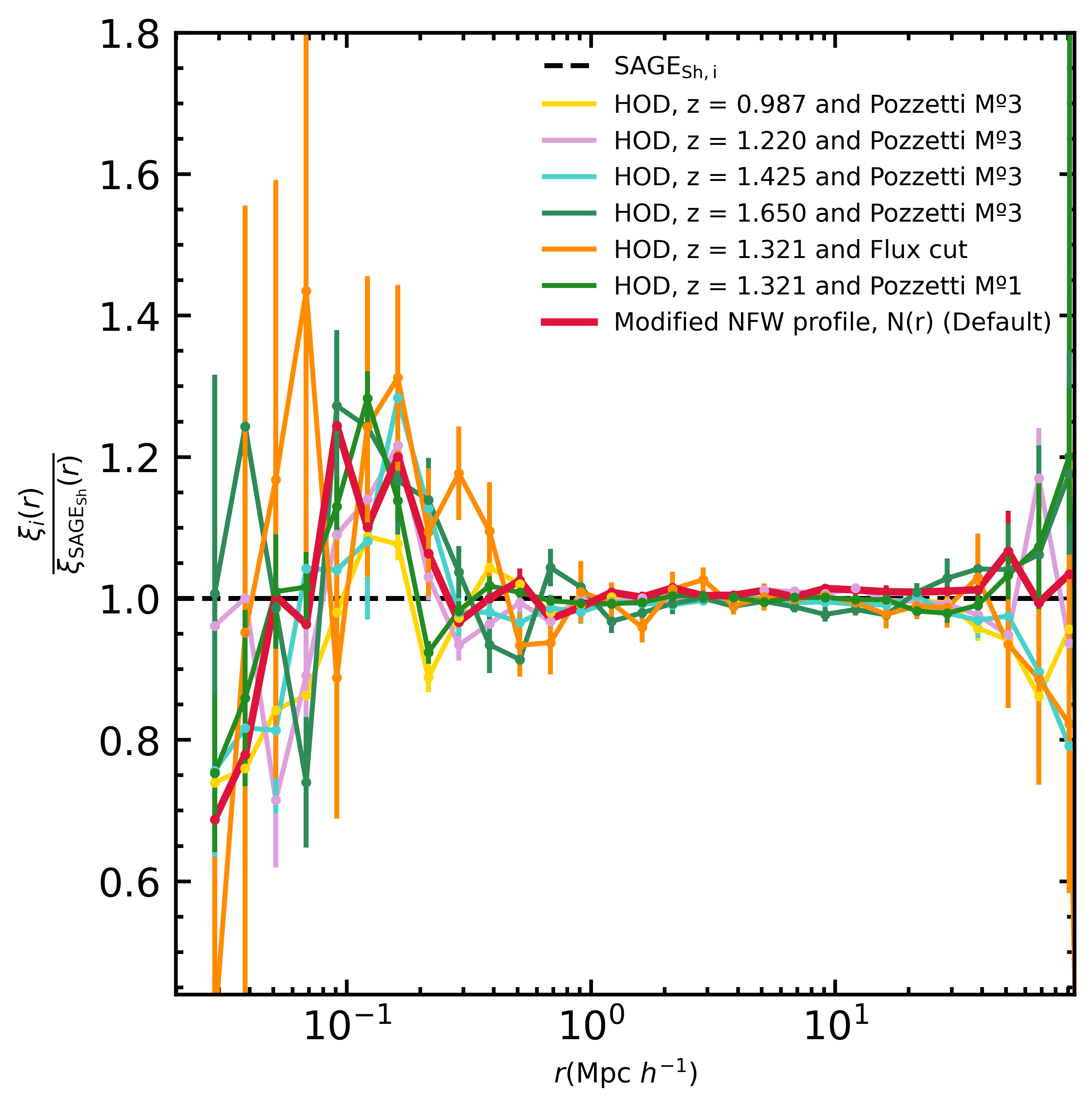}
    \caption{The ratio of two point correlation in real space for galaxies generated with our Default HOD models assuming different number densities and redshifts with the respective 2PCF of each \sagesh sample.}
    \label{fig:clusteringz}
\end{figure}

Our analysis is done, by default, for the number density corresponding to the Pozzetti model number 3 evaluated over the redshift range $0.9<z<1.8$, as indicated in \autoref{tab:nd_models}, and using the simulation snapshot corresponding to $z=1.3$. 
However, we also explore how our conclusions might change when a different number density or redshift is assumed when analysing the ELGs. For that, we construct alternative SAGE samples, as described in \autoref{subsec:euclidnd} and apply the same shuffling algorithm to them. We then apply the same steps to construct our default HOD as summarised in the bullet points above, and described in detail in the next sections. We then compare the clustering of the HOD-default model to that of \sagesh for alternative number densities or redshifts in \autoref{fig:clusteringz}. We find that the level of agreement between our default HOD to these alternative reference samples is as good as it was for our $z=1.3$ Pozzetti model 3 sample. We find sightly noisier results for these cases, but this is partly due to only using 20 random realisations, and also due to having larger error bars when having lower number densities.

\section{Modelling conformity}\label{sec:conformity}
\begin{figure}
	\includegraphics[width=\columnwidth]{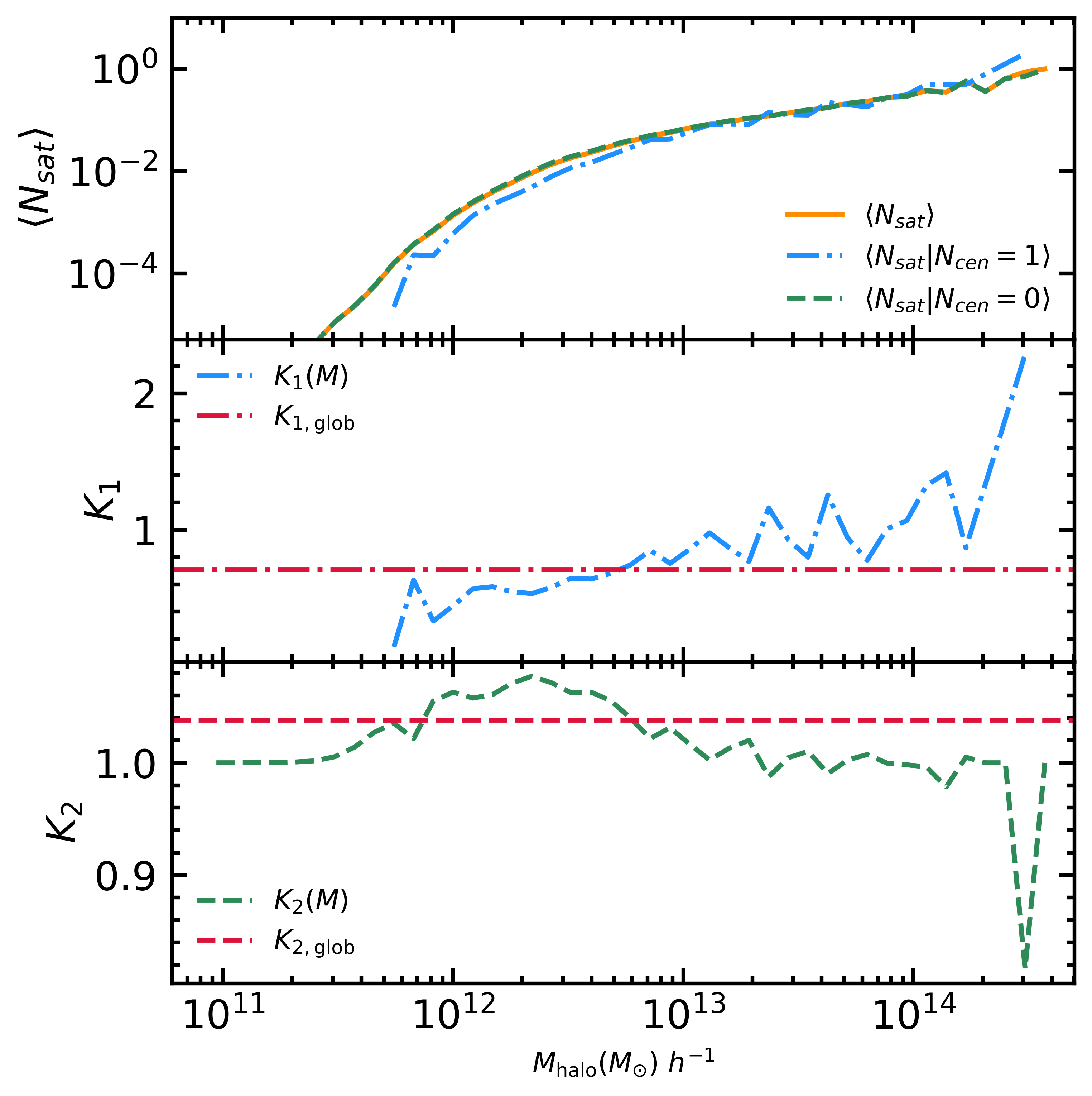}
    \caption{%\vgp{why the tick marks are different for the different panels? it seems you are plotting different ranges. Unidades no tienen que estar italizadas: ${\rm M}_{\odot}/h$. Mejor pinta el promedio total en negro, verde si hay central y naranja si no}
    {\it Top:} Mean number of satellite galaxies per halo as a function of halo mass: total (orange), with a companion central galaxy of the same type (blue), and without these centrals (green). %\vgp{para que esta gráfica sea más clara, creo que el global factor debería aparecer en el color correspondiente y en la leyenda debería aparecer $K_{1, \rm glob}$ y $K_{2, \rm glob}$}
    {\it Middle:} The conformity parameter $K_{1}(M)$ per mass bin (\autoref{eq:k1(m)}) and the global conformity factor $K_{1, \rm glob}$ (\autoref{eq:k1gf}), measured for \sagesh ELGs, as a function of halo mass. These parameters represent the ratio between the mean number of total satellite galaxies with a central galaxy (shown in blue in the top panel) with respect to the corresponding number of haloes (see \autoref{eq:themeannumberswc}). {\it Bottom:} Similar to the middle panel, but for $K_{2}(M)$ (\autoref{eq:k2(m)})  and the global factor, $K_{2,\rm glob}$} (\autoref{eq:k2gf}). In this case, these parameters represent the ratio between the mean number of total satellite galaxies without a central galaxy (shown in green in the top panel) with the number of haloes (see \autoref{eq:themeannumbersw/oc}).
    \label{fig:conformity}
\end{figure}

Usually, HOD models assume satellite and central galaxies of a given type to be independent events, i.e., the mean occupation of satellite galaxies does not depend whether or not the central is of the same type. Here we define the 1-halo conformity as deviations from independency, i.e., the mean occupation of satellite ELGs depend on having or not a central ELG in that halo.
%\vgp{párrafo breve comparando esta definición con la literatura con referencias}\vgp{siguiente texto a reducir y necesitas referencias}

%Lo cambio por algo que quizás tenga más sentido:  
%The phenomenon of galactic conformity refers to the correlation between the properties of galaxies and the large-scale structure of the universe \citep{Weinmann}. %\SA{Esto me parece una afirmación que va mucho más allá de lo que estábamos considerando nosotros. Aunque quizás sea cierta. Es decir, si hay más LRGs en clusters y más ELGs en filamentos, ¿esto se considera conformity? Quizás consular con Violeta}. 
%\vgp{has comentado lo que ha puesto Santi, pero no has hablado de esto conmigo???}

The phenomenon of galactic conformity refers to the correlation between the properties of neighbouring galaxies \citep{Weinmann}. 
This is thought to happen due to galaxies having a common evolution as baryonic matter collapses within dark matter haloes to form galaxies. Observed galaxies exhibit variations in mass, shape, star formation activity, colors, and ages, and these properties tend to be correlated. 
It is important to differentiate between one-halo conformity, which measures conformity at small separations within a single halo, and two-halo conformity, which measures conformity at larger separations between central galaxies and neighboring galaxies in adjacent haloes.

Conformity has been detected in simulations \citep[e.g][]{Lacerna,ayromlou2023} and there seems to be growing evidence of it from observations \citep[e.g.][]{Rocher2023, desi_conformity}.
%Some studies such as \citet{Lacerna}, investigate galactic conformity and its impact on galaxy clustering using semi-analytic models. These studies examine the correlation between star formation in central galaxies and neighboring galaxies at both small and large separations. However, reproducing this signal in semi-analytic models presents challenges, emphasizing the need for further research to understand the involved physical mechanisms. 
%\vgp{puedes añadir aquí un párrafo sobre las teorías del origen de la conformidad (con referencias), pero reducido porque no hacemos nada con esto}
%\SA{A mi me parece que con lo que hemos puesto arriba es más que suficiente. Sí faltaría añadir  referencias}
%Various models attempt to explain galactic conformity that includes gas chemistry, preheating physics, and the formation of more massive haloes containing red galaxies. However, different theories receive varying levels of support from observational data and simulation, highlighting the need for a better understanding of the underlying physical processes. 
%\SA{You mention "various models", "different theories", but you don't quote any. Need references.} 
Galactic conformity continues to be an active area of study and remains a significant puzzle in our understanding of galaxy formation and evolution.

%It is important to note that traditional "vanilla" Halo Occupation Distribution (HOD) models do not consider galactic conformity. In these models, central and satellite galaxies are assumed to be independent events with no connection between them. In this article, we utilise the concept of one-halo conformity, which is crucial for characterizing the small scale clustering of the satellite galaxy sample. A detailed explanation of this property is provided in the subsequent subsection.
%\grp{ Buscar mas bibliografía de este autor y mencionar el segundo: https://arxiv.org/pdf/1406.6058.pdf, mirar mas en detalle.}
%\vgp{Check the references in this paper for supporting observationally the "need" for comformity: https://arxiv.org/pdf/2210.05637.pdf. 
%\url{https://arxiv.org/pdf/2207.02218.pdf}, Creo que este es un buen sitio para comenzar la búsqueda bibliográfica sobre conformidad. Aquí se habla de large-scale y de quenched galaxies, que no es lo que hacemos, pero quizás se puedan mirar las referencias.}

%\vgp{no veo el sentido de comentar cosas que no has aplicado}

%\subsection{Computational implementation}

We define the mean value of satellite galaxies with, $\langle N_{\rm sat} | N_{\rm cen}=1\rangle$, and without a central galaxy, $\langle N_{\rm sat} | N_{\rm cen}=0\rangle$, as:

\begin{equation}
\langle N_{\rm sat} | N_{\rm cen}=1\rangle= \frac{N_{\rm sat,wc}}{N_{\rm h,w c}}= K_{1} \hspace{1mm} \langle N_{\rm sat}\rangle
\label{eq:themeannumberswc}
\end{equation}
\begin{equation}
\langle N_{\rm sat} | N_{\rm cen}=0\rangle= \frac{N_{\rm sat, w/o c}}{N_{\rm h, w/o c}}= K_{2}\hspace{1mm} \langle N_{\rm sat}\rangle \,,
\label{eq:themeannumbersw/oc}
\end{equation}
where $N_{\rm sat,wc}$ and $N_{\rm sat, w/o c}$ denote the  number of satellites with or without a central ELG. Note that the total number of satellite galaxies, $N_{\rm sat}$,  is the sum of these two values, $N_{\rm sat} = N_{\rm sat,wc} + N_{\rm sat,w/oc}$. The total number of dark matter haloes with and without a central \ha\ galaxy are $N_{\rm h,w c}$ and $N_{\rm h, w/o c}$, respectively. $\langle N_{\rm sat}\rangle$ is the average value of satellite galaxies in all types of haloes (with and without central galaxies of the given type). $K_{1}$ and $K_{2}$ are the {\it conformity} factors.

When there is no conformity, $K_{1}=K_{2}=1$, so the mean number of satellite galaxies is independent of having a central galaxy of the same type, $\langle N_{\rm sat}|N_{\rm cen} = 0,1 \rangle = \langle N_{\rm sat}\rangle$.

When there is conformity, $K_{1}\neq 1 \neq K_{2}$. To model the 1-halo conformity, the average number of satellites should depend on the presence or absence of a central galaxy of the same type.

%\vgp{describe top panel of \autoref{fig:conformity} and introduce the need for conformity. Adapt the text below}
To investigate conformity, we initially examine whether the average number of satellite galaxies, with or without a central galaxy, constitutes two completely independent events. To achieve this, we calculate the mean values $\langle N_{\rm sat} | N_{\rm cen}=1\rangle$ and $\langle N_{\rm sat} | N_{\rm cen}=0\rangle$ for our \sage galaxy sample and compare them to the mean value of $\langle N_{\rm sat} \rangle$. In the top panel of \autoref{fig:conformity}, we demonstrate that the total number of satellites with and without a central galaxy in our \sage sample is not reproduced when assuming independence. As a result, we can infer that introducing the concept of conformity is necessary to accurately reproduce these differences. 
Furthermore, it enables us to demonstrate the crucial role of conformity in replicating the distribution of satellites in \sage, thus marking it as one of the primary novel findings of our article. It is worth noting that in other studies, such as \citet{Lacerna} and \citet{DESIonehaloterm}, they have also found the need to include the conformity property.

%\vgp{comentario sobre la literatura, ¿qué se ha encontrado en otros estudios}

\subsection{Mass dependent conformity}
For a given bin of halo masses, $M_{i}$ and using the above expressions, it is possible to define $K_{1}(M_{i})$ and $K_{2}(M_{i})$ as a function of quantities that we can measure from the reference galaxy catalogue:
%Where using the equation \ref{eq:themeannumberswc} and \ref{eq:themeannumbersw/oc}, we can determine $N_{\rm sat,wc}$ and $N_{\rm sat,w/oc}$. So, substituting these two expressions in equation \ref{eq:ConservacionN} and dividing by the number of haloes we have that the expressions for $K_{1}$ and $K_{2}$ are given as follows:\\

\begin{equation}
K_{1}(M_{i}) = \frac{N_{\rm sat,wc}(M_{i})}{N_{\rm cen}(M_{i})\ \hspace{1mm} \langle N_{\rm sat}(M_{i}) \rangle} \,,
\label{eq:k1(m)}
\end{equation}
\begin{equation}
K_{2}(M_{i}) = \frac{1-K_{1}(M_{i})\hspace{1mm} \langle N_{\rm cen}(M_{i})\rangle}{1 - \langle N_{\rm cen}(M_{i})\rangle} \, .
\label{eq:k2(m)}
\end{equation}

The above conformity factors, $K_{1}(M_{i})$ and $K_{2}(M_{i})$, can be computed using these equations by inserting the galaxy/halo counts measured in our simulation. These factors are then used in a HOD prescription for each mass bin. In practice, what we need is to first sample the Bernoulli PDF for centrals in order to determine whether we have $N_{\rm cen}=0$ or  $N_{\rm cen}=1$ in a given halo (see PDF description in \autoref{sec:pdf}) and then modify $\langle N_{\rm sat}\rangle$ according to \autoref{eq:themeannumbersw/oc} or \autoref{eq:themeannumberswc}, respectively. Subsequently, we would sample the Poisson distribution (again \autoref{sec:pdf}) from the modified $\langle N_{\rm sat}\rangle$ ($=\langle N_{\rm sat} \lvert N_{\rm cen}\rangle$)  and continue with the rest of steps of the HOD (\autoref{sec:HOD_steps}).

In the middle panel of \autoref{fig:conformity}, we can observe the conformity parameter $K_{1}(M)$ (\autoref{eq:k1(m)}) for our \sagesh ELGs sample as a function of halo mass. These parameters represent the relationship between the average number of total satellite galaxies associated with a central galaxy and the overall average number of satellites, depicted in blue for $K_{1}(M)$. This comparison indicates the presence of conformity, i.e., how much it deviates from the scenario where the existence of a satellite galaxy and a central galaxy are two independent events (that would be represented by $K_{1}(M)=1$). Then, in the bottom panel, similar analyses are performed, but this time for $K_{2}(M)$ (\autoref{eq:k2(m)}). In this case, these parameters represent the ratio between the mean number of total satellite galaxies without a central galaxy and the overall average number of satellites, shown in green for $K_{2}(M)$. 

%\vgp{comment the middle and bottom panels of \autoref{fig:conformity}: how do they vary with halo mass? within what factor? Adapt the following text you already had}Then, we can observe this in the central panel and the bottom panel of \autoref{fig:conformity}, i.e., the values obtained for the corresponding global factors ($K_{\rm glob} =cte$) and the parameters $K_{1,2}(M)$ by mass bins, which differ from unity.

\subsection{Global conformity}

As an alternative, we consider global conformity factors, $K_{1,\rm glob}$ and $K_{2,\rm glob}$, that are assumed constant for a given reference galaxy catalogue and to be the same for all halo masses. 
$K_{1,\rm glob}$ and $K_{2,\rm glob}$ are then constants in \autoref{eq:themeannumberswc} and \autoref{eq:themeannumbersw/oc}. %
In this case, the global conformity factors are computed as:
\begin{equation}
K_{1,\rm glob} = \frac{\sum_{i=1}^{n}{N_{\rm sat,wc}(M_{i})} }{\sum_{i=1}^{n}\langle N_{\rm cen}(M_{i})\rangle\hspace{1mm} N_{\rm sat}(M_{i})} \,,
\label{eq:k1gf}
\end{equation}

\begin{equation}
K_{2,\rm glob} = \frac{\sum_{i=1}^{n}{N_{\rm sat,w/oc}(M_{i})}}{\sum_{i=1}^{n}(1 - \langle N_{\rm cen}(M_{i})\rangle)\hspace{1mm} N_{\rm sat}(M_{i})} \,,
\label{eq:k2gf}
\end{equation}
where $i$ indicates each of the different mass bins considered. The somewhat complicated expressions above are a consequence of populating the mean HOD $\langle N_{\rm sat}\rangle$ in bins of halo mass, hence, with  \autoref{eq:themeannumbersw/oc} and \autoref{eq:themeannumberswc} also applied in mass bins. 
This implies that we can not compute the global factors, $K_{1,\rm glob}$ and $K_{2,\rm glob}$, by simply dropping the mass dependence from \autoref{eq:k1(m)} and \autoref{eq:k2(m)}. Instead, we need to compute them in individual bins and then do the sum indicated in
\autoref{eq:k1gf} and \autoref{eq:k2gf}.

The middle panel of \autoref{fig:conformity} shows the global conformity factor $K_{1, \rm glob}$ (\autoref{eq:k1gf}) for our \sagesh ELGs sample. And the bottom panel the other global factor, $K_{2,\rm glob}$ (\autoref{eq:k2gf}). 
The global conformity factors, $K_{1,\rm glob}$ and $K_{2,\rm glob}$, roughly correspond to the respective mean values of the mass-dependent conformity ones, $K_{1}(M)$ and $K_{2}(M)$.
We find $K_{1, \rm glob}= 0.708$ and $K_{2, \rm glob}= 1.038$, implying that we detect conformity of the order of $\sim30$\%.

%SA->:
\subsubsection*{Variations with number density and redshift}

Since we will adopt global conformity in our default model, we now explore how the values reported above change when we choose a different reference sample by adopting a different reference number density (\autoref{tab:nd_models}) or a different redshift (\autoref{tab:densityz}). These values are summarised in \autoref{tab:ConformityParamiters}.
%, together with those derived for galaxy samples with the other two number densities considered in this study (see \autoref{tab:nd_models}).

The global conformity factors remain remarkably constant for samples of ELGs with different number densities. The variations are below $20$\% for $K_{1, \rm glob}$ and below $1$\% for $K_{2, \rm glob}$. It will be interesting to explore in the future if this is connected with conformity depending weakly with environment, in line with the proposal of spectral emission lines coming from star forming regions being triggered by merging events \citep{yuan23.conformity}.

The conformity global factors, $K_{1, \rm glob}$ and $K_{2, \rm glob}$, obtained for samples at different redshifts are summarised in  \autoref{tab:ConformityParamiters}. We find variations below $5\%$ for $K_{1, \rm glob}$ and below $1\%$ for $K_{2, \rm glob}$. The results suggest that there is no evolution in conformity in the explored redshift range, $0.987<z<1.650$.

%%%%%%%%Variations w nd and z
\begin{table}
\begin{center}
\begin{tabular}{ c c c c }
Sample & Redshift & $K_{1, \rm glob}$ & $K_{2, \rm glob}$ \\\hline
Pozzetti nº1 & $1.3$ & $0.799$ & $1.034$ \\
{\bf Pozzetti nº3} & $\mathbf{1.3}$ & $\mathbf{0.708}$ &  $\mathbf{1.038}$\\ 
Flux cut  & $1.3$ & $0.618$ & $1.033$ \\\hline
{\it dPozzetti nº3} & $0.987$  &  $0.741$ & $1.033$\\
{\it dPozzetti nº3} & $1.220$ &  $0.717$ & $1.035$ \\
%$\mathbf{1.321}$ & $\mathbf{0.708}$ &  $\mathbf{1.038}$ \\
{\it dPozzetti nº3} & $1.425$ &  $0.701$ & $1.040$ \\
{\it dPozzetti nº3} & $1.650$ &  $0.718$ & $1.042$ \\ \hline
\end{tabular}
\caption{ 
Global conformity parameters, $K_{1, \rm glob}$ (\autoref{eq:k1gf}) and $K_{2, \rm glob}$ (\autoref{eq:k2gf}) for galaxies produced with HOD models at different redshifts and with number densities obtained either in a range of redshifts, first three rows (\autoref{tab:nd_models}) or from a differential ({\it d})luminosity function (\autoref{tab:densityz}). Our default choice is the sample at $z=1.3$ with number density matching that of Pozzetti model number 3 in a range of redshifts. This is indicated by using bold face in the table.
}\label{tab:ConformityParamiters}
\end{center}
\end{table}

\subsection{Clustering}
%%%%%%%%%%%%%%
\begin{figure}
	\includegraphics[width=\columnwidth]{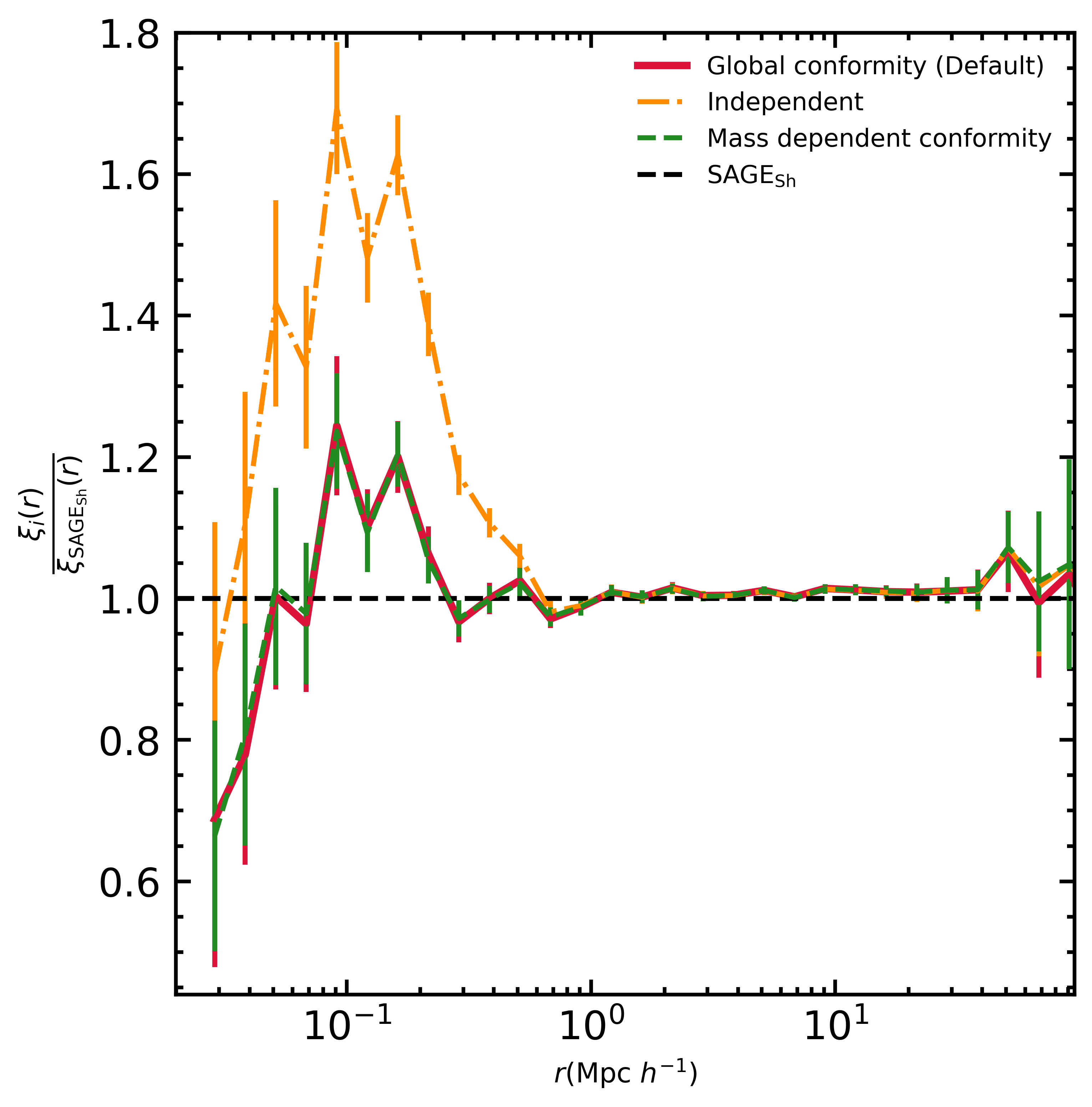}
    \caption{ 
    Ratios of the real space two-point correlation functions from galaxy catalogues generated with different HOD models with respect to that from the \sagesh reference galaxy. The result from the proposed HOD model (\S~\ref{sec:DefaultHOD}), which uses global conformity factors, is shown in red. Two modifications of this model are also shown: one with no conformity (centrals and satellites are treated as independent events), shown in orange; and one using mass dependent conformity factors, shown in green. Note that the three HOD catalogues use an analytical function to model the radial distribution of satellite galaxies (\S~\ref{eq:NFWM}). The error bars are calculated as the standard deviation of 100 realisations. 
    }\label{fig:xiconformity}
\end{figure}
%%%%%%%%%%%%%%%%%%

%\vgp{Tide up the text or remove as in principle, this has already been explained}
After computing the values of $K_{1}$ and $K_{2}$, either as a global factor or as a function of mass, we apply Equations \ref{eq:themeannumberswc} $\&$ \ref{eq:themeannumbersw/oc} to our HOD model, based on either the global conformity or the mass-dependent conformity, respectively.

Subsequently, we assess their clustering and compare, in \autoref{fig:xiconformity},  the results to a scenario without conformity  (orange dotted dashed). The dot-dashed green line illustrates the scenario where $K_{1}$ and $K_{2}$ are dependent on the mass of the bins, while the solid red line corresponds to the global factor conformity. First of all, it is noticeable that in the case of independence, the largest deviation from the reference sample in terms of clustering occurs around $r \sim 0.1 , \text{Mpc} , h^{-1}$, resulting in an approximately $\sim 60\%$ difference. However, by incorporating conformity, we mitigate this difference by $\sim 40\%$ at this scale. Moreover, we observe that both implementations of conformity yield similar outcomes, significantly superior to the scenario without considering conformity. 
The clustering is found nearly indistinguishable between the two models proposed for conformity. Consequently, for the sake of simplicity, we opt for using global conformity factors as our default modeling approach. It is important to emphasize that the results regarding the necessity of including the conformity property align with what has been found in other studies, as mentioned in \citet{Lacerna} or \citet{DESIonehaloterm}. In the latter, a mean satellite occupation function was obtained that aligns with physically motivated models of \OII emitters only if conformity between centrals and satellites is introduced. This implies that the occupation of satellites is conditioned by the presence of central galaxies of the same type. It turns out that this aligns with what we obtain in this article, while using very different angles.

\section{The radial profile of satellite galaxies}\label{sec:radialprofile}

The distribution of substructure in dark matter haloes has been accurately described by either a Navarro-Frenk-White profile~\citep[NFW, ][]{NFW} or an Einasto profile~\citep{Einasto}. However, the distribution of galaxies within haloes could be biased with respect to that of dark matter~\citep[e.g][]{Yuan2023,Rocher2023,Hadzhiyska2023}. 

Central galaxies are placed in the center of potential of the dark matter halo, so we focus here on satellite galaxies and how they are distributed with respect to their center of halo.

In this work, we first study the spatial distribution of \sagesh ELGs that are satellite galaxies within dark matter haloes. Then we explore how to best use the measured distribution to inform HOD models with the aim to reproduce the \sagesh 1-halo term clustering.

\begin{figure}
	\includegraphics[width=\columnwidth]{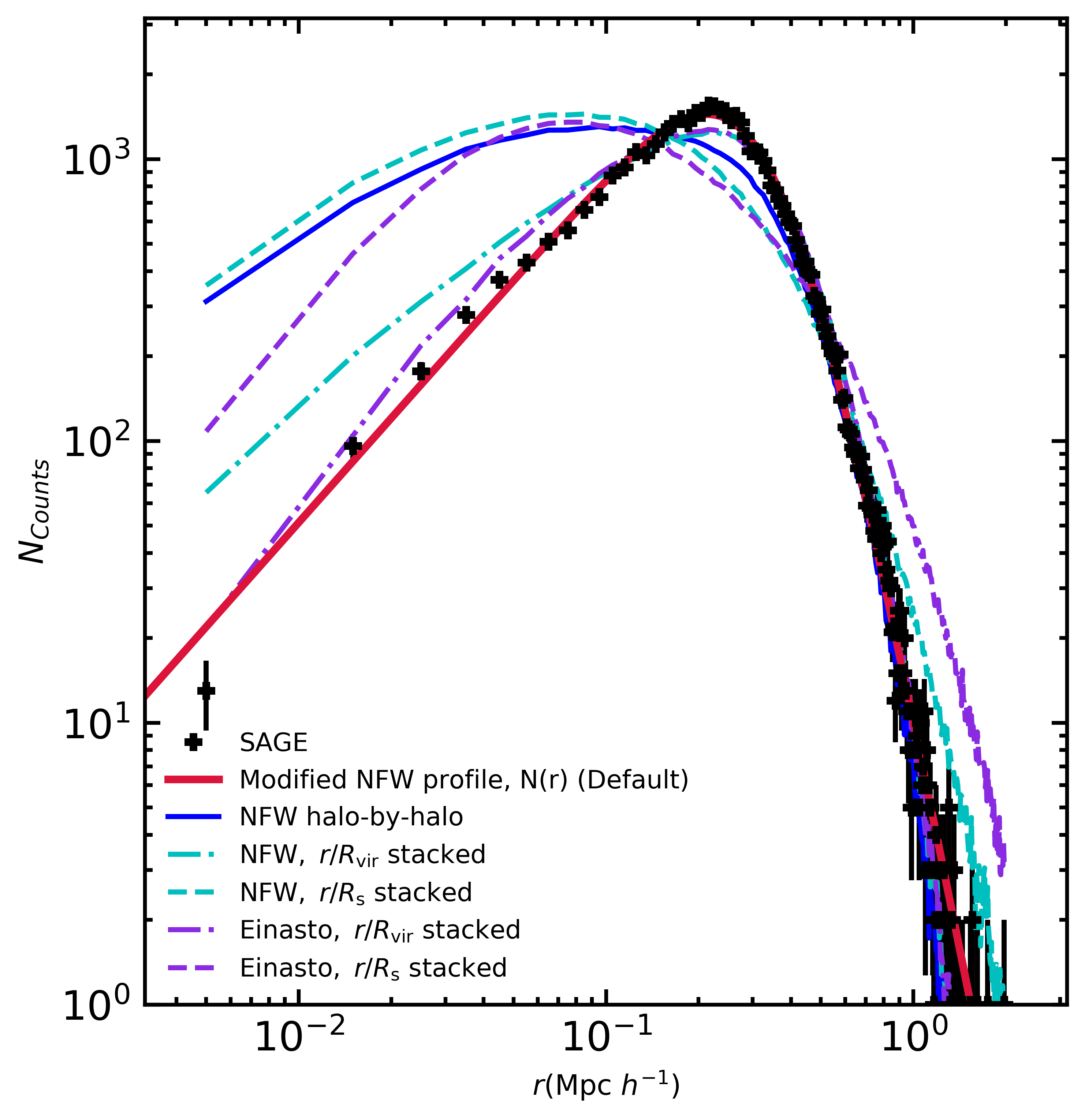}
    \caption{
    Radial profile for satellite galaxies as a function of their distance, $r$, to the center of their host halo. ELGs from \sagesh are shown as filled symbols with Poisson error bars. The lines correspond to catalogues produced with different HOD models, as indicated in the legend. The results from our proposed HOD model (\S~\ref{sec:DefaultHOD}), which uses an analytical function (\autoref{eq:NFWM}) to fit the radial distribution of \sagesh satellite galaxies, is shown in red. Five modifications of this model are also shown in the plot: a model assuming a NFW distribution given the concentration of each halo (solid blue line); models assuming a NFW density profile fitted to reprocude the average \sagesh one as a function of $r/R_{\rm vir}$ (dash-dotted blue line)  and $r/R_{\rm s}$ (dashed blue line); models assuming an Einasto profile fitted to reproduce the average \sagesh one as a function of $r/R_{\rm vir}$ (dash-dotted purple line) and $r/R_{\rm s}$ (dashed purple line). Note that the binning is linear but we have plot in logarithmic scale to properly appreciate the change of slopes.
    }\label{fig:radialprofile}
\end{figure}

The radial profile of \sagesh satellite ELGs is shown in \autoref{fig:radialprofile} as filled circles. This plot shows the number of satellite galaxies as a function of their distance to the central galaxy, $r =|\Vec{r}| = |\Vec{r}_{\rm sat} - \Vec{r}_{\rm h}|$. Note that we have verified the existence of a displacement between the position of the central galaxy and the center of the halo. However, we have studied these differences, and they are smaller than $1\times 10^{-5}$ for all central galaxies, so we can assume that $\Vec{r}_{\rm cen} = \Vec{r}_{\rm h}$. Satellite galaxies in \sagesh are preferentially placed at a distance of  
$\sim0.2{\rm Mpc}h^{-1}$ from their central galaxy. 

\begin{figure}
	\includegraphics[width=\columnwidth]{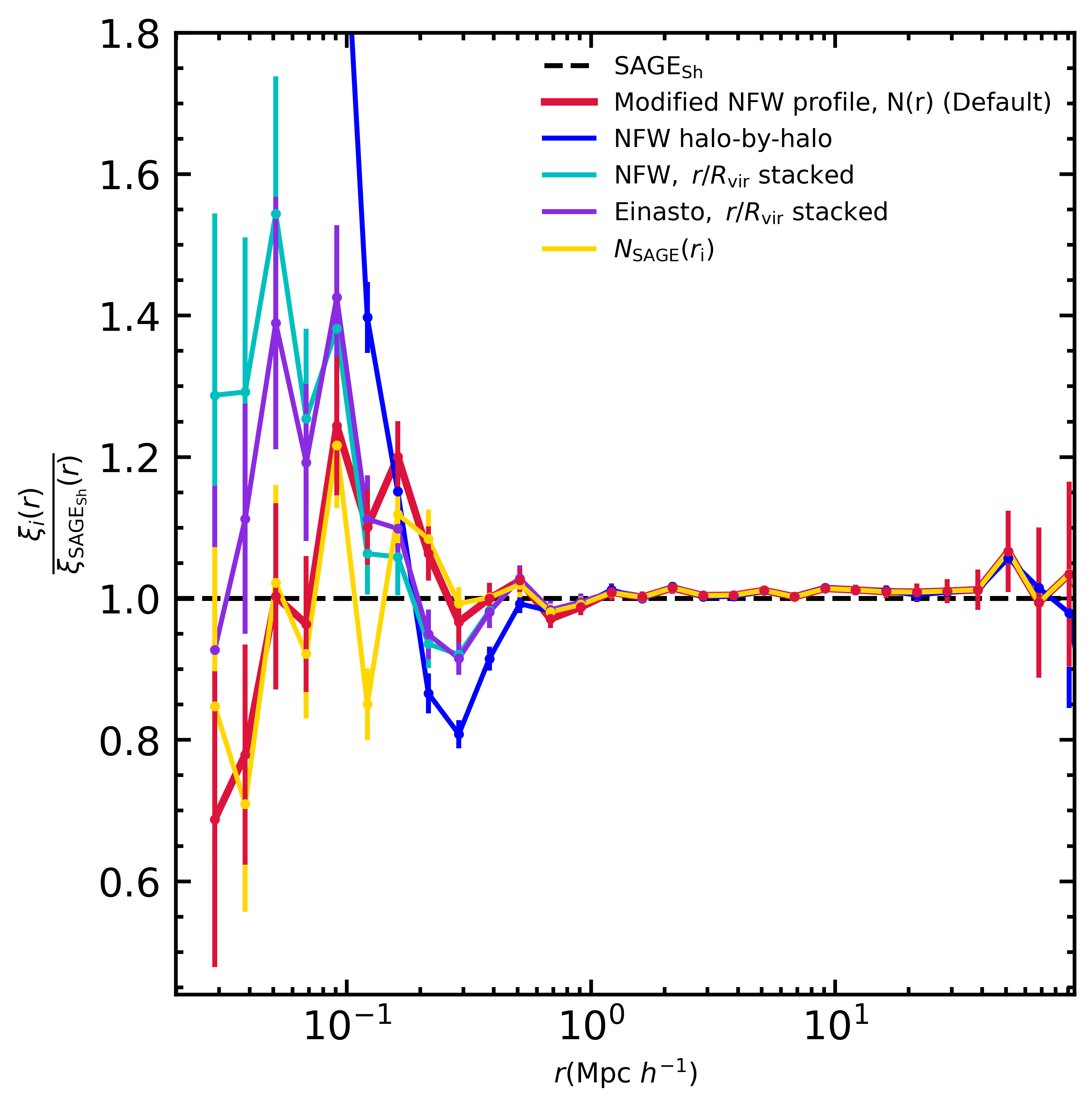}
    \caption{Ratios of the real space two-point correlation functions from galaxy catalogues generated with different HOD models with respect to that from the \sagesh reference galaxy sample. 
    The results from our proposed HOD model (\S~\ref{sec:DefaultHOD}), which uses an analytical function (\autoref{eq:NFWM}) to fit the radial distribution of \sagesh satellite galaxies, is shown in red. Four modifications of this model are also shown in the plot, with similar colours to those shown in \autoref{fig:radialprofile}.
    %: a model that directly uses the values of the \sagesh satellite radial distribution in mass bins (yellow line); 
    %a model assuming a NFW distribution given the concentration of each halo (dark blue line); a model assuming a NFW density profile fitted to reprocude the average \sagesh one as a function of $r/R_{\rm vir}$ (light blue line); and a model assuming an Einasto profile fitted to reproduce the average \sagesh one as a function of $r/R_{\rm vir}$ (purple line).
    Error bars are calculated as the standard deviation of 100 realisations with changing seeds for the generators of random numbers. %(maybe add this info in the main text once, when describing "realisations)".
    %\SA{Cambiar Fittedc to "stacked fit"}
    }
    \label{fig:radialprof2PCF}
\end{figure}

In \autoref{fig:radialprofile}, we present the radial profile as a function of the relative distance to the halo, $r$, because we have found that this approach provides an HOD model clustering closer to the reference \sagesh sample. This result is shown in \ref{fig:radialprof2PCF}. It is clear from this figure that the 1-halo clustering is better recovered when the radial profile as a function of $r$ is best matched, instead of $r/R_{\rm s}$ or $r/R_{\rm vir}$. This is a surprising result as haloes of different sizes and masses are being mixed in the radial profile shown in \autoref{fig:radialprofile}, as a function of $r$.

%So, we present the radial profile of our sample of satellite galaxies (black dots) in comparison with several HOD prescriptions described below. To obtain this profile, we identify the dark matter haloes that contain these satellite galaxies and determine the corresponding radial distances, $r =|\Vec{r}| = |\Vec{r}_{\rm sat} - \Vec{r}_{\rm h}|$, of each satellite galaxy from the center of its host halo. These points correspond to a histogram in which we count the number of satellite galaxies located at a distance r from the center of the halo to which they belong. We have used a linear binning for this purpose, although we represent it on a logarithmic scale. Additionally, it is worth noting that in \autoref{fig:radialprofile}, we chose to create the histogram for SAGE and the various mocks we obtained using the corresponding adjustments discussed in the following subsections based on r. This allows for a better comparison of all the prescriptions made and a more straightforward connection to the results obtained in terms of clustering.

%\vgp{luego hacer un review de lo que se suele hacer en modelos de HOD y después explicar qué métodos se van a utilizar}
%\vgp{esta descripción entiendo que debería venir más tarde}The outcome of this study can be seen 

Typically, HOD models place satellite galaxies in dark matter haloes assuming the NFW profile given the concentration (or mass) of the halo~\citep[e.g.][]{Zheng2005}. This is typically done halo-by-halo. An other commonly used option is to use the positions of dark matter particles~\citep[e.g.][]{avila2020,Rocher2023}. 

Our aim is to have an HOD model with galaxies clustered as close as possible to that from \sagesh, in small scales. To achieve this, we explore HOD models with  different ways of placing satellite galaxies within haloes: (i) NFW 
profiles using the concentration from each halo
(\S~\ref{sec:radialhalo}, {\it NFW halo-by-halo}); (ii) NFW and Einasto curves fitted to the $r/R_{\rm s}$-stacked profiles (\S~\ref{subsec:r/Rs}); (iii) NFW and Einasto curves fitted to the $r/R_{\rm vir}$-stacked profile (\S~\ref{subsec:r/Rvir}); (iv) radial profile from \sagesh, as a function of $r$, using it either directly as a discrete distribution function or to fit an analytical function (\S~\ref{subsec:analytical}).

\autoref{fig:radialprofile} presents the radial distribution of galaxies modelled with all the approaches described above, together with the \sagesh distribution. 
From this figure, it is clear that the first three approaches cannot reproduce the radial profile of satellite galaxies in \sagesh. %Moreover, the 1-halo term clustering derived from the first three options is worse than using the \sagesh distribution, either directly or through the analytical extension to a NFW profile proposed in \autoref{subsec:analytical}. 
A similar conclusion was reached by \citet{Parkinson2} also using the \sage SAM to study the distribution of galaxies. In their work they applied their proposed extended profile to SSDS DR10 group catalogue galaxies. 

The different HOD approaches described above are implemented following a similar method.
For each halo, we generate a number of random values equal to the number of satellite galaxies assigned to that halo, which we then transform to a distance $x$ using an inverse cumulative distribution sampling of the profile (in terms of $N(x)\propto x^2\rho(x)$).
This distance $x$ can be $r/R_{\rm vir}$, $r/R_{\rm s}$ or $r$ depending on the chosen independent variable in each prescription. The first two distances can be transformed into $r$, given the halo $R_{\rm vir}$ or $R_s$. Once we have $r$, we place the satellite galaxies by randomly sampling the angular position  with respect to the halo center.

%SA: Estos detalles no son necesraios, no?->
%It is worth noting that for the case of the extended NFW density profile we have implemented, essentially only the last step needs to be applied, as using our analytical function as the Inverse Cumulative Distribution Function directly yields the value of r. However, in the standard option used in the literature case, the procedure involves "adjusting the satellite galaxies within each halo" \SA{<- What do you mean by this? I don't think it is true.} to the associated NFW profile with the corresponding halo concentration. Then, we apply the Inverse Cumulative Distribution Function in the same manner, but this time using the halo-specific adjusted NFW profile as the cumulative probability distribution. In this way, we assign positions to the reproduced satellite galaxies.
%<- Repasalo, pero me parece que tanto detalle no aporta nada nueo y solo embrolla. 
%Si acaso merece la pena mencionar alguna de estas diferencias, se puede hacer después en las subsecciones

%\vgp{for all the following sections: obtain the ratios or the differences ar 0.1, 0.3 and 1, so you can quantify the discussions}

%%%%%%%%%%%%%%%%%%%%%%%%%%%%%%%%%%%%%%%%%%%

%%%%%%%%%%%%%%%%%%%%%%%%%%%%%%%%%%%%%%%%%%%

\subsection{{\it NFW halo-by-halo}: Sampling individual halo profiles}\label{sec:radialhalo} 
HOD models usually place satellite galaxies within haloes following the distribution of dark matter itself~\citep[e.g.][]{S.Avila2018}. This is often described by a Navarro–Frenk–White (NFW) profile, which was derived from dark matter only N-body simulations~\citep{NFW}. The NFW profile describes the density of dark matter, $\rho_{\rm NFW}$, as a function of distance to the center of a halo, $r$:
\begin{equation}
\rho_{\rm NFW}(r) =  \frac{4\, \rho_{\rm s}}{\frac{r}{R_{\rm s}}\left(1+\frac{r}{R_{\rm s}}\right)^{2}} \ , 
\label{eq:NFW}
\end{equation} 
where $R_{\rm s}$ is the scale radius provided by \textsc{Rockstar}, and $\rho_{\rm s}$ is the mass density at this radius. The halo concentration, $C$, relates the virial and scale radii of haloes, $R_{\rm vir}=CR_{\rm s}$.

Given a halo concentration, the NFW can be sampled in a halo-by-halo basis. We have followed this method to construct catalogues of ELGs. In \autoref{fig:radialprofile} we compare the global radial profile for satellite galaxies from this catalogue (dark blue solid line, {\it NFW halo-by-halo}) with that from \sagesh. The differences are large, in particular below $0.1 h^{-1}{\rm Mpc}$. At these scales, we obtain differences over $70\%$ compared to our \sagesh sample. The difference increases for smaller values of $r$.

The {\it NFW halo-by-halo} clearly fails to reproduce the global radial profile of \sagesh satellite galaxies. This difference results in a very strong clustering at scales below $\sim 0.2 h^{-1}{\rm Mpc}$. The {\it NFW halo-by-halo} results in galaxy clustering over a factor of $2$ above the clustering measured for \sagesh galaxies at scales below $0.1 h^{-1}{\rm Mpc}$ (\autoref{fig:radialprof2PCF}). This is of great importance, as the {\it NFW halo-by-halo} approach is very commonly used one in the literature to populate halo catalogues from large simulations. 

\subsection{\texorpdfstring{$r/R_{\rm s}$}{r/Rs}-stacked profiles}
\label{subsec:r/Rs}

We have fitted the \sagesh $r/R_{\rm s}$-stacked satellite profile with single NFW\citep{NFW} and Einasto~\citep{Einasto} parametrisations. To construct the stacked profile from the \sagesh sample we start by converting the counts of satellite galaxies as a function of distance to the central galaxy, $N(r)$, into densities, with $\rho = N(r)/(4\pi r^{2}{\rm d}r)$. We use a binning of ${\rm d}r=0.1 h^{-1}$Mpc. 
%\vgp{have you tried to change the binning, do results change?}

In the previous section, we introduced the NFW parametrisation of radial profiles (\autoref{eq:NFW}). The Einasto parametrisation can be written as a function of the scale radius, $R_{\rm s}$, as follows:
\begin{equation}
\rho_{\rm Ein}(r) =  \rho_{\rm s}\hspace{1mm} e^{ -\frac{2}{\alpha}[(\frac{r}{R_{\rm s}})^{\alpha}-1]} \,,
\label{eq:Einasto} 
\end{equation}
where $\rho_{\rm s}$ is the mass density at $r =R_{\rm s}$ and $\alpha$ determines the curvature of the profile. 

When fitting the  \sagesh $r/R_{\rm s}$-stacked profiles we aim to recover the total number of satellite galaxies. To achieve this, the NFW (\autoref{eq:NFW}) and Einasto (\autoref{eq:Einasto}) parametrisations must be integrated up to $r=R_{\rm vir}$, which is considered the edge of the halo. Thus, we ensure the total number of satellite galaxies, $N_{\rm sat}$, is preserved by imposing:
\begin{equation}
\label{eq:NRsNFW}
    N_{\rm sat} = R_{\rm s}^3\int_{0}^{\left( \frac{r}{R_{\rm s}}\right)=\frac{R_{\rm vir}}{R_{\rm s}}\equiv C}4\pi \left( \frac{r}{R_{\rm s}}\right)^{2}\rho\left(\frac{r}{R_{\rm s}}\right) {\rm d}\left( \frac{r}{R_{\rm s}}\right) \,.
    %N_{\rm sat} = R_{\rm s}^3\int_{0}^{C}4\pi \left( \frac{r}{R_{\rm s}}\right)^{2}\rho\left(\frac{r}{R_{\rm s}}\right) {\rm d}\left( \frac{r}{R_{\rm s}}\right) \,.
\end{equation}

When fitting the NFW and Einasto parametrisations, we find the halo concentration, $C = R_{\rm vir}/R_{\rm s}$, that gives the closest value of $N_{\rm sat}$ to the \sagesh one. Note that the concentration appears as an integration limit in \autoref{eq:NRsNFW}. For both parametrisations, $\rho_{0}$ provides the normalisation of the fit to the \sagesh $r/R_{\rm s}$-stacked satellite profile. The NFW profile has the halo concentration as the only free parameter, while the Einasto one also has $\alpha$ as a second free parameter.

%To conduct the fitting procedure, we initially select a range of concentration values spanning from 0.1 to 10. As previously mentioned, the concentration corresponds to the integration limit of the $r/R_{\rm s}$ variable, which is compatible with the cutoff we made due to the lack of objects. For each concentration value ($C$), we construct a histogram of the galaxies within each interval between $C_0$ and $C_i$. The counts obtained are then transformed into densities. Then, we take the midpoint of the bins in which we perform the satellite histogram, and evaluate both the NFW and Einasto functions at those points for subsequent comparison. 
For performing the fit to the NFW profile, we allow the halo concentration to vary from $0.1$ to $10$. This truncation is chosen based on the scarcity of satellite galaxies, $N \lesssim 1$, beyond this limit. %which adversely affects the fitting process. SA: Diria  que no hay problemas porque el best fit parece que sale a Cs más pequeños. 
This choice is conservative, compared with the results from \citet{sujatha2019} where they used dark matter simulations to measure the distribution of different halo properties. In their study, the halo concentration is found between $0$ and $\sim 30$.

To determine the parameters that yield the best fit to the \sagesh $r/R_{\rm s}$-stacked profiles, we iterate through different values of halo concentration, $C_i$, which gives the upper limit of the integration in \autoref{eq:NRsNFW}. 
%We perform the integration using  \autoref{eq:NRsNFW} %SA: unneeded: (for both profiles) to obtain the normalization parameter $\rho_{s}$  correct in all the cases, and subsequently calculate the $\chi^{2}$ statistic. 
The parameters yielding the lowest $\chi^{2}$ per data point (see below) value are selected as the optimal fit. The $\chi^{2}$ value is determined using the following expression:

\begin{equation}
\label{eq:Chi2}
\chi^{2} = \sum \frac{(\rho_{\rm SAGE_{sh}}-\rho)^{2}}{\sigma^{2}} \,, 
\end{equation}
where we assume a Poisson error on the counts, which translates to an uncertainty on the density profile of 
\begin{equation}
\sigma = \frac{\sqrt{N}}{4\pi \left(\frac{r}{R_{\rm s}}\right)^{2}\Delta \left(\frac{r}{R_{\rm s}}\right)} \ .
\end{equation}

We perform the integration within $\Delta(r/R_{\rm s}) = 0.05$ bins. Thus, for smaller $C_i$ values, the number of data points used to evaluate the functions will be fewer compared to larger ones, due to the profile cutoff set by this integration limit (see \autoref{eq:NRsNFW}). To address this differences, we normalize the $\chi^{2}$ value dividing it by the total number of bins within each interval. We minimize this normalised quantity.

%This procedure is primarily applied to fitting the NFW function since it only has one free parameter and one normalization parameter. 
In the case of an Einasto profile, a combination of $C_i$ and $\alpha_i$ values are sampled from a grid. 
For $C_i$, we follow the same approaches as for NFW, and consider $\alpha_i$ values from $0.1$ to $2$, incremented by $\Delta \alpha=0.01$
%SA:rm: ${\rm d}\alpha = 0.01$.
%Dejar claro que se hace con sage antes de hablar de fits, tengo que decir que para el caso r/rvir nos hemos quedado solo hasta con los casos de r = rvir, debido que a partir de aqui hay muy pocos objetos. Para el caso r/rs cortar en 10 donde ya hay muy pocas galaxias.
%Plot Xi square frente C, variendo los delta C, mirar si se obtienen cosas extrañas en la Xisquare. Mirar directamente sumatorios en lugar de integrales, multiplciar por el ancho. Añadir detalles sobre como calculamos la chisquare, el rango de C

\begin{figure*}
	\includegraphics[width=\columnwidth]{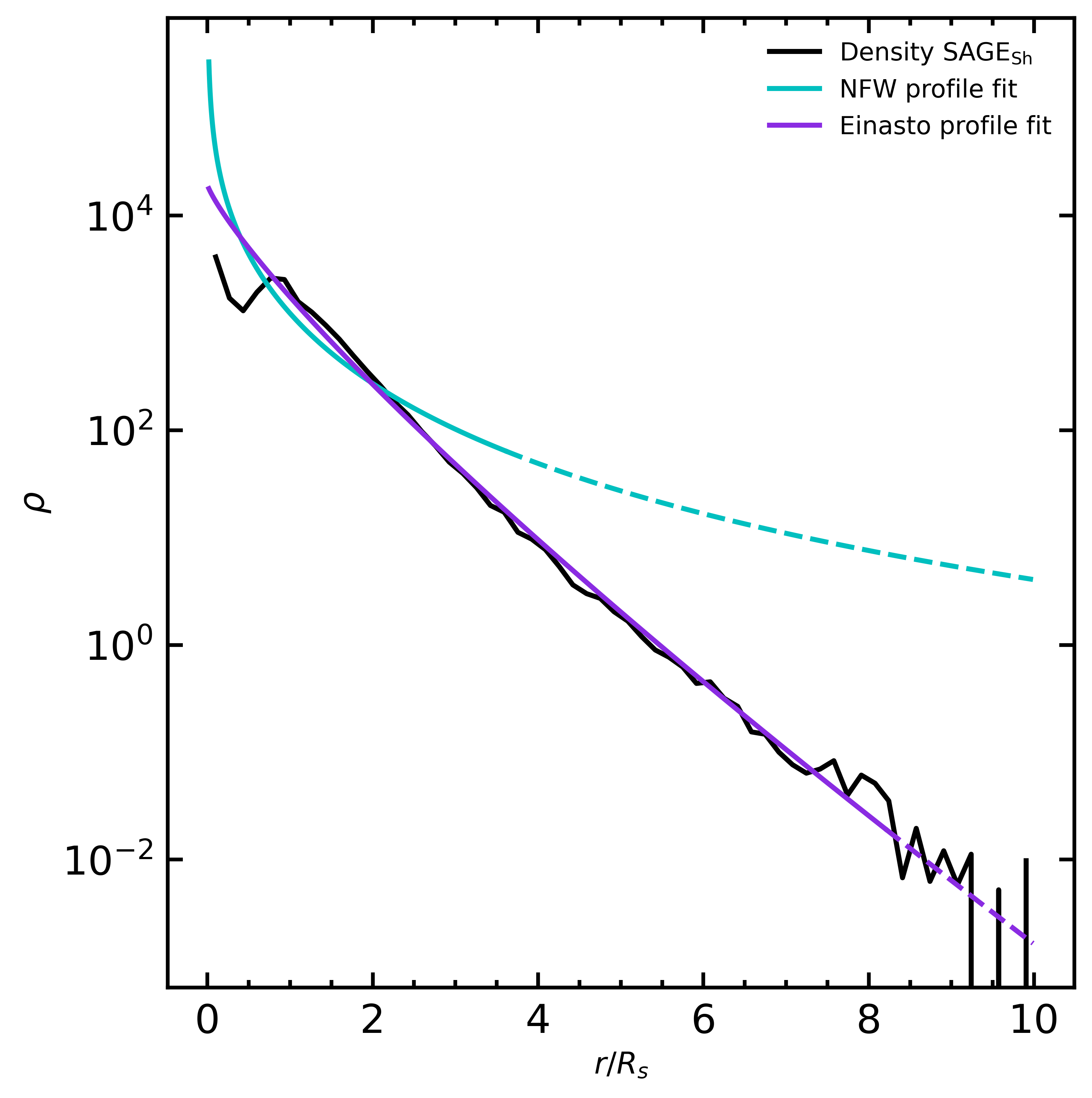}\includegraphics[width=\columnwidth]{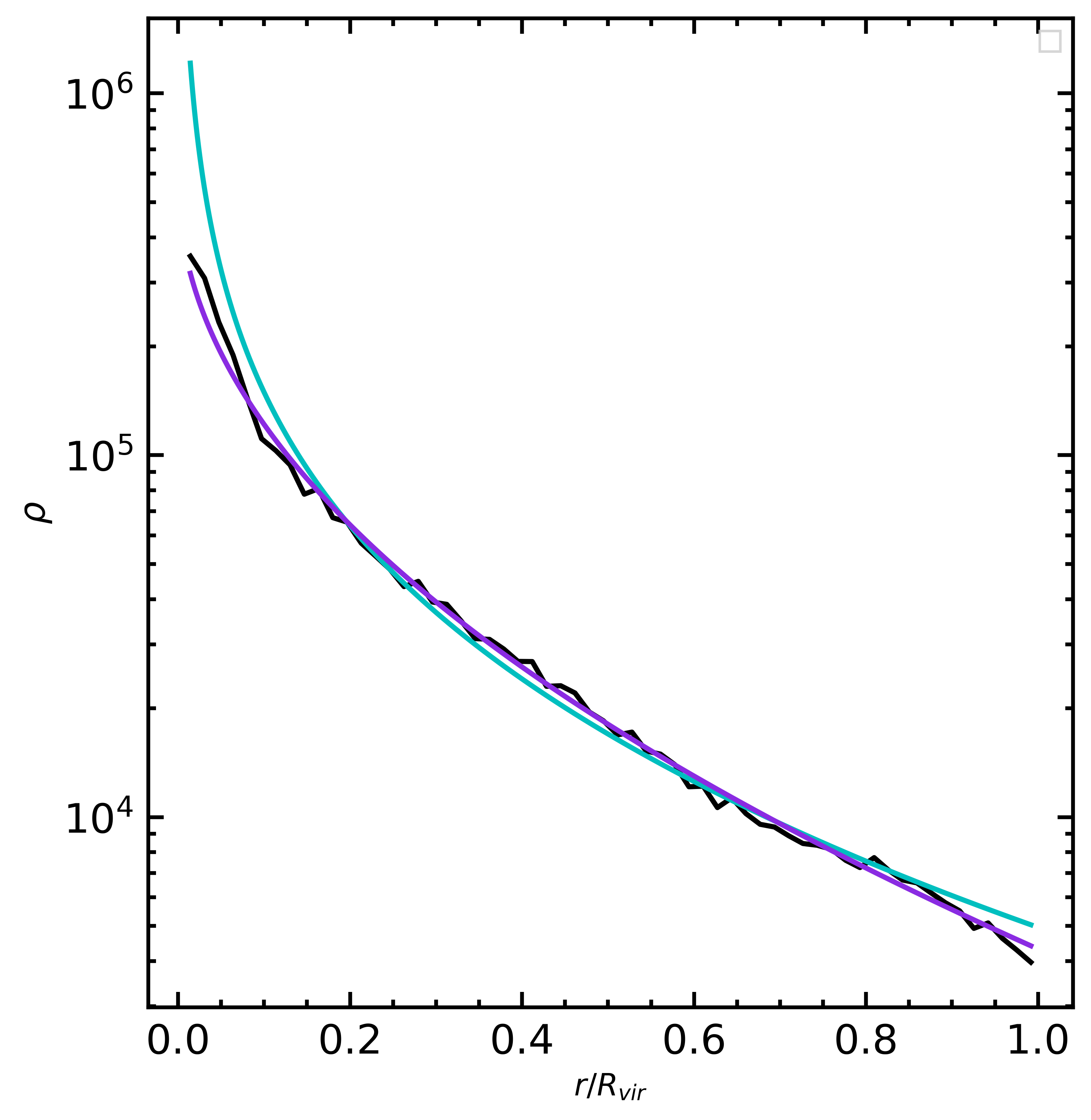}
    \caption{Density profiles of the \sagesh ELGs as a function of $r/R_{\rm s}$, left, and $r/R_{\rm vir}$, right, in black, together with fits using either NFW, in blue, or Einasto profiles, in purple.
    In the case of density profiles as a function of $r/R_{\rm s}$, the average concentration, $\langle C\rangle$, is a free parameter that appears as an integration limit (see \autoref{eq:NRvNFW}). We show as solid lines the fit up to $\langle C\rangle$, beyond this value we show dashed lines. Note that in this case we are effectively averaging different parts of haloes with different sizes.
    }\label{fig:EinastoNFW}
\end{figure*}

\autoref{tb:parametros} presents the best fit parameters to the \sagesh $r/R_{\rm s}$-stacked satellite profile for both NFW and Einasto profiles. The left panel of \autoref{fig:EinastoNFW} show the best fit profiles compared to the ELGs. We can observe that the Einasto density profile fits reasonably well, whereas the NFW profile is far from being a good fit.
%\vgp{this needs to be quantified}

\begin{table}
\begin{center}
\begin{tabular}{| c c c |}
\hline
$r/R_{\rm s}$ & NFW & Einasto  \\ 
\hline
\\
$\rho_{0}\left(\frac{ h^{3}}{\rm Mpc^{3}}\right)$ & 1227 &  1742\\
\\
C & 3.60 &  8.20\\
$\alpha$ & - & 0.83\\\hline
\\
$r/R_{\rm vir}$ \\ 
\\
\hline
\\
$\rho_{0}\left(\frac{ h^{3}}{\rm Mpc^{3}}\right)$ & 3941 &  10593\\
\\
C & 0.89 & 1.49 \\
$\alpha$ & - & 0.51 \\\hline

\end{tabular}
\caption{Best fit parameters to the \sagesh $r/R_{\rm s}$ and $r/R_{\rm vir}$-stacked satellite profile for both NFW and Einasto profiles.
It should be noted that the best fit of NFW for C is very similar to what we obtain by averaging the $C_{i}$ values of haloes containing satellite galaxies.
%\vgp{is this for Rs, Rvir or both???}
}
\label{tb:parametros}
\end{center}
\end{table}
 
Finally, we generate the corresponding mocks using these fitted $r/R_S$ profiles. In  \autoref{fig:radialprofile}, we can observe the profiles in terms of $r$ (dashed violet and cyan). %SA: reprtitive: that we obtain for the fits of the NFW and Einasto density profiles in terms of $r/R_{\rm s}$ as described in this subsection. 
We find that even when utilizing our fitted $r/R_{\rm s}$-stacked profiles, we still achieve a poor $N(r)$ curve, akin to the one obtained through the halo-by-halo sampling approach (blue curve). This is particularly striking for the Einasto profile, which showed a good fit in the stacked $\rho(r/R_s)$ profile.

\subsection{\texorpdfstring{$r/R_{\rm vir}$}{r/Rvir}-stacked profiles}
\label{subsec:r/Rvir}

In order to address the next scenario, we begin by rewriting equations \ref{eq:NFW} and \ref{eq:Einasto} in terms of the virial radius of the haloes with $R_{\rm vir} = C\hspace{1mm} R_{\rm s}$. The viral radius is defined as the radius that contains a mass with a virial overdensity, defined by $\rho$ from \citet{Bryan} and internally computed by \textsc{Rockstar}. 
It is considered the size of the halo, hence objects beyond this radius are considered outside the halo. By now stacking the profile in terms of $r/R_{\rm vir}$ we are encapsulating the information of the size of each halo (instead of the concentration as in \autoref{subsec:r/Rs}).

Following a similar procedure to the previous case, we observe that both classical models have the same number of free parameters and normalization parameters for fitting using r/$R_{\rm vir}$. However, in this situation, the free parameter C does not appear as an integration limit, but rather as a component within the density profile functions for both models. To ensure the conservation of the total number of satellites, we integrate these theoretical expressions up to r = $R_{\rm vir}$, leading to the following equivalences:
\begin{equation}
N_{\rm sat} = R_{\rm vir}^3\int_{0}^{\left( \frac{r}{R_{\rm vir} }\right)= \frac{R_{\rm vir}}{R_{\rm vir}}=1 }4\pi \left( \frac{r}{R_{\rm vir}}\right)^{2}\rho\left(\frac{r}{R_{\rm vir}}\right) {\rm d}\left( \frac{r}{R_{\rm vir}}\right) \ 
\label{eq:NRvNFW}
\end{equation}
It is important to note that, in this case, when performing the fit, we truncate the density profile obtained from our sample of satellite galaxies up to  $r = R_{\rm vir}$, as there are very few satellite galaxies beyond this value.
Technically, there should not be any. By definition, there are no subhaloes outside the virial radius, but there are some exceptions, likely due to how the merger trees are constructed with \textsc{ConsistentTrees} \citep{MergerTrees}.

For the fitting, we use the same procedure as in the previous section, with a few differences. Now, C does not appear as an integration limit but is explicit present in the functions of both models. Likewise, in this case we always evaluate the number of data points from our histograms ($0\leq r \leq R_{\rm vir}$ with $\Delta(r/R_{\rm vir})$ = 0.01), eliminating the need for normalization by the total number of bins.

In the right panel of \autoref{fig:EinastoNFW}, we observe the fittings achieved by the parameter set that minimizes the $\chi^{2}$ of the classical models (Einasto and NFW) with respect to our sample of \sage satellite galaxies, when stacking the profiles on the variable $r/R_{\rm vir}$. The resulting parameters for both models are shown in \autoref{tb:parametros}. For this case, we can confirm that a decent qualitative fit has been achieved for both the NFW and Einasto profiles, unlike the $r/R_{\rm s}$ case. However, the Einasto profile still provides a better fit, as expected, given that it includes one more free parameter.

It is worth noting that even though both models share the same physical parameters ($\rho_s$, $C$, $\alpha$), the values obtained differ when using $r/R_{\rm s}$ or $r/R_{\rm vir}$ as our variable (see \autoref{tb:parametros}). These differences stem from the lack of a direct correlation between the variables $R_{\rm s}$ and $R_{\rm vir}$, with a Pearson correlation coefficient of only $r_{\rm R_{\rm vir}, R_{\rm s}}$ = 0.36. As a result, we find a great dispersion between these variables, between these and the concentration $C$ and also between all those quantities and the $r$ found in the ELGs.
This non-univocal relation between the variables considered, makes possible the differences observed in the goodness of fit and best fit parameters when considering a fit to an  $r/R_{\rm S}$-stacked profile or an $r/R_{\rm vir}$-stacked profile. 

This level of dispersion is consistent with what has been previously presented in \citet{Salvador}. In the aforementioned article, various models with distinct simulations and redshift values, showing an existence of a concentration dispersion in relation to the mass of haloes. The resulting $C-M_{200c}$ diagrams can be extrapolated using the dispersion we discovered for the variables $R_{\rm s}$ and $R_{\rm vir}$, which are correlated with mass. 

In \autoref{fig:radialprofile}, we can observe the profiles as a function of distance ({$r$}) that we have obtained from the generated mocks using the average fits of the NFW and Einasto density profiles in terms of $r/R_{\rm vir}$, as described in this subsection. As observed, both Einasto and NFW models show better results when the fits are performed in terms of $r/R_{\rm vir}$ (dotted dashed)  compared to when we use $r/R_{\rm s}$ (dashed), in comparison with our reference, \sage (black dots). It is also possible to appreciate a better adjustment of the Einasto model with respect to the NFW model, at least on the low-$r$ side, in line with the findings in \autoref{fig:EinastoNFW}. In light of these findings, when studying the clustering of the different samples, we will only show results of the $r/R_{\rm vir}$-fitted classical profiles (and not the $r/R_{\rm S}$ ones), to avoid overcrowding our figure.

%It is important to note that, in terms of clustering, differences arise since it is not possible to accurately reproduce the global radial profile of \sagesh satellite galaxies using these adjustments. 
The differences observed in the profile as a function of distance $N(r)$ translate to differences in the observed galaxy clustering.
These differences are reflected in \autoref{fig:radialprof2PCF}, where a strong clustering is observed at scales below $\sim 0.1 h^{-1}{\rm Mpc}$. This difference reaches almost $\sim40\%$ for both the NFW and Einasto cases at $r \sim 0.1 h^{-1}{\rm Mpc}$. Furthermore, we can confirm that the Einasto fit proves to be superior to the NFW concerning small scales ($<0.1 h^{-1}{\rm Mpc}$), consistent with the results obtained earlier in this subsection. 
%SA: remove. Otherwise it is very easy to get lost. it is simpler to just look at the figure. Also it sounded like it is doing better than the otehr curves, but it is not.

\subsection{The inherent \texorpdfstring{\sagesh}{SAGEsh} distribution \texorpdfstring{$\rho_{\rm SAGE_{sh}}(r)$}{RhoSAGEsh} profile.}
\label{subsec:The inherent}

In order to achieve a better fit on the small scale in terms of clustering compared to the various prescriptions discussed in the previous subsections, we make use of the inherent \sagesh distribution $\rho_{\sage}(r)$ profile. This allows us to determine the discrete distribution profile (in a histogram) of our sample of satellite galaxies from \sagesh. We use linear binning with a bin size of $\Delta r = 0.01 h^{-1}$Mpc. We then use this histogram as a discrete distribution function to generate new catalogs of satellite galaxies with their respective positions. In \autoref{fig:radialprof2PCF}, we can see the clustering we obtain for this case (gold curve). Where it can be observed that it corresponds to the best results in terms of fitting the small scale, below $r = 1  h^{-1}{\rm Mpc}$. It is worth noting that for $r \sim 0.1 h^{-1}{\rm Mpc}$, this difference is below $25\%$.

\subsection{Extending NFW for an average  \texorpdfstring{$\rho(r)$}{Rho(r)}profile.}\label{subsec:analytical}

In this section, we present an analytical expression that describes the \sagesh profile with greater accuracy. Since the previous prescriptions did not perform adequately in reproducing the radial profile ($N(r)$) of our \sagesh satellite sample, we now attempt to directly fit the profile based on the variable $r$. This variable is the most closely related one to observable quantities and during our study we have found that the profiles in $r$ are a better indicator of the measured clustering. %First and foremost, it is worth highlighting that we also used the 
For example, as we see in \autoref{subsec:The inherent}, when we use the $N(r)$ measured directly from \sagesh, we achieved good results in terms of clustering. Motivated by this, we now perform an analytical fit to $N(r)$ that may make more {\it portable} our results to other models and easily implementable in different simulations.
After some search, we found that the following analytical function  describes well the $N(r)$ profile of \sagesh: 
\begin{equation}
N(r) = N_{0} \cdot \left(\frac{r}{r_{0}}\right)^{\alpha} \left(1 + \left(\frac{r}{r_{0}}\right)^{\beta}\right)^{\kappa} \,, 
\label{eq:NFWM}
\end{equation}
with $N_{0} = 3928.273$ , $r_{0}= 0.34h^{-1}{\rm Mpc}$, $\alpha = 1.23$, $\beta = 3.19$ and $\kappa = -2.1$. Note that these parameters are free, and this case is applicable to our binning in $\Delta r=0.01$, thus the normalization would change with the binning.

In \autoref{fig:radialprofile} it is evident that our implemented analytical function (solid red line) offers a better fit to describe the \sage profile compared to all the other used prescriptions. 

The derived analytic expression can be interpreted as a generalization of the NFW density profile:

\begin{equation}
\rho(r) = 4\rho_{0}\cdot \left(\frac{r}
{r_{0}}\right)^{\alpha -2} \left(1 + \left(\frac{r}{r_{0}}\right)^{\beta }\right)^{\kappa } \ ,
\label{eq:mNFW}
\end{equation}
where $\rho_{0} = N_{0}/16\pi r_{0}^{2}$. When $\alpha = 3$, $\beta = 1$ and $\kappa = -2$, the expresion above is reduced to the classical NFW profile (\ref{eq:NFW}). In this case, the free parameter $r_{0}$ would correspond to the $R_{s}$ variable in the NFW density profile, and our definition of $\rho_{0}$ with the $\rho_{s}$ variable.

\autoref{fig:radialprof2PCF} shows how reproducing the global radial profile of \sagesh satellite galaxies using an analytical extension of the NFW density profile yields similar results to those we obtain if we use the histogram of the distribution profile of satellite galaxies in \sagesh (gold curve). For small scales, below $r = 1 h^{-1}{\rm Mpc}$, our analytical expression (red curve) achieves a clustering that reproduces that of \sagesh within $\sim25\%$.

\subsubsection*{Variations with number density and redshift}

Since we will adopt this extended NFW profile in our default model, we now explore how this model performs when we choose a different reference sample by adopting a different reference number density (\autoref{tab:nd_models}) or a different redshift (\autoref{tab:densityz}). 

%%%%%Variations with nd and z
\begin{table}
\begin{center}
\begin{tabular}{ c c c c c c}
Sample & Redshift & $\alpha$ & $\beta$ &$\kappa$ & $r_{0}$ \\\hline
Pozzetti nº1 & $1.3$ & $1.29$ & $2.98$ & $-2.21$ & $0.31$\\
{\bf Pozzetti nº3} & $\mathbf{1.3}$ & $\mathbf{1.23}$ &  $\mathbf{3.19}$ & $\mathbf{-2.1}$  &  $\mathbf{0.34}$ \\
Flux cut  & $1.3$ & $1.1$ & $3.17$ & $-2.49$  &  $0.44$ \\\hline
{\it dPozzetti nº3} & $0.987$  & $1.35$ &  $2.53$ & $-2.56$ & $0.33$  \\
{\it dPozzetti nº3} & $1.220$ & $1.24$ & $3.05$ & $-2.16$ & $0.34$ \\
%$\mathbf{1.321}$ & $\mathbf{1.23}$ &  $\mathbf{3.19}$ & $\mathbf{-2.1}$ & $\mathbf{0.344}$ \\ 
{\it dPozzetti nº3} & $1.425$  & $1.17$ & $3.27$ & $-2.09$ & $0.35$ \\
{\it dPozzetti nº3} & $1.650$  & $1.16$ & $3.05$ & $-2.63$ & $0.39$ \\ \hline
\end{tabular}
\caption{Values of the free parameter in \autoref{eq:NFWM} and \autoref{eq:mNFW} that best fit the radial profiles of our reference samples of ELGs for samples at different redshifts and with number densities obtained either in a range of redshifts, first three rows (\autoref{tab:nd_models}) or from a differential luminosity function (\autoref{tab:densityz}). Our default choice is the sample at $z=1.3$ with number density matching that of Pozzetti model number 3 in a range of redshifts. This is indicated by using bold face in the table.
}\label{tab:NWFMparameterz}
\end{center}
\end{table}

\begin{figure}
	\includegraphics[width=\columnwidth]{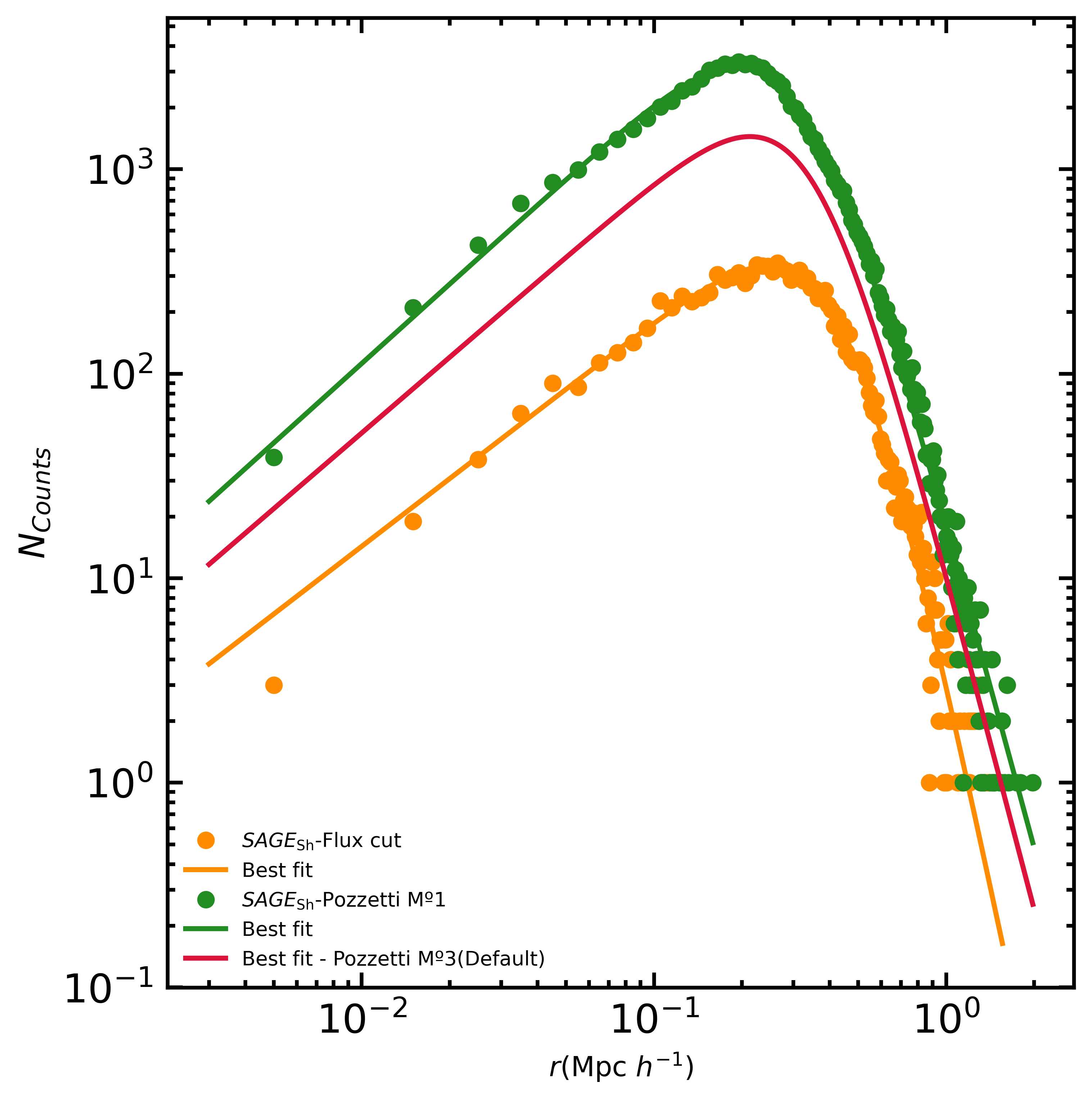}
    \caption{
    Radial profile for satellites at $z=1.3$ from our reference sample of  ELGs modelled by \sagesh (filled symbols) assuming two number densities: {\it Pozzetti model number 1} and {\it Flux cut} (\autoref{tab:nd_models}), as indicated in the legend. These reference profiles are compared with their best fits to the analytical function for the radial profiles described by \autoref{eq:NFWM} (\autoref{tab:NWFMparameterz}). For reference, the profile from the {\it default} case, {\it Pozzetti model number 3}, is shown as a red line.
    }
    \label{fig:radialprofiledensity}
\end{figure}

\begin{figure}
	\includegraphics[width=\columnwidth]{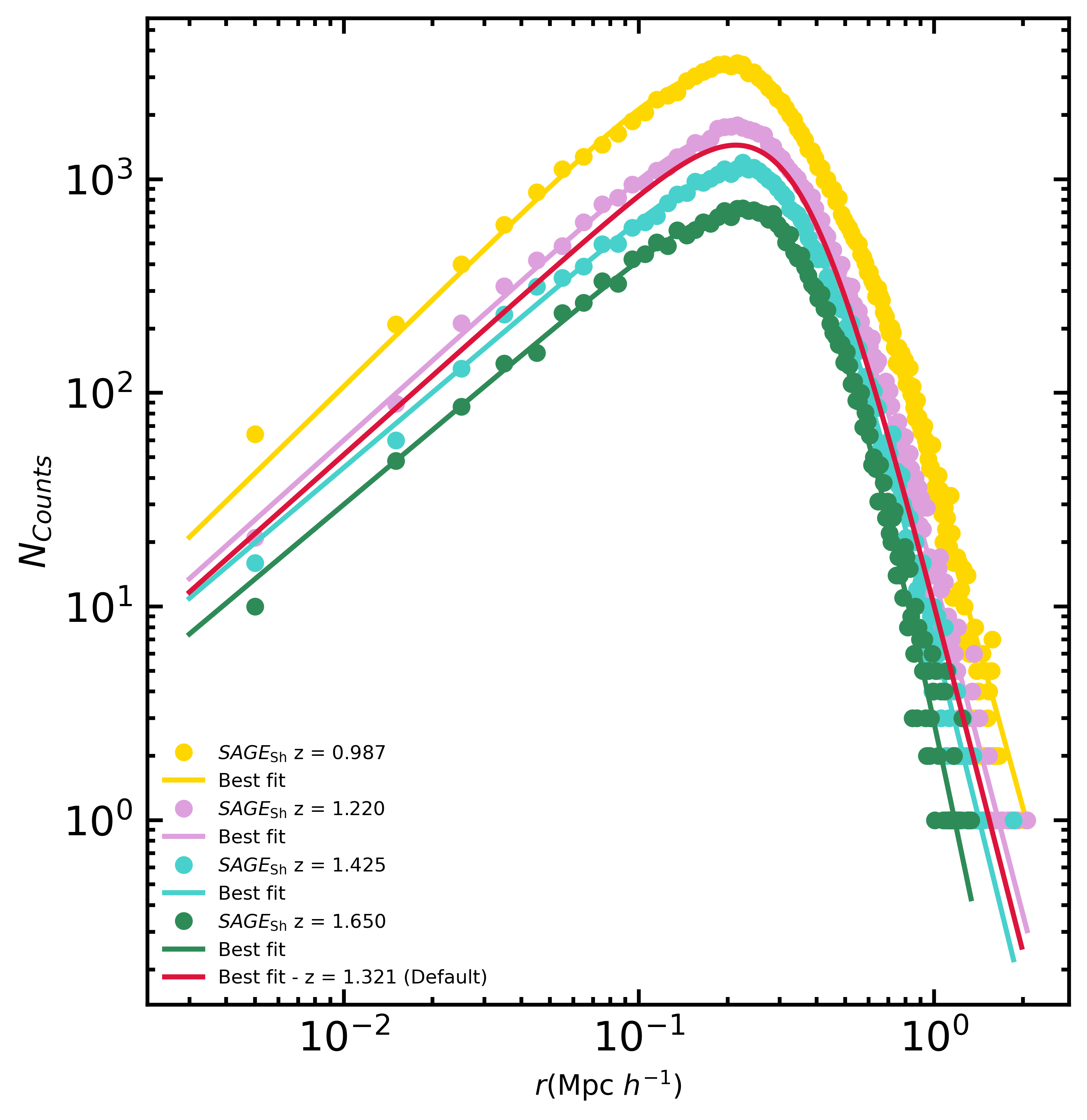}
    \caption{
    Radial profile for satellites at different redshift from our reference sample of  ELGs modelled by \sagesh (filled symbols) assuming number densities matching the differential luminosity function number 3 from \citeauthor{Pozzetti} (\autoref{tab:densityz}). These are compared with the corresponding best fits to the analytical function described in \autoref{eq:NFWM} (\autoref{tab:NWFMparameterz}). For reference, the profile from the {\it default} case, {\it Pozzetti model number 3} in a range of redshifts centered at $z=1.3$, is shown as a red line.
}\label{fig:radialprofilez}
\end{figure}

In \autoref{fig:radialprofiledensity} we compare the radial profile of satellites from the three \sagesh ELG samples at $z=1.3$ with different number densities (\autoref{tab:nd_models}). The radial profiles for the three \sagesh samples exhibit the same shape with varying normalisation, as expected from the change in number density. At $r=0.21{\rm Mpc}h^{-1}$, the value associated with the maximum number of counts for our reference sample, the count number increases by a factor of $2.19$ from the reference sample to that with the highest density. This factor practically coincides with the ratio of total satellite galaxies between both samples, $2.04$. The count number increases by a factor of $4.15$ from the sample with the lowest density to the reference one; approaching the value corresponding to the ratio of total satellite galaxies for that case, which is $3.73$. 

The three samples with varying number density at $z=1.3$ are well described by our proposed extended NFW profile (\autoref{eq:NFWM}). The best fit parameters corresponding to each sample are summarised in \autoref{tab:NWFMparameterz}. We find variations of a factor of $1.17$ for $\alpha$, $1.07$ for $\beta$, $1.19$ for $\kappa$, and $1.42$ for $r_{0}$. Thus, there is not much variation in the shape of the profiles for these three samples.

In \autoref{fig:radialprofilez} we show the radial profile of \sagesh satellite samples at different redshifts (see \autoref{subsec:evoz} for details on the construction of the samples). Following the decrease in number density of the samples (\autoref{tab:densityz}), the normalisation of the curves decrease with increasing redshift. At $r=0.21{\rm Mpc}h^{-1}$, the value associated with the maximum number of counts for our default sample, the count number increases by a factor of $4.52$ from the lowest value at $z = 1.650$ to that at $z = 0.987$. This factor is close to the ratio of total satellite galaxies between both extreme samples, which is $4.63$. 

At all redshifts, the shape of the profiles remain similar and well described by \autoref{eq:NFWM}. \autoref{tab:NWFMparameterz} summarises the best fit parameters to this equation for each redshift. The parameter $r_{0}$ remains nearly unchanged with redshift. We find the other parameters to change by at most a factor of $1.16$ for $\alpha$, $1.29$ for $\beta$, and $1.26$ for $\kappa$. 

Therefore, we can conclude, that \autoref{eq:NFWM} is a good description of the radial profiles of ELGs that are satellite galaxies at different redshifts.

\subsection{Clustering}\label{subsec:2PCF}

%\vgp{individual comparissons to their corresponding sections, here global comments and conclusion for this section}

In this subsection, we provide a global discussion of the results obtained for the clustering, comparing the 2-point correlation function of all previously mentioned radial profile prescriptions and concluding the respective section. These include sampling individual halo profiles assuming an NFW profile based on halo concentration, using NFW and Einasto density profiles as a function of $R_{\rm vir}$ to fit our data, utilizing the normalized histogram of \sagesh satellite galaxies as a function of $r$, and proposing an extension to NFW that fits the overall distribution observed for \sagesh galaxies. It is important to note that we generated 100 mocks with different seeds for each of the aforementioned descriptions.

In terms of clustering, we show in the \autoref{fig:radialprof2PCF} how the classical density profiles (Einasto and NFW) do not correctly reproduce the small scales clustering of \sage satellite galaxies for any of the prescriptions described above (halo-by-halo, $r/R_{\rm S}$-stacked-fit, $r/R_{\rm vir}$-stacked-fit,). Moving on, we see that the stacked fits show some improvement, but still exhibit a deviation of around 40$\%$ at 0.1 $h^{-1}{\rm Mpc}$. However, it is worth noting that among the stacked fits, the Einasto profile (purple curve) yields better results than the NFW profile (cyan curve), which shows a maximum difference of 60$\%$ in the range $r<$0.1$h^{-1}{\rm Mpc}$. Finally, we have the curves corresponding to the radial profile of our \sagesh sample (yellow curve) and the extended NFW profile that we have implemented (red curve). We can verify that these fits provide the best results for small scales compared to the previous prescriptions, with differences less than 22$\%$ to $r\sim 0.1h^{-1}{\rm Mpc}$ and ratios closer to 1 than the other profiles.

%\SA{Falta que comentes esta figura más, línea por línea. Ejemplo (a explicar bien):
%La que peor ajusta es la azul, se va mas de 80\% a 0.1... Esto es muy importante porque es lo más utilizado en la literatura... 
%Después vemos que los ajustes stacked mejoran algo, pero fallan al \% 
%Finalmente la linea roja... y la amarilla...}
%\SA{No tienes que motivar esto, aquí->}
The great improvement in the small scale clustering (when compared to our reference \sagesh sample) introduced by the \textit{modified} NFW in \autoref{eq:mNFW} is one of the main results of this paper. 
It should be noted that when we analyzed the $\chi^2$ of all the fits shown in the \autoref{fig:radialprof2PCF}, the one that presented the smallest value for the small scale was the one in which we used the own distribution profile of \sage ($N(r_i)$): $\chi^2(r<1 h^{-1}{\rm Mpc})= 36.35$, followed by our Default model (\autoref{sec:DefaultHOD}, using Eq. \ref{eq:mNFW}): $\chi^2(r<1 h^{-1}{\rm Mpc} )= 42.34$ compared to $\chi^2(r<1 h^{-1}{\rm Mpc} )= 654.32$  and $\chi^2(r<1 h^{-1}{\rm Mpc} )= 54.01$ for {\it halo-by-halo} and $r/R_{\rm vir}$ Einasto, respectively. 
However, we have chosen \autoref{eq:mNFW} our default model because it is more practical to implement an analytical expression that can be used in other simulations without resorting to the actual profile of the sample or running \sage again. 
%Specifically, to compare the performance of those prescriptions in small scale clustering, we only take into account the values of the 2PCF have you defined what is the 2PCF at $r<1{\rm Mpc} \,h^{-1}$. 
In accordance with the results obtained in terms of clustering, we can observe that it is interesting how galaxies do not seem to follow the classic dark matter density profiles, despite the different prescriptions made. This result aligns with various current studies \citep{Parkinson2}.%\vgp{ojo, te he corregido el formatio de las unidades}

%\begin{table}[t]
%\begin{center}
% \begin{tabular}{ c c c c c|}
% Small scale (r < 1) & NFW & Einasto & Analytical function & \sage profile \\\hline
% $\chi^{2}(r) $ & 49.47 & 54.02 & 42.34 & 36.35\\\hline

% \end{tabular}
% \caption{.}
% \label{tab:nd_models}
% \end{center}
% \end{table}

%\section{The Velocity profile}

\section{Summary and Conclusions}
\label{sec:conclusion}
%\grp{dejar claro que tenemos un codigo( enlace del codigo) dado un input de galaxias satellites nos produce un HOD siguiendo los perfiles y la conformdiad correspodiente. Codigo es general para culquier tipo de modelo y ademas es modular ( se puede hacer modular, extensiones velocidades)}
%Elimination of assembly bias: The study focuses on Euclid-like model galaxies from shuffled SAGE, removing the influence of assembly bias.
%SA: Yo no hablaría del código aún. No lo hemos revisado para nada. Y creo que recordar que estaba como dividido en muchos códigos. Podemos esperar con esto 

We have explored the essential components needed in a HOD prescription to reproduce the clustering of ELGs.
In particular, we aim to reproduce the small-scale of the real-space two-point correlation function (2PCF) from our reference sample, which matches the  abundance of ELGs expected to be detected by Euclid. In this work we have proposed a Halo Occupation Distribution (HOD) prescription capable of significantly improving the clustering of model emission line galaxies compared to HOD models without conformity and/or assuming a radial NFW profile based on individual halo concentration.  

We have generated our reference catalogue of model galaxies by running the semi-analytical model (SAM) of galaxy formation \sage on the UNITsims dark matter simulations \citep{2019Unitsims}, as outlined in \citet{Knebe2022}.
The reference sample is obtained by selecting a UNITsim-\sage galaxy sample at $z = 1.321$ with a cut in \ha flux to match 
the density predicted by the \cite{Pozzetti} model number 3 over the entire Euclid redshift range ($0.9<z<1.8$).
%a target number density. 
%The chosen redshift, $z = 1.321$, approximately corresponds to the  mean value of the redshift range of the Euclid spectroscopic survey, . The target number density corresponds to that predicted by the \cite{Pozzetti} model number 3 over the entire Euclid redshift range. 
%SA: innnecesario: This selection of \ha galaxies is specific to the objectives of this article, unlike the samples selected in \citet{Knebe2022}. 
%SA: no lo veo necesario aqui: We will focus solely on studying the conformity and radial profile properties to accurately reproduce the 1-halo term. 
We then \textit{shuffle} the \sage sample (see \autoref{subsec:shuffle}), to remove assembly bias from our reference sample, \sagesh (\autoref{sec:sage}).  
%SA: too much detail, refer to section.: We shuffle the positions of haloes with similar masses by carrying together all the galaxies living in that halo and  preserving the galaxy relative positions and velocities within each halo. Hence, the resulting catalogue, \sagesh, will be our reference sample for \ha ELG clustering. 
This last step allows us to focus on the main goal of the paper: the influence of conformity and the radial profiles on galaxy clustering without {\it contamination} from assembly bias. 

Our HOD prescription is very modular, containing a series of ingredients that we have study in comparisson to the clustering measured for the \sagesh reference sample (\autoref{sec:HODm}). We focus our analysis on the effect that conformity and the radial profiles have on the final galaxy clustering. However, we outline here all the ingredients of our HOD model. We do this following the order in the code and highlighting the choice made for our {\it default} model: 

\begin{itemize}[wide, labelwidth=!,itemindent=!,labelindent=0pt, leftmargin=0em,parsep=0pt]
    \item  \underline{Mean halo occupation shape} for centrals, $\langle N_{\rm cen}(M) \rangle$, and satellites, $\langle N_{\rm sat}(M) \rangle$. Here, we fix these properties to what is directly measured in \sage and shown in \autoref{fig:HMF}.

%To achieve this, we analyze the shape of the mean HOD of our \ha emission galaxy sample and make the standard selections: a Poisson distribution for satellites and a Bernoulli distribution for centrals.

%Subsequently, we delve into studying the spatial distribution of satellite galaxies within dark matter haloes in \sagesh. By implementing these two properties into an HOD model, we want to modelling of the 1-halo conformity and the radial placement of satellite galaxies within haloes.

%"\grp{informacion sobre el valor medio y probabilidad del HOD}

%\grp{mencionar comentario del código}

%We have focused on the modelling of the 1-halo conformity and the radial placement of satellite galaxies within haloes.\vgp{explica sucintamente conformity y radial profile}\vgp{as there should be ony 2 points, remove bullet points}

%\item The shape of the mean HOD $\langle N(M_h)\rangle$: we fixed it and determined it by measuring the average occupancy of central and satellite \sage galaxies in mass bins of dark matter, as shown in \autoref{fig:HMF}. \vgp{remove, not relevant as you keep it fixed}

%\item Probability Distribution Function (P(N\lvert$\langle N\rangle)$). Here we take the standard choices: a Poisson distribution for the satellites and a Bernoulli distribution for centrals (see \autoref{sec:pdf}).  \vgp{I don't think we need this here}

\item \underline{Conformity}. We investigate whether satellite and central galaxies of a given type are independent events. We define the 1-halo conformity as deviations from independence and we quantify this with two factors $K_1$ and $K_2$ (\autoref{eq:k1(m)}, \autoref{eq:k2(m)}). We consider two cases for modeling conformity: 
one where the modification of satellite occupation is performed in mass bins (mass dependent conformity), and the other with global factors, constant for all mass bins (global conformity). Introducing conformity greatly  improves the match to the 2PCF of the reference sample at small scales, with respect to the independent case.  \autoref{fig:xiconformity} shows that the 2PCF of HOD models assuming independence, present a $\sim 60\%$ difference at $r\sim 0.1 {\rm Mpc}\,h^{-1}$ with respect to the reference sample. This difference is reduced to $\sim 20\%$ when conformity is introduced in the HOD prescriptions. Both models of conformity result in similar 2PCF, and thus, we adopt the {\bf Global conformity} as our {\it default} model for simplicity. 

%\vgp{termina la parte de conformity con algo de contexto: Lacerna, DESI, etc. Puedes utilizar la frase siguiente que tenías más abajo}We emphasize the importance of conformity in reproducing the satellite distribution in \sage.

\item \underline{Probability Distribution Function}. We assume a Bernoulli distribution for central galaxies and a Poisson one for satellites. These are good descriptions for our reference sample (\autoref{sec:pdf}).

\item \underline{Radial profile.} These are the prescriptions for the radial profiles of satellite galaxies we have implemented in our HOD models (\autoref{sec:radialprofile}): 
\begin{enumerate}[wide, labelwidth=!,itemindent=!,labelindent=0pt, leftmargin=0em, label=(\roman*), itemsep=0.2cm, parsep=0pt]
    \item Sampling individual halo profiles assuming an NFW given by the concentration of each halo. This is usually the default in the literature.
    \item Adjusting the NFW and Einasto curves to the $r/R_{\rm s}$-stacked profile from our reference sample. With $r$ being the distance between a satellite galaxy and the center of its host halo. $R_{\rm S}$ is the scale radius of the halo. 
    \item Adjusting the NFW and Einasto  curves to the $r/R_{\rm vir}$-stacked profile from our reference sample. $R_{\rm vir}$ is the viral radius of the halo. 
    \item The {\it inherent} distribution profile of satellite galaxies from our reference sample, measured directly as a normalised histogram of satellite counts as a function of $r$.
    \item {\bf Modified NFW profile}. We introduce a generalized version of the NFW density profile that effectively models the stacked profile of our reference sample of galaxies as a function of $r$. We find an excellent fit with the following modified NFW curve (\autoref{eq:mNFW}) and we adopt it for our {\it default} HOD model:
    \begin{equation*}
\rho(r) = 4\rho_{0}\cdot \left(\frac{r}
{r_{0}}\right)^{\alpha -2 }\left(1 + \left(\frac{r}{r_{0}}\right)^{\beta }\right)^{\kappa } \, .
    %\label{eq:mNFW}
    \end{equation*}
\end{enumerate} 
%Where $\rho_{0} = N_{0}/16\pi r_{0}^{2}$, $r_{0}= 0.34(Mpc/h)$, $\alpha = 1.23$, $\beta = 3.19$ and $\kappa = -2.1$. 
\end{itemize}

The NFW and Einasto density profiles are unable to replicate the small-scale clustering of our sample of \sagesh ELGs, with any of the implementations tried here. It is particularly striking the case of the Einasto $r/R_{\rm vir}$-stacked curve, which shows a very good fit in $\rho(r/R_{\rm vir})$, but does not recover the clustering from the reference sample. We argue that this is due to the low correlation between $r$ and $R_{\rm vir}$ in the reference sample. Therefore, we provide an analytical expression for the positional profile (v above) of this type of satellite galaxies, which can be interpreted as an extension of the NFW profile (\autoref{eq:NFWM} and \autoref{eq:mNFW}). 

We find that a good fit to the radial profile, $N(r)$, of our \sagesh reference sample is a good predictor for how good the clustering of galaxies generated with an HOD model will be reproducing the reference one (\autoref{fig:radialprof2PCF}).
The goodness of the two last  presciptions above (inherent, iv, or fitted, v) stand out with respect to all the others (\autoref{fig:radialprofile}). Directly using the inherent $N(r)$ profile (iv above) provides a clustering that matches that from the reference sample slightly better than using the analytical fitted expression (v above). However, the gain is  small compared to the error bars. As we find difficult to use the inherent \sagesh $N(r)$ for different simulations or reference samples, we decide to use as our {\it default}  the analytical expression above (\autoref{eq:mNFW}).

Our proposed ({\it default}) HOD model includes a model for the 1-halo conformity and a modified NFW radial profile for satellite galaxies. The conformity model is implemented by computing two constants: $K_{1,\, \rm glob}=0.708$ (\autoref{eq:k1gf}) and $K_{2,\, \rm glob}=1.038$ (\autoref{eq:k2gf}); and then applying them within the HOD model using \autoref{eq:themeannumberswc} and  \autoref{eq:themeannumbersw/oc}.
We assume that satellite galaxies occupy haloes following the modified NFW profile described by \autoref{eq:mNFW}. This equation has 4 free parameters: $\alpha$, $\beta$, $\kappa$ and $r_0$ (\autoref{subsec:analytical}). 

The proposed {\it default} HOD model improves significantly the small scale clustering when compared to the benchmark model we started from, the {\it vanilla} HOD. This HOD model does not include conformity and assumed a radial profile for satellite galaxies based on a halo-by-halo NFW profile (\ref{sec:VDefaultHOD}). This is clearly depicted in \autoref{fig:vanilla2PCF}, where we see a $\sim15\%$ improvement at $r\sim0.3h^{-1}{\rm Mpc}$ and more than $50\%$ improvement below $\sim 0.1 h^{-1}{\rm Mpc}$.

There are four \sage-UNITsim simulation boxes. To understand how much of what we learn from one simulation box can be extrapolated to the other ones, we have applied our {\it default} HOD model to the other simulation boxes. The measured clustering is consistent among the boxes, with noise levels similar to those found for the corresponding variations in \sage. Hence, we conclude that any noise in our inferred parameters would not affect our conclusions.

The main conclusion of this work is that modelling the clustering of ELGs requires the inclusion of conformity and a radial distribution of satellite galaxies as a function of the distance to the corresponding central galaxy (\autoref{eq:mNFW}). This conclusion is robust to different number densities (those summarised in \autoref{tab:nd_models}) and redshifts between $0.987$ and $1.650$. 

The work presented here can have a large impact in the way mock catalogues of emission line galaxies are generated in the future. This can be of special relevance for Euclid, but also for DESI, Roman or 4MOST, which will study the clustering of galaxies with strong spectral emission lines. 
As a next step, we aim to implement the velocity profile of the satellites to investigate its impact on the study of redshift space distortions. 
A natural follow up to this paper would be to study the assembly bias for ELGs, by exploring the differences in clustering between \sage and \sagesh (\autoref{fig:vanilla2PCF}).

As we enter the fourth stage of dark energy experiments, with sub-percent precision cosmology, controlling the origin of any contribution to galaxy clustering is fundamental. Hence, studies like this one will help us in the future to understand the different contributions of galaxy formation physics to the observed galaxy clustering. %\vgp{no entiendo qué quieres decir}

\section*{Acknowledgements}

We would like to thank the anonymous referee for their insightful comments, that pushed us to explore how our results change with varying number density and redshift. GRP is supported by the FPI SEVERO OCHOA (SEV-2016-0597-18-2) program from the the Ministry of Science, Innovation and Universities, with reference PRE2018-087035. 
This work has been supported by Ministerio de Ciencia e Innovaci\'{o}n (MICINN) under the following research grants:
PID2021-122603NB-C21 (VGP, AK and GY), 
PID2021-123012NB-C41 (SA) and PGC2018-094773-B-C32 (GRP, SA). 
GRP, SA and VGP have been or are supported by the Atracci\'{o}n de Talento Contract no. 2019-T1/TIC-12702 granted by the Comunidad de Madrid in Spain. 
IFAE is partially funded by the CERCA program of the Generalitat de Catalunya. AK further thanks Brighter for the `around the world in eighty days' EP. The UNIT simulations have been run in the MareNostrum
Supercomputer, hosted by the Barcelona Supercomputing Center,
Spain, under the PRACE project number 2016163937.

%%%%%%%%%%%%%%%%%%%%%%%%%%%%%%%%%%%%%%%%%%%%%%%%%%
\section*{Data Availability}

The data used in this study are publicly available
at \url{http://www.unit sims.org}, as described in \citep{Knebe2022}. The code developed during this analysis will be released at a later stage. In the meantime it can be shared upon request.



\bibliographystyle{mnras}
\bibliography{HODSAGE} % if your bibtex file is called example.bib

\begin{thebibliography}{}
\makeatletter
\relax
\def\mn@urlcharsother{\let\do\@makeother \do\$\do\&\do\#\do\^\do\_\do\%\do\~}
\def\mn@doi{\begingroup\mn@urlcharsother \@ifnextchar [ {\mn@doi@}
  {\mn@doi@[]}}
\def\mn@doi@[#1]#2{\def\@tempa{#1}\ifx\@tempa\@empty \href
  {http://dx.doi.org/#2} {doi:#2}\else \href {http://dx.doi.org/#2} {#1}\fi
  \endgroup}
\def\mn@eprint#1#2{\mn@eprint@#1:#2::\@nil}
\def\mn@eprint@arXiv#1{\href {http://arxiv.org/abs/#1} {{\tt arXiv:#1}}}
\def\mn@eprint@dblp#1{\href {http://dblp.uni-trier.de/rec/bibtex/#1.xml}
  {dblp:#1}}
\def\mn@eprint@#1:#2:#3:#4\@nil{\def\@tempa {#1}\def\@tempb {#2}\def\@tempc
  {#3}\ifx \@tempc \@empty \let \@tempc \@tempb \let \@tempb \@tempa \fi \ifx
  \@tempb \@empty \def\@tempb {arXiv}\fi \@ifundefined
  {mn@eprint@\@tempb}{\@tempb:\@tempc}{\expandafter \expandafter \csname
  mn@eprint@\@tempb\endcsname \expandafter{\@tempc}}}

\bibitem[\protect\citeauthoryear{{Abbott} et~al.,}{{Abbott}
  et~al.}{2018}]{Abbott2018b}
{Abbott} T.~M.~C.,  et~al., 2018, \mn@doi [\apjs] {10.3847/1538-4365/aae9f0},
  \href {https://ui.adsabs.harvard.edu/abs/2018ApJS..239...18A} {239, 18}

\bibitem[\protect\citeauthoryear{{Adame} et~al.,}{{Adame}
  et~al.}{2024}]{DESI2016}
{Adame} A.~G.,  et~al., 2024, \mn@doi [\aj] {10.3847/1538-3881/ad0b08}, \href
  {https://ui.adsabs.harvard.edu/abs/2024AJ....167...62A} {167, 62}

\bibitem[\protect\citeauthoryear{{Alam} et~al.,}{{Alam}
  et~al.}{2017}]{Alam2017}
{Alam} S.,  et~al., 2017, \mn@doi [\mnras] {10.1093/mnras/stx721}, \href
  {https://ui.adsabs.harvard.edu/abs/2017MNRAS.470.2617A} {470, 2617}

\bibitem[\protect\citeauthoryear{{Alam} et~al.,}{{Alam} et~al.}{2021a}]{eBoss}
{Alam} S.,  et~al., 2021a, \mn@doi [\prd] {10.1103/PhysRevD.103.083533}, \href
  {https://ui.adsabs.harvard.edu/abs/2021PhRvD.103h3533A} {103, 083533}

\bibitem[\protect\citeauthoryear{{Alam} et~al.,}{{Alam}
  et~al.}{2021b}]{alam2021}
{Alam} S.,  et~al., 2021b, \mn@doi [\prd] {10.1103/PhysRevD.103.083533}, \href
  {https://ui.adsabs.harvard.edu/abs/2021PhRvD.103h3533A} {103, 083533}

\bibitem[\protect\citeauthoryear{{Alam}, {Paranjape}  \& {Peacock}}{{Alam}
  et~al.}{2023}]{2023arXiv230501266A}
{Alam} S.,  {Paranjape} A.,   {Peacock} J.~A.,  2023, \mn@doi [arXiv e-prints]
  {10.48550/arXiv.2305.01266}, \href
  {https://ui.adsabs.harvard.edu/abs/2023arXiv230501266A} {p. arXiv:2305.01266}

\bibitem[\protect\citeauthoryear{{Alonso}}{{Alonso}}{2012}]{Alonso2012}
{Alonso} D.,  2012, \mn@doi [arXiv e-prints] {10.48550/arXiv.1210.1833}, \href
  {https://ui.adsabs.harvard.edu/abs/2012arXiv1210.1833A} {p. arXiv:1210.1833}

\bibitem[\protect\citeauthoryear{{Amendola} et~al.,}{{Amendola}
  et~al.}{2018}]{Amendola}
{Amendola} L.,  et~al., 2018, \mn@doi [Living Reviews in Relativity]
  {10.1007/s41114-017-0010-3}, \href
  {https://ui.adsabs.harvard.edu/abs/2018LRR....21....2A} {21, 2}

\bibitem[\protect\citeauthoryear{{Angulo} \& {Pontzen}}{{Angulo} \&
  {Pontzen}}{2016}]{2016Angulo}
{Angulo} R.~E.,  {Pontzen} A.,  2016, \mn@doi [\mnras] {10.1093/mnrasl/slw098},
  \href {https://ui.adsabs.harvard.edu/abs/2016MNRAS.462L...1A} {462, L1}

\bibitem[\protect\citeauthoryear{{Aric{\`o}}, {Angulo},
  {Hern{\'a}ndez-Monteagudo}, {Contreras}, {Zennaro}, {Pellejero-Iba{\~n}ez}
  \& {Rosas-Guevara}}{{Aric{\`o}} et~al.}{2020}]{baryons_arico2020}
{Aric{\`o}} G.,  {Angulo} R.~E.,  {Hern{\'a}ndez-Monteagudo} C.,  {Contreras}
  S.,  {Zennaro} M.,  {Pellejero-Iba{\~n}ez} M.,   {Rosas-Guevara} Y.,  2020,
  \mn@doi [\mnras] {10.1093/mnras/staa1478}, \href
  {https://ui.adsabs.harvard.edu/abs/2020MNRAS.495.4800A} {495, 4800}

\bibitem[\protect\citeauthoryear{{Avila} et~al.,}{{Avila}
  et~al.}{2018}]{S.Avila2018}
{Avila} S.,  et~al., 2018, \mn@doi [\mnras] {10.1093/mnras/sty1389}, \href
  {https://ui.adsabs.harvard.edu/abs/2018MNRAS.479...94A} {479, 94}

\bibitem[\protect\citeauthoryear{{Avila} et~al.,}{{Avila}
  et~al.}{2020}]{avila2020}
{Avila} S.,  et~al., 2020, \mn@doi [\mnras] {10.1093/mnras/staa2951}, \href
  {https://ui.adsabs.harvard.edu/abs/2020MNRAS.499.5486A} {499, 5486}

\bibitem[\protect\citeauthoryear{{Avila}, {Vos-Gin{\'e}s}, {Cunnington},
  {Stevens}, {Yepes}, {Knebe}  \& {Chuang}}{{Avila} et~al.}{2022}]{Avila22}
{Avila} S.,  {Vos-Gin{\'e}s} B.,  {Cunnington} S.,  {Stevens} A. R.~H.,
  {Yepes} G.,  {Knebe} A.,   {Chuang} C.-H.,  2022, \mn@doi [\mnras]
  {10.1093/mnras/stab3406}, \href
  {https://ui.adsabs.harvard.edu/abs/2022MNRAS.510..292A} {510, 292}

\bibitem[\protect\citeauthoryear{{Ayromlou}, {Kauffmann}, {Anand}  \&
  {White}}{{Ayromlou} et~al.}{2023}]{ayromlou2023}
{Ayromlou} M.,  {Kauffmann} G.,  {Anand} A.,   {White} S. D.~M.,  2023, \mn@doi
  [\mnras] {10.1093/mnras/stac3637}, \href
  {https://ui.adsabs.harvard.edu/abs/2023MNRAS.519.1913A} {519, 1913}

\bibitem[\protect\citeauthoryear{{Baugh}}{{Baugh}}{2006}]{Baugh2006}
{Baugh} C.~M.,  2006, \mn@doi [Reports on Progress in Physics]
  {10.1088/0034-4885/69/12/R02}, \href
  {https://ui.adsabs.harvard.edu/abs/2006RPPh...69.3101B} {69, 3101}

\bibitem[\protect\citeauthoryear{{Behroozi}, {Wechsler}  \& {Wu}}{{Behroozi}
  et~al.}{2013a}]{Bezi}
{Behroozi} P.~S.,  {Wechsler} R.~H.,   {Wu} H.-Y.,  2013a, \mn@doi [\apj]
  {10.1088/0004-637X/762/2/109}, \href
  {https://ui.adsabs.harvard.edu/abs/2013ApJ...762..109B} {762, 109}

\bibitem[\protect\citeauthoryear{{Behroozi}, {Wechsler}, {Wu}, {Busha},
  {Klypin}  \& {Primack}}{{Behroozi} et~al.}{2013b}]{MergerTrees}
{Behroozi} P.~S.,  {Wechsler} R.~H.,  {Wu} H.-Y.,  {Busha} M.~T.,  {Klypin}
  A.~A.,   {Primack} J.~R.,  2013b, \mn@doi [\apj]
  {10.1088/0004-637X/763/1/18}, \href
  {https://ui.adsabs.harvard.edu/abs/2013ApJ...763...18B} {763, 18}

\bibitem[\protect\citeauthoryear{{Benson}, {Cole}, {Frenk}, {Baugh}  \&
  {Lacey}}{{Benson} et~al.}{2000}]{benson2000}
{Benson} A.~J.,  {Cole} S.,  {Frenk} C.~S.,  {Baugh} C.~M.,   {Lacey} C.~G.,
  2000, \mn@doi [\mnras] {10.1046/j.1365-8711.2000.03101.x}, \href
  {https://ui.adsabs.harvard.edu/abs/2000MNRAS.311..793B} {311, 793}

\bibitem[\protect\citeauthoryear{{Berlind} et~al.,}{{Berlind}
  et~al.}{2003}]{berlind2003}
{Berlind} A.~A.,  et~al., 2003, \mn@doi [\apj] {10.1086/376517}, \href
  {https://ui.adsabs.harvard.edu/abs/2003ApJ...593....1B} {593, 1}

\bibitem[\protect\citeauthoryear{{Bryan} \& {Norman}}{{Bryan} \&
  {Norman}}{1998}]{Bryan}
{Bryan} G.~L.,  {Norman} M.~L.,  1998, \mn@doi [\apj] {10.1086/305262}, \href
  {https://ui.adsabs.harvard.edu/abs/1998ApJ...495...80B} {495, 80}

\bibitem[\protect\citeauthoryear{{Cardelli}, {Clayton}  \& {Mathis}}{{Cardelli}
  et~al.}{1989}]{Cardelli1989}
{Cardelli} J.~A.,  {Clayton} G.~C.,   {Mathis} J.~S.,  1989, \mn@doi [\apj]
  {10.1086/167900}, \href
  {https://ui.adsabs.harvard.edu/abs/1989ApJ...345..245C} {345, 245}

\bibitem[\protect\citeauthoryear{{Carretero}, {Castander}, {Gazta{\~n}aga},
  {Crocce}  \& {Fosalba}}{{Carretero} et~al.}{2015}]{Carretero}
{Carretero} J.,  {Castander} F.~J.,  {Gazta{\~n}aga} E.,  {Crocce} M.,
  {Fosalba} P.,  2015, \mn@doi [\mnras] {10.1093/mnras/stu2402}, \href
  {https://ui.adsabs.harvard.edu/abs/2015MNRAS.447..646C} {447, 646}

\bibitem[\protect\citeauthoryear{{Chaves-Montero}, {Angulo}  \&
  {Contreras}}{{Chaves-Montero} et~al.}{2023}]{chaves2023}
{Chaves-Montero} J.,  {Angulo} R.~E.,   {Contreras} S.,  2023, \mn@doi [\mnras]
  {10.1093/mnras/stad243}, \href
  {https://ui.adsabs.harvard.edu/abs/2023MNRAS.521..937C} {521, 937}

\bibitem[\protect\citeauthoryear{{Christodoulou} et~al.,}{{Christodoulou}
  et~al.}{2012}]{christodoulou2012}
{Christodoulou} L.,  et~al., 2012, \mn@doi [\mnras]
  {10.1111/j.1365-2966.2012.21434.x}, \href
  {https://ui.adsabs.harvard.edu/abs/2012MNRAS.425.1527C} {425, 1527}

\bibitem[\protect\citeauthoryear{{Chuang} et~al.,}{{Chuang}
  et~al.}{2019}]{2019Unitsims}
{Chuang} C.-H.,  et~al., 2019, \mn@doi [\mnras] {10.1093/mnras/stz1233}, \href
  {https://ui.adsabs.harvard.edu/abs/2019MNRAS.487...48C} {487, 48}

\bibitem[\protect\citeauthoryear{{Cochrane}, {Best}, {Sobral}, {Smail}, {Wake},
  {Stott}  \& {Geach}}{{Cochrane} et~al.}{2017}]{Cochrane}
{Cochrane} R.~K.,  {Best} P.~N.,  {Sobral} D.,  {Smail} I.,  {Wake} D.~A.,
  {Stott} J.~P.,   {Geach} J.~E.,  2017, \mn@doi [\mnras]
  {10.1093/mnras/stx957}, \href
  {https://ui.adsabs.harvard.edu/abs/2017MNRAS.469.2913C} {469, 2913}

\bibitem[\protect\citeauthoryear{{Cole}, {Lacey}, {Baugh}  \& {Frenk}}{{Cole}
  et~al.}{2000}]{cole00}
{Cole} S.,  {Lacey} C.~G.,  {Baugh} C.~M.,   {Frenk} C.~S.,  2000, \mnras, 319,
  168

\bibitem[\protect\citeauthoryear{{Cole} et~al.,}{{Cole} et~al.}{2005}]{2dFGRS}
{Cole} S.,  et~al., 2005, \mn@doi [\mnras] {10.1111/j.1365-2966.2005.09318.x},
  \href {https://ui.adsabs.harvard.edu/abs/2005MNRAS.362..505C} {362, 505}

\bibitem[\protect\citeauthoryear{{Contreras}, {Chaves-Montero}, {Zennaro}  \&
  {Angulo}}{{Contreras} et~al.}{2021}]{Cosmologicalassemblybias}
{Contreras} S.,  {Chaves-Montero} J.,  {Zennaro} M.,   {Angulo} R.~E.,  2021,
  \mn@doi [\mnras] {10.1093/mnras/stab2367}, \href
  {https://ui.adsabs.harvard.edu/abs/2021MNRAS.507.3412C} {507, 3412}

\bibitem[\protect\citeauthoryear{{Cora} et~al.,}{{Cora}
  et~al.}{2018}]{cora2018}
{Cora} S.~A.,  et~al., 2018, \mn@doi [\mnras] {10.1093/mnras/sty1131}, \href
  {https://ui.adsabs.harvard.edu/abs/2018MNRAS.479....2C} {479, 2}

\bibitem[\protect\citeauthoryear{{Croton}, {Gao}  \& {White}}{{Croton}
  et~al.}{2007}]{Shuffled}
{Croton} D.~J.,  {Gao} L.,   {White} S. D.~M.,  2007, \mn@doi [\mnras]
  {10.1111/j.1365-2966.2006.11230.x}, \href
  {https://ui.adsabs.harvard.edu/abs/2007MNRAS.374.1303C} {374, 1303}

\bibitem[\protect\citeauthoryear{Croton et~al.,}{Croton
  et~al.}{2016}]{Croton_2016}
Croton D.~J.,  et~al., 2016, \mn@doi [The Astrophysical Journal Supplement
  Series] {10.3847/0067-0049/222/2/22}, 222, 22

\bibitem[\protect\citeauthoryear{{Dawson} et~al.,}{{Dawson}
  et~al.}{2013}]{Dawson2012}
{Dawson} K.~S.,  et~al., 2013, \mn@doi [\aj] {10.1088/0004-6256/145/1/10},
  \href {https://ui.adsabs.harvard.edu/abs/2013AJ....145...10D} {145, 10}

\bibitem[\protect\citeauthoryear{{Dawson} et~al.,}{{Dawson}
  et~al.}{2016}]{Dawson2016}
{Dawson} K.~S.,  et~al., 2016, \mn@doi [\aj] {10.3847/0004-6256/151/2/44},
  \href {https://ui.adsabs.harvard.edu/abs/2016AJ....151...44D} {151, 44}

\bibitem[\protect\citeauthoryear{{Drinkwater} et~al.,}{{Drinkwater}
  et~al.}{2010}]{Drinkwater}
{Drinkwater} M.~J.,  et~al., 2010, \mn@doi [\mnras]
  {10.1111/j.1365-2966.2009.15754.x}, \href
  {https://ui.adsabs.harvard.edu/abs/2010MNRAS.401.1429D} {401, 1429}

\bibitem[\protect\citeauthoryear{{Einasto}}{{Einasto}}{1969}]{Einasto}
{Einasto} J.,  1969, \mn@doi [Astronomische Nachrichten]
  {10.1002/asna.19682910303}, \href
  {https://ui.adsabs.harvard.edu/abs/1969AN....291...97E} {291, 97}

\bibitem[\protect\citeauthoryear{{Eisenstein} et~al.,}{{Eisenstein}
  et~al.}{2005}]{Eisenstein}
{Eisenstein} D.~J.,  et~al., 2005, \mn@doi [\apj] {10.1086/466512}, \href
  {https://ui.adsabs.harvard.edu/abs/2005ApJ...633..560E} {633, 560}

\bibitem[\protect\citeauthoryear{{Favole} et~al.,}{{Favole}
  et~al.}{2020}]{favole2020}
{Favole} G.,  et~al., 2020, \mn@doi [\mnras] {10.1093/mnras/staa2292}, \href
  {https://ui.adsabs.harvard.edu/abs/2020MNRAS.497.5432F} {497, 5432}

\bibitem[\protect\citeauthoryear{{Favole} et~al.,}{{Favole}
  et~al.}{2024}]{favole2023}
{Favole} G.,  et~al., 2024, \mn@doi [\aap] {10.1051/0004-6361/202346443}, \href
  {https://ui.adsabs.harvard.edu/abs/2024A&A...683A..46F} {683, A46}

\bibitem[\protect\citeauthoryear{{Gao} et~al.,}{{Gao}
  et~al.}{2023}]{desi_conformity}
{Gao} H.,  et~al., 2023, \mn@doi [arXiv e-prints] {10.48550/arXiv.2309.03802},
  \href {https://ui.adsabs.harvard.edu/abs/2023arXiv230903802G} {p.
  arXiv:2309.03802}

\bibitem[\protect\citeauthoryear{{Gonzalez-Perez} et~al.,}{{Gonzalez-Perez}
  et~al.}{2018}]{Violeta2018}
{Gonzalez-Perez} V.,  et~al., 2018, \mn@doi [\mnras] {10.1093/mnras/stx2807},
  \href {https://ui.adsabs.harvard.edu/abs/2018MNRAS.474.4024G} {474, 4024}

\bibitem[\protect\citeauthoryear{{Gonzalez-Perez} et~al.,}{{Gonzalez-Perez}
  et~al.}{2020}]{VGP2020}
{Gonzalez-Perez} V.,  et~al., 2020, \mn@doi [\mnras] {10.1093/mnras/staa2504},
  \href {https://ui.adsabs.harvard.edu/abs/2020MNRAS.498.1852G} {498, 1852}

\bibitem[\protect\citeauthoryear{{Hadzhiyska} et~al.,}{{Hadzhiyska}
  et~al.}{2023}]{Hadzhiyska2023}
{Hadzhiyska} B.,  et~al., 2023, \mn@doi [\mnras] {10.1093/mnras/stad279}, \href
  {https://ui.adsabs.harvard.edu/abs/2023MNRAS.524.2524H} {524, 2524}

\bibitem[\protect\citeauthoryear{{Hirschmann}, {De Lucia}  \&
  {Fontanot}}{{Hirschmann} et~al.}{2016}]{Hirschmann2016}
{Hirschmann} M.,  {De Lucia} G.,   {Fontanot} F.,  2016, \mn@doi [\mnras]
  {10.1093/mnras/stw1318}, \href
  {https://ui.adsabs.harvard.edu/abs/2016MNRAS.461.1760H} {461, 1760}

\bibitem[\protect\citeauthoryear{{Jim{\'e}nez}, {Contreras}, {Padilla},
  {Zehavi}, {Baugh}  \& {Gonzalez-Perez}}{{Jim{\'e}nez}
  et~al.}{2019}]{Jimenez2019}
{Jim{\'e}nez} E.,  {Contreras} S.,  {Padilla} N.,  {Zehavi} I.,  {Baugh} C.~M.,
    {Gonzalez-Perez} V.,  2019, \mn@doi [\mnras] {10.1093/mnras/stz2790}, \href
  {https://ui.adsabs.harvard.edu/abs/2019MNRAS.490.3532J} {490, 3532}

\bibitem[\protect\citeauthoryear{{Jim{\'e}nez}, {Padilla}, {Contreras},
  {Zehavi}, {Baugh}  \& {Orsi}}{{Jim{\'e}nez} et~al.}{2021}]{jimenez2021}
{Jim{\'e}nez} E.,  {Padilla} N.,  {Contreras} S.,  {Zehavi} I.,  {Baugh} C.~M.,
    {Orsi} {\'A}.,  2021, \mn@doi [\mnras] {10.1093/mnras/stab1819}, \href
  {https://ui.adsabs.harvard.edu/abs/2021MNRAS.506.3155J} {506, 3155}

\bibitem[\protect\citeauthoryear{{Knebe} et~al.,}{{Knebe}
  et~al.}{2022}]{Knebe2022}
{Knebe} A.,  et~al., 2022, \mn@doi [\mnras] {10.1093/mnras/stac006}, \href
  {https://ui.adsabs.harvard.edu/abs/2022MNRAS.510.5392K} {510, 5392}

\bibitem[\protect\citeauthoryear{{Lacerna} \& {Padilla}}{{Lacerna} \&
  {Padilla}}{2011}]{Natureassembly}
{Lacerna} I.,  {Padilla} N.,  2011, \mn@doi [\mnras]
  {10.1111/j.1365-2966.2010.17988.x}, \href
  {https://ui.adsabs.harvard.edu/abs/2011MNRAS.412.1283L} {412, 1283}

\bibitem[\protect\citeauthoryear{{Lacerna}, {Contreras}, {Gonz{\'a}lez},
  {Padilla}  \& {Gonzalez-Perez}}{{Lacerna} et~al.}{2018}]{Lacerna}
{Lacerna} I.,  {Contreras} S.,  {Gonz{\'a}lez} R.~E.,  {Padilla} N.,
  {Gonzalez-Perez} V.,  2018, \mn@doi [\mnras] {10.1093/mnras/stx3253}, \href
  {https://ui.adsabs.harvard.edu/abs/2018MNRAS.475.1177L} {475, 1177}

\bibitem[\protect\citeauthoryear{{Lagos} et~al.,}{{Lagos}
  et~al.}{2019}]{lagos2019}
{Lagos} C. d.~P.,  et~al., 2019, \mn@doi [\mnras] {10.1093/mnras/stz2427},
  \href {https://ui.adsabs.harvard.edu/abs/2019MNRAS.489.4196L} {489, 4196}

\bibitem[\protect\citeauthoryear{{Laureijs} et~al.,}{{Laureijs}
  et~al.}{2011}]{Laureijs}
{Laureijs} R.,  et~al., 2011, \mn@doi [arXiv e-prints]
  {10.48550/arXiv.1110.3193}, \href
  {https://ui.adsabs.harvard.edu/abs/2011arXiv1110.3193L} {p. arXiv:1110.3193}

\bibitem[\protect\citeauthoryear{{Manera} et~al.,}{{Manera}
  et~al.}{2013}]{Manera}
{Manera} M.,  et~al., 2013, \mn@doi [\mnras] {10.1093/mnras/sts084}, \href
  {https://ui.adsabs.harvard.edu/abs/2013MNRAS.428.1036M} {428, 1036}

\bibitem[\protect\citeauthoryear{{McCarthy}, {Schaye}, {Bird}  \& {Le
  Brun}}{{McCarthy} et~al.}{2017}]{Bahamas}
{McCarthy} I.~G.,  {Schaye} J.,  {Bird} S.,   {Le Brun} A. M.~C.,  2017,
  \mn@doi [\mnras] {10.1093/mnras/stw2792}, \href
  {https://ui.adsabs.harvard.edu/abs/2017MNRAS.465.2936M} {465, 2936}

\bibitem[\protect\citeauthoryear{{Merson}, {Wang}, {Benson}, {Faisst},
  {Masters}, {Kiessling}  \& {Rhodes}}{{Merson} et~al.}{2018}]{merson2018}
{Merson} A.,  {Wang} Y.,  {Benson} A.,  {Faisst} A.,  {Masters} D.,
  {Kiessling} A.,   {Rhodes} J.,  2018, \mn@doi [\mnras]
  {10.1093/mnras/stx2649}, \href
  {https://ui.adsabs.harvard.edu/abs/2018MNRAS.474..177M} {474, 177}

\bibitem[\protect\citeauthoryear{{Navarro}, {Frenk}  \& {White}}{{Navarro}
  et~al.}{1997}]{NFW}
{Navarro} J.~F.,  {Frenk} C.~S.,   {White} S. D.~M.,  1997, \mn@doi [\apj]
  {10.1086/304888}, \href
  {https://ui.adsabs.harvard.edu/abs/1997ApJ...490..493N} {490, 493}

\bibitem[\protect\citeauthoryear{{Obuljen}, {Percival}  \& {Dalal}}{{Obuljen}
  et~al.}{2020}]{2020JCAP...10..058O}
{Obuljen} A.,  {Percival} W.~J.,   {Dalal} N.,  2020, \mn@doi [\jcap]
  {10.1088/1475-7516/2020/10/058}, \href
  {https://ui.adsabs.harvard.edu/abs/2020JCAP...10..058O} {2020, 058}

\bibitem[\protect\citeauthoryear{Orsi \& Angulo}{Orsi \&
  Angulo}{2018}]{Orsi_2018}
Orsi {\'{A} }.~A.,  Angulo R.~E.,  2018, \mn@doi [Monthly Notices of the Royal
  Astronomical Society] {10.1093/mnras/stx3349}, 475, 2530

\bibitem[\protect\citeauthoryear{Orsi, Padilla, Groves, Cora, Tecce, Gargiulo
  \& Ruiz}{Orsi et~al.}{2014}]{orsi_2014}
Orsi {\'{A} }.,  Padilla N.,  Groves B.,  Cora S.,  Tecce T.,  Gargiulo I.,
  Ruiz A.,  2014, \mn@doi [Monthly Notices of the Royal Astronomical Society]
  {10.1093/mnras/stu1203}, 443, 799

\bibitem[\protect\citeauthoryear{{Parkinson} et~al.,}{{Parkinson}
  et~al.}{2012}]{Parkinson}
{Parkinson} D.,  et~al., 2012, \mn@doi [\prd] {10.1103/PhysRevD.86.103518},
  \href {https://ui.adsabs.harvard.edu/abs/2012PhRvD..86j3518P} {86, 103518}

\bibitem[\protect\citeauthoryear{{Pillepich} et~al.,}{{Pillepich}
  et~al.}{2018}]{TNG_clustering}
{Pillepich} A.,  et~al., 2018, \mn@doi [\mnras] {10.1093/mnras/stx2656}, \href
  {https://ui.adsabs.harvard.edu/abs/2018MNRAS.473.4077P} {473, 4077}

\bibitem[\protect\citeauthoryear{{Pozzetti} et~al.,}{{Pozzetti}
  et~al.}{2016}]{Pozzetti}
{Pozzetti} L.,  et~al., 2016, \mn@doi [\aap] {10.1051/0004-6361/201527081},
  \href {https://ui.adsabs.harvard.edu/abs/2016A&A...590A...3P} {590, A3}

\bibitem[\protect\citeauthoryear{{Qin}, {Parkinson}, {Stevens}  \&
  {Howlett}}{{Qin} et~al.}{2023}]{Parkinson2}
{Qin} F.,  {Parkinson} D.,  {Stevens} A. R.~H.,   {Howlett} C.,  2023, \mn@doi
  [arXiv e-prints] {10.48550/arXiv.2308.03298}, \href
  {https://ui.adsabs.harvard.edu/abs/2023arXiv230803298Q} {p. arXiv:2308.03298}

\bibitem[\protect\citeauthoryear{{Ramakrishnan}, {Paranjape}, {Hahn}  \&
  {Sheth}}{{Ramakrishnan} et~al.}{2019}]{sujatha2019}
{Ramakrishnan} S.,  {Paranjape} A.,  {Hahn} O.,   {Sheth} R.~K.,  2019, \mn@doi
  [\mnras] {10.1093/mnras/stz2344}, \href
  {https://ui.adsabs.harvard.edu/abs/2019MNRAS.489.2977R} {489, 2977}

\bibitem[\protect\citeauthoryear{{Rocher} et~al.,}{{Rocher}
  et~al.}{2023a}]{Rocher2023}
{Rocher} A.,  et~al., 2023a, \mn@doi [arXiv e-prints]
  {10.48550/arXiv.2306.06319}, \href
  {https://ui.adsabs.harvard.edu/abs/2023arXiv230606319R} {p. arXiv:2306.06319}

\bibitem[\protect\citeauthoryear{{Rocher} et~al.,}{{Rocher}
  et~al.}{2023b}]{DESIonehaloterm}
{Rocher} A.,  et~al., 2023b, \mn@doi [arXiv e-prints]
  {10.48550/arXiv.2306.06319}, \href
  {https://ui.adsabs.harvard.edu/abs/2023arXiv230606319R} {p. arXiv:2306.06319}

\bibitem[\protect\citeauthoryear{{Salvador-Sol{\'e}}, {Manrique}, {Canales}  \&
  {Botella}}{{Salvador-Sol{\'e}} et~al.}{2023}]{Salvador}
{Salvador-Sol{\'e}} E.,  {Manrique} A.,  {Canales} D.,   {Botella} I.,  2023,
  \mn@doi [\mnras] {10.1093/mnras/stad642}, \href
  {https://ui.adsabs.harvard.edu/abs/2023MNRAS.521.1988S} {521, 1988}

\bibitem[\protect\citeauthoryear{{Schaye} et~al.,}{{Schaye}
  et~al.}{2015}]{schaye2015}
{Schaye} J.,  et~al., 2015, \mn@doi [\mnras] {10.1093/mnras/stu2058}, \href
  {https://ui.adsabs.harvard.edu/abs/2015MNRAS.446..521S} {446, 521}

\bibitem[\protect\citeauthoryear{{Schneider} \& {Teyssier}}{{Schneider} \&
  {Teyssier}}{2015}]{baryons_schneider}
{Schneider} A.,  {Teyssier} R.,  2015, \mn@doi [\jcap]
  {10.1088/1475-7516/2015/12/049}, \href
  {https://ui.adsabs.harvard.edu/abs/2015JCAP...12..049S} {2015, 049}

\bibitem[\protect\citeauthoryear{{Somerville} \& {Dav{\'e}}}{{Somerville} \&
  {Dav{\'e}}}{2015}]{somerville15}
{Somerville} R.~S.,  {Dav{\'e}} R.,  2015, \mn@doi [\araa]
  {10.1146/annurev-astro-082812-140951}, 53, 51

\bibitem[\protect\citeauthoryear{{Spergel} et~al.,}{{Spergel}
  et~al.}{2013}]{Spergel2013}
{Spergel} D.,  et~al., 2013, \mn@doi [arXiv e-prints]
  {10.48550/arXiv.1305.5422}, \href
  {https://ui.adsabs.harvard.edu/abs/2013arXiv1305.5422S} {p. arXiv:1305.5422}

\bibitem[\protect\citeauthoryear{{Spergel} et~al.,}{{Spergel}
  et~al.}{2015}]{Spergel2015}
{Spergel} D.,  et~al., 2015, \mn@doi [arXiv e-prints]
  {10.48550/arXiv.1503.03757}, \href
  {https://ui.adsabs.harvard.edu/abs/2015arXiv150303757S} {p. arXiv:1503.03757}

\bibitem[\protect\citeauthoryear{{Springel}}{{Springel}}{2005}]{Gat}
{Springel} V.,  2005, \mn@doi [\mnras] {10.1111/j.1365-2966.2005.09655.x},
  \href {https://ui.adsabs.harvard.edu/abs/2005MNRAS.364.1105S} {364, 1105}

\bibitem[\protect\citeauthoryear{{Springel} et~al.,}{{Springel}
  et~al.}{2018}]{TNG_method_2}
{Springel} V.,  et~al., 2018, \mn@doi [\mnras] {10.1093/mnras/stx3304}, \href
  {https://ui.adsabs.harvard.edu/abs/2018MNRAS.475..676S} {475, 676}

\bibitem[\protect\citeauthoryear{{The Dark Energy Survey Collaboration}}{{The
  Dark Energy Survey Collaboration}}{2005}]{DES}
{The Dark Energy Survey Collaboration} 2005, \mn@doi [arXiv e-prints]
  {10.48550/arXiv.astro-ph/0510346}, \href
  {https://ui.adsabs.harvard.edu/abs/2005astro.ph.10346T} {pp
  astro--ph/0510346}

\bibitem[\protect\citeauthoryear{{Vos-Gin{\'e}s}, {Avila}, {Gonzalez-Perez}  \&
  {Yepes}}{{Vos-Gin{\'e}s} et~al.}{2023}]{VosGines24}
{Vos-Gin{\'e}s} B.,  {Avila} S.,  {Gonzalez-Perez} V.,   {Yepes} G.,  2023,
  \mn@doi [arXiv e-prints] {10.48550/arXiv.2310.18189}, \href
  {https://ui.adsabs.harvard.edu/abs/2023arXiv231018189V} {p. arXiv:2310.18189}

\bibitem[\protect\citeauthoryear{{Wang}, {Mao}, {Zentner}, {Guo}, {Lange}, {van
  den Bosch}  \& {Mezini}}{{Wang} et~al.}{2022}]{2022MNRAS.516.4003W}
{Wang} K.,  {Mao} Y.-Y.,  {Zentner} A.~R.,  {Guo} H.,  {Lange} J.~U.,  {van den
  Bosch} F.~C.,   {Mezini} L.,  2022, \mn@doi [\mnras]
  {10.1093/mnras/stac2465}, \href
  {https://ui.adsabs.harvard.edu/abs/2022MNRAS.516.4003W} {516, 4003}

\bibitem[\protect\citeauthoryear{{Wechsler} \& {Tinker}}{{Wechsler} \&
  {Tinker}}{2018}]{Wechsler2018}
{Wechsler} R.~H.,  {Tinker} J.~L.,  2018, \mn@doi [\araa]
  {10.1146/annurev-astro-081817-051756}, \href
  {https://ui.adsabs.harvard.edu/abs/2018ARA&A..56..435W} {56, 435}

\bibitem[\protect\citeauthoryear{{Weinmann}, {van den Bosch}, {Yang}  \&
  {Mo}}{{Weinmann} et~al.}{2006}]{Weinmann}
{Weinmann} S.~M.,  {van den Bosch} F.~C.,  {Yang} X.,   {Mo} H.~J.,  2006,
  \mn@doi [\mnras] {10.1111/j.1365-2966.2005.09865.x}, \href
  {https://ui.adsabs.harvard.edu/abs/2006MNRAS.366....2W} {366, 2}

\bibitem[\protect\citeauthoryear{{Xu}, {Zehavi}  \& {Contreras}}{{Xu}
  et~al.}{2021}]{GAB}
{Xu} X.,  {Zehavi} I.,   {Contreras} S.,  2021, \mn@doi [\mnras]
  {10.1093/mnras/stab100}, \href
  {https://ui.adsabs.harvard.edu/abs/2021MNRAS.502.3242X} {502, 3242}

\bibitem[\protect\citeauthoryear{{York} et~al.,}{{York} et~al.}{2000}]{York}
{York} D.~G.,  et~al., 2000, \mn@doi [\aj] {10.1086/301513}, \href
  {https://ui.adsabs.harvard.edu/abs/2000AJ....120.1579Y} {120, 1579}

\bibitem[\protect\citeauthoryear{{Yuan}, {Hadzhiyska}, {Bose}, {Eisenstein}  \&
  {Guo}}{{Yuan} et~al.}{2021}]{2021MNRAS.502.3582Y}
{Yuan} S.,  {Hadzhiyska} B.,  {Bose} S.,  {Eisenstein} D.~J.,   {Guo} H.,
  2021, \mn@doi [\mnras] {10.1093/mnras/stab235}, \href
  {https://ui.adsabs.harvard.edu/abs/2021MNRAS.502.3582Y} {502, 3582}

\bibitem[\protect\citeauthoryear{{Yuan} et~al.,}{{Yuan}
  et~al.}{2023a}]{Yuan2023}
{Yuan} S.,  et~al., 2023a, \mn@doi [arXiv e-prints]
  {10.48550/arXiv.2306.06314}, \href
  {https://ui.adsabs.harvard.edu/abs/2023arXiv230606314Y} {p. arXiv:2306.06314}

\bibitem[\protect\citeauthoryear{{Yuan} et~al.,}{{Yuan}
  et~al.}{2023b}]{yuan23.conformity}
{Yuan} S.,  et~al., 2023b, \mn@doi [arXiv e-prints]
  {10.48550/arXiv.2310.09329}, \href
  {https://ui.adsabs.harvard.edu/abs/2023arXiv231009329Y} {p. arXiv:2310.09329}

\bibitem[\protect\citeauthoryear{{Zehavi} et~al.,}{{Zehavi}
  et~al.}{2005}]{zehavi2005}
{Zehavi} I.,  et~al., 2005, \mn@doi [\apj] {10.1086/431891}, \href
  {https://ui.adsabs.harvard.edu/abs/2005ApJ...630....1Z} {630, 1}

\bibitem[\protect\citeauthoryear{{Zhai}, {Benson}, {Wang}, {Yepes}  \&
  {Chuang}}{{Zhai} et~al.}{2019}]{zhai2019}
{Zhai} Z.,  {Benson} A.,  {Wang} Y.,  {Yepes} G.,   {Chuang} C.-H.,  2019,
  \mn@doi [\mnras] {10.1093/mnras/stz2844}, \href
  {https://ui.adsabs.harvard.edu/abs/2019MNRAS.490.3667Z} {490, 3667}

\bibitem[\protect\citeauthoryear{{Zhai}, {Wang}, {Benson}, {Colbert}, {Bagley},
  {Henry}  \& {Baronchelli}}{{Zhai} et~al.}{2021a}]{zhai2021}
{Zhai} Z.,  {Wang} Y.,  {Benson} A.,  {Colbert} J.,  {Bagley} M.,  {Henry} A.,
   {Baronchelli} I.,  2021a, \mn@doi [arXiv e-prints]
  {10.48550/arXiv.2109.12216}, \href
  {https://ui.adsabs.harvard.edu/abs/2021arXiv210912216Z} {p. arXiv:2109.12216}

\bibitem[\protect\citeauthoryear{{Zhai}, {Wang}, {Benson}, {Chuang}  \&
  {Yepes}}{{Zhai} et~al.}{2021b}]{zhai_hod}
{Zhai} Z.,  {Wang} Y.,  {Benson} A.,  {Chuang} C.-H.,   {Yepes} G.,  2021b,
  \mn@doi [\mnras] {10.1093/mnras/stab1539}, \href
  {https://ui.adsabs.harvard.edu/abs/2021MNRAS.505.2784Z} {505, 2784}

\bibitem[\protect\citeauthoryear{{Zheng} et~al.,}{{Zheng}
  et~al.}{2005}]{Zheng2005}
{Zheng} Z.,  et~al., 2005, \mn@doi [\apj] {10.1086/466510}, \href
  {https://ui.adsabs.harvard.edu/abs/2005ApJ...633..791Z} {633, 791}

\bibitem[\protect\citeauthoryear{{de Jong} et~al.,}{{de Jong}
  et~al.}{2012}]{DeJong}
{de Jong} R.~S.,  et~al., 2012, in {McLean} I.~S.,  {Ramsay} S.~K.,   {Takami}
  H.,  eds,  Society of Photo-Optical Instrumentation Engineers (SPIE)
  Conference Series Vol. 8446, Ground-based and Airborne Instrumentation for
  Astronomy IV. p. 84460T (\mn@eprint {arXiv} {1206.6885}),
  \mn@doi{10.1117/12.926239}

\makeatother
\end{thebibliography}

\bsp	% typesetting comment
\label{lastpage}
\end{document}